\pdfoutput=1
\documentclass[fleqn,usenatbib,useAMS]{mnras}

\usepackage{graphicx}	
\usepackage{amsmath}	
\usepackage{amssymb}	
\usepackage{multicol}        
\usepackage{bm}		
\usepackage{pdflscape}	
\usepackage[capitalise,noabbrev]{cleveref}

\usepackage{float} 
\usepackage{xspace}
\usepackage{multirow}
\usepackage{multicol}
\usepackage{threeparttable}
\usepackage{comment}
\usepackage[normalem]{ulem}
\usepackage{caption}
\usepackage{subcaption}
\usepackage{verbatim} 
\usepackage{xcolor}
\graphicspath{{./figures/}}

\newcommand{\planck}{\emph{Planck}\xspace}

\newcommand{\soliket}{\texttt{SOLikeT}\xspace}
\newcommand{\cobaya}{\texttt{Cobaya}\xspace}
\newcommand{\flask}{\texttt{FLASK}\xspace}
\newcommand{\camb}{\texttt{CAMB}\xspace}
\newcommand{\ccl}{\texttt{CCL}\xspace}
\newcommand{\hmcode}{\texttt{HMCODE}\xspace}
\newcommand{\halofit}{\texttt{halofit}\xspace}
\newcommand{\namaster}{\texttt{NAMASTER}\xspace}
\newcommand{\pixell}{\texttt{pixell}\xspace}
\newcommand{\healpix}{\texttt{HEALPix}\xspace}
\newcommand{\metacalibration}{\texttt{METACALIBRATION}\xspace}

\newcommand{\lcdm}{\ensuremath{\Lambda}CDM\xspace}
\newcommand{\nside}{\texttt{Nside}\xspace}
\newcommand{\Omegam}{\ensuremath{\Omega_{\rm m}}\xspace}
\newcommand{\kcmb}{\ensuremath{\kappa_{\rm C}}\xspace}
\newcommand{\gE}{\ensuremath{\gamma_{\rm E}}\xspace}
\newcommand{\clkg}{\ensuremath{{C}^{\kappa_{\rm C} \gamma_{\rm E} }_{\ell}}\xspace}
\newcommand{\barclkg}{\ensuremath{\bar{C}^{\kappa_{\rm C} \gamma_{\rm E} }_{\ell}}\xspace}

\newcommand{\wcmb}{\ensuremath{W^{\rm CMB}_{\kappa}}\xspace}
\newcommand{\wgalaxy}{\ensuremath{W^{\rm g}_{\gamma}}\xspace}

\newcommand{\neff}{\ensuremath{n_{\rm eff}} \xspace}

\newcommand{\lmin}{\ensuremath{\ell_{\rm min}}\xspace}
\newcommand{\lmax}{\ensuremath{\ell_{\rm max}}\xspace}

\usepackage[T1]{fontenc}
\usepackage{ae,aecompl}

\usepackage{eso-pic}
\AddToShipoutPictureBG*{%
  \AtPageUpperLeft{%
    \hspace{0.7\paperwidth}%
    \raisebox{-4.5\baselineskip}{%
      \makebox[0pt][l]{\textnormal{DES-2023-0764}}
}}}%

\AddToShipoutPictureBG*{%
  \AtPageUpperLeft{%
    \hspace{0.7\paperwidth}%
    \raisebox{-5.5\baselineskip}{%
      \makebox[0pt][l]{\textnormal{FERMILAB-PUB-23-432-PPD}}
}}}%

\title[ACT-DR4 lensing $\times$ DES-Y3 shear]{Cosmology from Cross-Correlation of ACT-DR4 CMB Lensing and  DES-Y3 Cosmic Shear}

\author[Shaikh et al.]{S.~Shaikh,$^{1}$\thanks{E-mail: sshaik14@asu.edu}
I.~Harrison,$^{2}$\thanks{E-mail: harrisoni@cardiff.ac.uk}
A.~van Engelen,$^{1}$
G.~A.~Marques,$^{3,4}$
T.~M.~C.~Abbott,$^{5}$
M.~Aguena,$^{6}$\newauthor
O.~Alves,$^{7}$
A.~Amon,$^{8,9}$
R.~An,$^{10}$
D.~Bacon,$^{11}$
% ...
(The ACT and DES Collaborations)
\\
Author affiliations are listed at the end of the paper.
\vspace{-1.4cm}
}
\pubyear{2023}
\begin{document}
\label{firstpage}
\pagerange{\pageref{firstpage}--\pageref{lastpage}}
\maketitle

\begin{abstract}
Cross-correlation between weak lensing of the Cosmic Microwave Background (CMB) and weak lensing of galaxies offers a way to place robust constraints on cosmological and astrophysical parameters with reduced sensitivity to certain systematic effects affecting individual surveys. 
We measure the angular cross-power spectrum between the Atacama Cosmology Telescope (ACT) DR4 CMB lensing and the galaxy weak lensing measured by the Dark Energy Survey (DES) Y3 data.
Our baseline analysis uses the CMB convergence map derived from ACT-DR4 and \emph{Planck} data, where most of the contamination due to the thermal Sunyaev Zel'dovich effect is removed, thus avoiding important systematics in the cross-correlation. In our modelling, we consider the nuisance parameters of the photometric uncertainty, multiplicative shear bias and intrinsic alignment of galaxies. The resulting cross-power spectrum has a signal-to-noise ratio $= 7.1$ and passes a set of null tests. We use it to infer the amplitude of the fluctuations in the matter distribution ($S_8 \equiv \sigma_8 (\Omega_{\rm m}/0.3)^{0.5} = 0.782\pm 0.059$) with informative but well-motivated priors on the nuisance parameters. We also investigate the validity of these priors by significantly relaxing them and checking the consistency of the resulting posteriors, finding them consistent, albeit only with relatively weak constraints. This cross-correlation measurement will improve significantly with the new ACT-DR6 lensing map and form a key component of the joint 6x2pt analysis between DES and ACT.
\end{abstract}




\section{Introduction}
\label{sec:intro}
Observations of the $z \sim 1100$ Cosmic Microwave Background (CMB) and the Large Scale Structure (LSS) at $z \lesssim 3$ give a remarkably consistent picture of the physics and contents of the Universe. Measurements of the primary CMB temperature and polarization anisotropies from \planck 2018 \citep{2020A&A...641A...6P}, ACT Data Release 4 (DR4) \citep{2020JCAP...12..047A} and SPT-3G \citep{2021PhRvD.104b2003D} achieve sub-percent precision on the six main parameters of the spatially flat Lambda Cold Dark Matter (\lcdm) cosmological model. This model allows us to predict several derived parameters, which can be measured using different probes at lower redshifts. One such derived parameter is the matter clustering parameter $\sigma_8$, which describes the amplitude of fluctuations in the over-density of matter on scales of $8\,h^{-1}\,$Mpc. Large photometric and spectroscopic surveys of galaxies have recently begun to place constraints on this parameter comparable in precision to those obtained from CMB predictions. The most recent results from the Dark Energy Survey \citep[DES-Y3,][]{2022PhRvD.105b3520A}, the Kilo-Degree Survey \citep[KiDS-1000,][]{2021A&A...646A.140H} and the Hyper-Suprime Cam survey \citep[HSC-Y3,][]{More:2023knf, Miyatake:2023njf, Sugiyama:2023fzm} all combine galaxy clustering and galaxy weak lensing measurements to infer the value of $\sigma_8$ and the total matter abundance \Omegam, with the best-constrained parameter combination given by $S_8 \equiv \sigma_8 \left( \Omegam / 0.3  \right)^{0.5}$.

As the statistical uncertainty from these two different sets of experiments shrank, a discrepancy emerged: high redshift CMB observations favour a value scattering around $S_8 \approx 0.83$ (ACT-DR4: $0.830\pm0.043$; \planck PR3: $0.834\pm0.016$; SPT-3G 2018: $0.797\pm0.041$), whilst low redshift galaxy and lensing observations appear close to a lower value of $S_8 \approx 0.77$ (DES-Y3: $0.776\pm0.017$; KiDS-1000: $0.766^{+0.020}_{-0.014}$, HSC-Y3: $0.775^{+0.043}_{-0.038}$). This disagreement is marginally statistically significant but remains consistent when comparing different experiments \citep[see][for a review; here, we attempt to include a representative sub-sample of the latest results]{2022JHEAp..34...49A}. This disagreement could be due to unaccounted-for systematics in one (or both) types of experiment or due to a missing piece of physics affecting structure growth at different redshifts and/or physical scales. The prospect of modifications to the current understanding of non-linear structure formation and baryonic feedback contributions is pointed to by \cite*{2023MNRAS.518..477A}, \cite{2022MNRAS.516.5355A}, \cite{2023MNRAS.tmp.2346G} and references therein. A number of other explanations include new dark sector physics, including interacting dark energy and dark matter \citep[e.g.][]{2023PhRvD.107l3538P}, and ultra-light axions \citep[e.g.][]{2023JCAP...06..023R}.

Along with these two principal probes, several other probes are sensitive to an intermediate range of redshifts. Gravitational lensing of the primary CMB is sensitive to a broad range of redshifts and large angular scales. It agrees largely on the value of $S_8$ with the primary CMB itself: the latest ACT results from the newly produced DR6 lensing map \citep{ACT:2023dou, ACT:2023kun, ACT:2023ubw} find $S_8 = 0.840\pm0.028$. Cross-correlations of this CMB lensing signal with galaxy surveys are beginning to be detected at increasing signal-to-noise, hence their ability to provide useful constraints. These cross-correlations are sensitive to lower redshifts and smaller scales compared to the CMB lensing auto-spectrum and generally prefer values of $S_8 < 0.8$, in agreement with the galaxy clustering and weak lensing measurements (e.g. \citealt{2021A&A...649A.146R}: $0.64\pm0.08$, \citealt{2021JCAP...12..028K}: $0.784\pm0.015$, \citealt{2023PhRvD.107b3530C}: $0.74^{+0.034}_{-0.029}$,  \citealt{2023arXiv230617268M}: $0.75^{+0.04}_{-0.05}$.). 

Here, we focus specifically on one of these cross-correlations: the one between CMB lensing ($\kappa_{\rm C}$) and galaxy weak lensing ($\gamma_{\rm E}$), which we will refer to as $\clkg$. To measure this, we use a combination of the ACT-DR4 CMB lensing map \citep{2021MNRAS.500.2250D} and the DES-Y3 galaxy shape catalogue \citep*{2021MNRAS.504.4312G}. The cross-correlation lensing kernel peaks between those of each probe individually (see lower panel of \cref{fig:dndz_wz}) and hence probes somewhat different redshift range than the galaxy weak lensing alone. CMB lensing-galaxy weak lensing cross-correlations are not sensitive to galaxy bias and also provide useful information on the systematics of both probes. Specifically, the extra high-redshift lensing bin from the CMB has long been proposed as a useful way of calibrating multiplicative biases in the difficult measurement of galaxy lensing shear and shift biases in the estimated mean photometric redshift of the galaxy samples \citep[e.g.][]{2013arXiv1311.2338D}.

A number of analyses have already detected this cross-correlation signal \citep{2015PhRvD..91f2001H,2015PhRvD..92f3517L,2016MNRAS.459...21K,2017MNRAS.464.2120S,2016MNRAS.460..434H,2017MNRAS.471.1619H,2019PhRvD.100d3501O, 2020ApJ...904..182M, 2021A&A...649A.146R, 2023PhRvD.107b3530C}. Some of these early works focus the signal-to-noise available from their data onto a single phenomenological parameter $A_{\rm cross}$, which is the amplitude of the cross-correlation power spectrum relative to that predicted by primary CMB data. Note that we denote this parameter by $A_{\rm cross}$ to distinguish it from the parameter measuring the smearing of the peaks in the primary CMB power spectrum, $A_{\rm lens}$, as introduced in \cite{2008PhRvD..77l3531C}. \cite{2021A&A...649A.146R} also explicitly measure $S_8$ jointly with other cosmological and systematics parameters, finding a 1D marginalised constraint of $S_8 = 0.64 \pm 0.08$, which is consistent with low redshift weak lensing only constraints but inconsistent with results derived from high redshift CMB measurements. \cite{2023PhRvD.107b3529O} and \cite{2023PhRvD.107b3530C} measure the real-space equivalents of the \clkg data vector and the CMB lensing-galaxy clustering cross-correlation between SPT and DES-Y3, finding $S_8 = 0.74^{+0.034}_{-0.029}$. They then combine these cross-correlations with the three DES-Y3 data vectors and one SPT lensing data vector for a full `6x2pt' \footnote{So called because it involves six combinations of the two-point correlation functions of CMB lensing $\kappa$, galaxy lensing $\gamma$ and galaxy positions $g$: $\langle \kappa \kappa \rangle, \langle \gamma \gamma \rangle, \langle g g \rangle, \langle \kappa \gamma \rangle, \langle \kappa g \rangle, \langle \gamma g \rangle$.} analysis using information from this wide range of kernels spanning a large range of redshifts, finding $S_8 = 0.792 \pm 0.012$ \citep{2023PhRvD.107b3531A}.

In addition, \cite{2021A&A...649A.146R} and \cite{2020ApJ...904..182M} also assess the consistency of their \clkg only data with the priors on multiplicative shear and redshift calibration biases, which are derived by the weak lensing experiments using a combination of simulations and deep ancillary observational data. For these two types of parameters, there is very little constraining power available from current \clkg data, but the results are indeed consistent with the priors derived without the assistance of the high redshift CMB lensing bin (which is independent of the calibration parameters).

Another physical effect that affects the amplitude of the \clkg signal is the intrinsic alignment of galaxies (IA), which can mimic the alignment caused by the weak lensing cosmic shear signal \citep[for a review see][]{2015PhR...558....1T}. The amplitude of the power spectrum of intrinsic alignments is highly degenerate with the lensing amplitude and forms a contribution to the observed power spectrum of $\mathcal{O}(10 \%)$ \citep{2014MNRAS.443L.119H}. Models for the power spectrum of IAs motivated by galaxy formation physics are relatively uncertain but are expected to have redshift and scale dependencies which help to break this degeneracy \citep[][and references therein]{2020JCAP...01..025V}.

With $450\,$deg$^2$ of overlapping ACT-DR4 and DES-Y3 data, we have the necessary ingredients to perform a full tomographic analysis using the four redshift bins defined by DES-Y3. We include a set of four redshift calibration parameters, four shear calibration parameters, and two parameters describing the IA amplitude and redshift dependence. The current signal-to-noise from the ACT-DR4 and DES-Y3 allows us to put constraints on $S_8$ from $\kappa_{\rm C} \gamma_{\rm E}$ which, although weaker than those of \cite{2023PhRvD.107b3531A} (primarily due to a smaller available overlapping sky area) provides an opportunity to favour or disfavour the somewhat inconsistent values for $S_8$ from \clkg currently found in the literature. Furthermore, we are careful to ensure our methods are adequate for the incoming three-fold increase in constraining power available from the ACT-DR6 lensing map relative to ACT-DR4. The ACT-DR6 lensing map covers most of the DES survey footprint, allowing for a factor of $\approx 9$ increase in the area of overlap between the two surveys, compared to the ACT-DR4 lensing map considered in this work. This will bring our constraining power up to a level comparable to the best current measurements of \clkg from \cite{2023PhRvD.107b3530C}. Our analysis is performed in harmonic space, rather than real space as in that work, and thus has different sensitivity to behaviour at different redshifts and scales, and may thus provide useful verification of earlier results.

We have structured the paper in the following manner:
\begin{itemize}
    \item In \cref{sec:theory}, we describe the theory predicting our observable: the angular cross-power spectrum between CMB weak lensing and galaxy weak lensing.
    \item In \cref{sec:data}, we briefly describe the overall features of the ACT and DES surveys. We discuss the ACT-DR4 lensing map and DES-Y3 cosmic shear catalogue, which we use as inputs to our analysis.
    \item In \cref{sec:method}, we describe cross-power spectrum estimation from these inputs, including the generation of the simulations we use for pipeline validation and estimating the covariance matrix for the data.
    \item In \cref{sec:inference}, we describe the framework in which we compare the data vector to theory predictions, including the parameterisation of the cosmological model and galaxy weak lensing nuisance model. We also describe our inference pipeline in terms of likelihood, prior, and sampling methodology choices.
    \item In \cref{sec:validation}, we describe the validation of this pipeline. We conduct a series of null tests on the blinded data vector to ensure there is no significant detectable contamination from un-modelled observational and astrophysical effects. We also inject simulated data into our inference pipeline and show we can recover the input model parameters in an unbiased way. We demonstrate the stability of our measurement of the cosmological parameters to different choices of the underlying modelling and splitting our data vector into sub-samples in a number of ways.
    \item In \cref{sec:results}, we show our constraints on cosmological and weak lensing galaxy nuisance parameters. We first infer the value of the lensing amplitude $A_{\rm cross}$ with respect to the prediction from a standard \lcdm cosmology. We then show our measurement of the parameters in the full model, including cosmology and galaxy weak lensing nuisance parameters. We also explore our constraining power on the nuisance parameters when DES simulation- and deep data-derived priors are relaxed and when using only high- and low-redshift sub-samples of our data.
    \item In \cref{sec:conclusions}, we review our conclusions and discuss their implications.
\end{itemize}

\section{Theory}\label{sec:theory}
Gravitational lensing of the light from cosmic sources such as the CMB and galaxies 
allows us to probe
the distribution of matter intervening between these sources and the observer. Weak lensing convergence $(\kappa)$ is the weighted integral of the matter density contrast $\delta(z, \hat{n})$ \citep[e.g.][and references therein]{2005astro.ph..9252S}
\begin{equation}
    \kappa(\hat{n}) = \int W(z) \delta(z, \hat{n} ) dz,
\end{equation}
where $W(z)$ is the lensing weight as a function of redshift and $\hat{n}$ is the direction on the sky. $W(z)$ represents the lensing efficiency of the matter distribution along the line of sight. Weak lensing shear ($\bm{\gamma}$), which is a spin-2 quantity with two components, ($\gamma_1, \gamma_2$), is related to $\kappa$ through the following harmonic space relation
\begin{equation}
    \gamma^{\rm E}_{\ell m} = -\sqrt{\frac{(\ell-1)(\ell+2)}{\ell(\ell+1)}} \kappa_{\ell m},
\end{equation}
where $\gamma^{\rm E}_{\ell m}$ are the E-mode spherical harmonic coefficients of the $\bm{\gamma}(\hat{n})$ map \citep{2005PhRvD..72b3516C}. At linear order in deflection, weak lensing by large-scale structure only contributes to the E-mode signal in the shear. This work uses the correlation between the convergence reconstructed from the observed CMB ($\kcmb$) and the weak lensing shear measured by galaxy imaging surveys ($\bm{\gamma}$). \kcmb is reconstructed from the observed CMB maps using quadratic estimators \citep{2021MNRAS.500.2250D}, whereas $\bm{\gamma}$ is estimated from the measurement of galaxy ellipticities $\bm{e} \equiv (e_1, e_2)$, with $e_1$ and $e_2$ being two components of the galaxy ellipticities \citep*{2021MNRAS.504.4312G}. Even though, in principle, shear can be estimated from a simple average of ellipticities, DES-Y3 analysis uses the \metacalibration method \citep{Huff:2017qxu}:
\begin{equation}\label{eq:response}
    \langle \bm{\gamma} \rangle \approx \langle \bm{R} \rangle^{-1} \langle \bm{e} \rangle,
\end{equation}
where the matrix $\bm{R}$ is the shear response for the galaxies, measured by repeating the ellipticity measurement on sheared versions of the galaxy images:
\begin{equation}\label{eq:response_ij}
    R_{i,j} = \frac{e_i^+ - e_i^-}{\Delta \gamma_j},
\end{equation}
where $e^\pm$ is the measurement on an image sheared by a small amount $\pm\gamma$ and $\Delta \gamma = 2\gamma$.

\begin{figure}
 \centering
 \includegraphics[width = 0.45\textwidth]{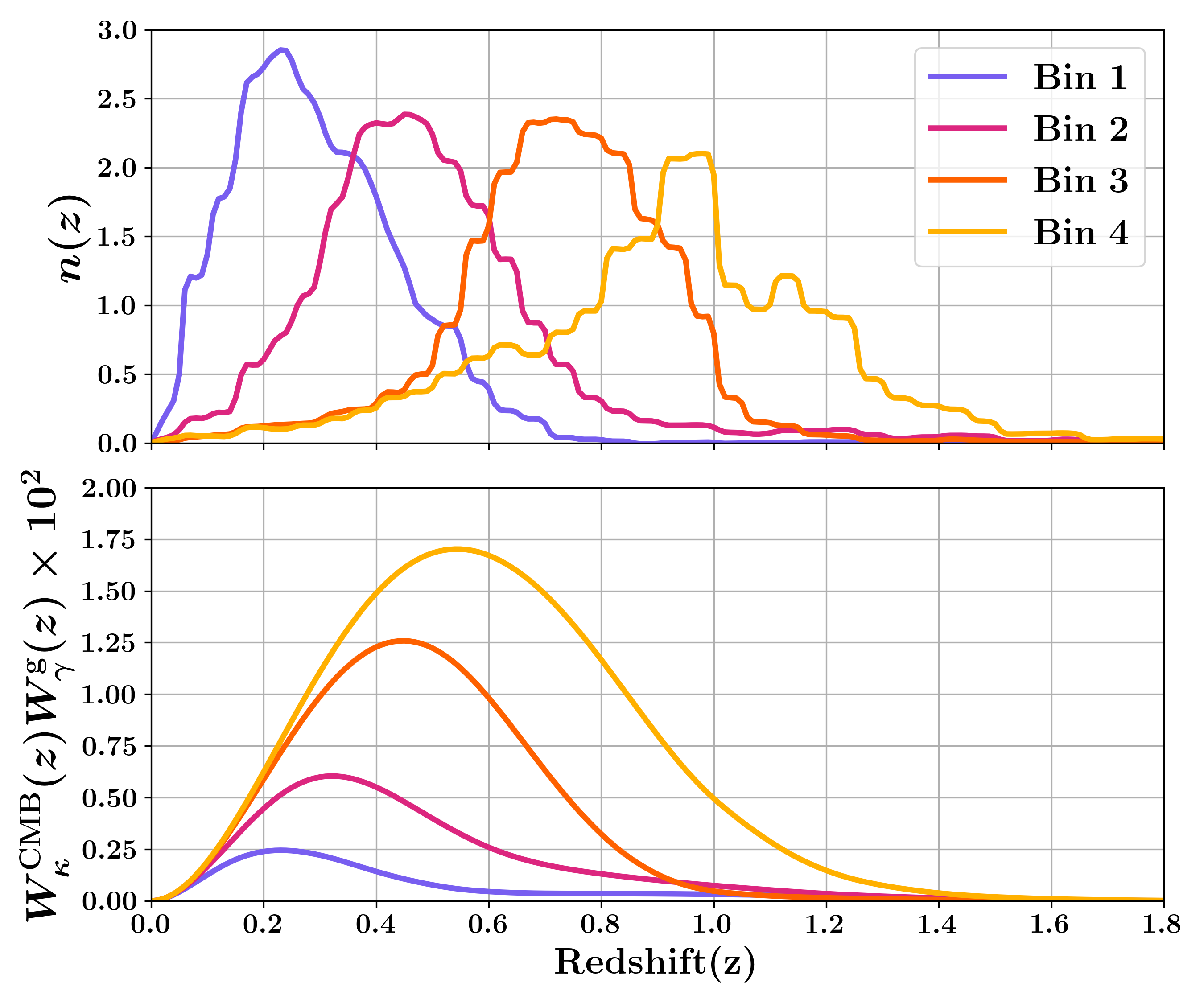}
 \caption{Top panel shows the DES-Y3 source galaxy redshift distribution, $n(z)$, in four tomographic bins. The bottom panel shows the product of the respective galaxy weak lensing kernel with the CMB weak lensing kernel.}
 \label{fig:dndz_wz}
\end{figure}

We model the correlation between $\kcmb$ and $\gE$ in spherical harmonic space. The angular power spectrum between the CMB convergence $\kcmb$ and the E-mode of the galaxy shear $\gE$ at multipole $\ell$, under the Limber approximation \citep{Limber1953, 2008PhRvD..78l3506L}, is \citep[e.g.][]{1992ApJ...388..272K}
\begin{equation}\label{eq:clkg}
\clkg = \int^{z_H}_{0} dz \frac{H(z)}{\chi^2(z) c} \wcmb(z) \wgalaxy(z)  P_{\delta \delta}\Big( k = \frac{\ell + 0.5}{\chi(z)}, z \Big),
\end{equation}
where $P_{\delta \delta}(k, z)$ is the matter power spectrum at redshift $z$, $\chi(z)$ and $a(z)$ denote the comoving distance and the scale factor at $z$, respectively, $c$ is the speed of light, and $H(z)$ is the Hubble parameter as a function of $z$. $\wcmb(z)$ and $\wgalaxy(z)$ are the lensing weights for the CMB and the source galaxies, respectively. The lensing weight for the CMB is given by:\begin{equation}\label{eq:wcmb}
 \wcmb(z) = \frac{3 H^2_0 \Omega_{\rm{m},0}}{2 H(z) c} \frac{\chi(z)}{a(z)} \frac{\chi(z^*) - \chi(z)}{\chi(z^*)},
\end{equation}
where $z^*$ is the redshift of the surface of the last scattering of the CMB, $\Omega_{\rm{m},0}$ and $H_0$ are matter density and Hubble parameters at the current epoch. The lensing weight for the source galaxies depends on their redshift distribution, $n(z)$:
\begin{equation}\label{eq:wgal}
 \wgalaxy(z) = \frac{3 H^2_0 \Omega_{\rm{m},0}}{2 H(z) c} \frac{\chi(z)}{a(z)} \int^{z_H}_{z} dz' n(z') \frac{\chi(z') - \chi(z)}{\chi(z')}.
\end{equation}
We use the Core Cosmology Library \citep[\ccl, ][]{2019ApJS..242....2C} to compute \clkg.\footnote{\url{https://github.com/LSSTDESC/CCL}} We model the non-linear contributions to $P_{\delta \delta}(k)$ using the \halofit model \citep{2003MNRAS.341.1311S, 2012ApJ...761..152T}. We also include contributions to the observed power spectrum from astrophysical and experimental effects, which we fully describe in \cref{subsec:nuisance_model}.

In \cref{fig:dndz_wz}, we show the source redshift distribution $n(z)$ used in this work and the product of the lensing weight function $\wcmb(z) \wgalaxy(z)$. The latter shows the redshift range of the matter distribution that contributes to the cross-correlation \clkg. 

\section{Data}\label{sec:data}
We use overlapping CMB weak lensing and galaxy weak lensing data from the ACT and DES, respectively. We extensively use the individual work of these collaborations in reducing their raw data and preparing science-ready CMB lensing maps and cosmic shear catalogues, but we perform our own analyses to generate the cross-correlation \clkg data vector.
\begin{figure}
 \centering
 \includegraphics[scale=0.38]{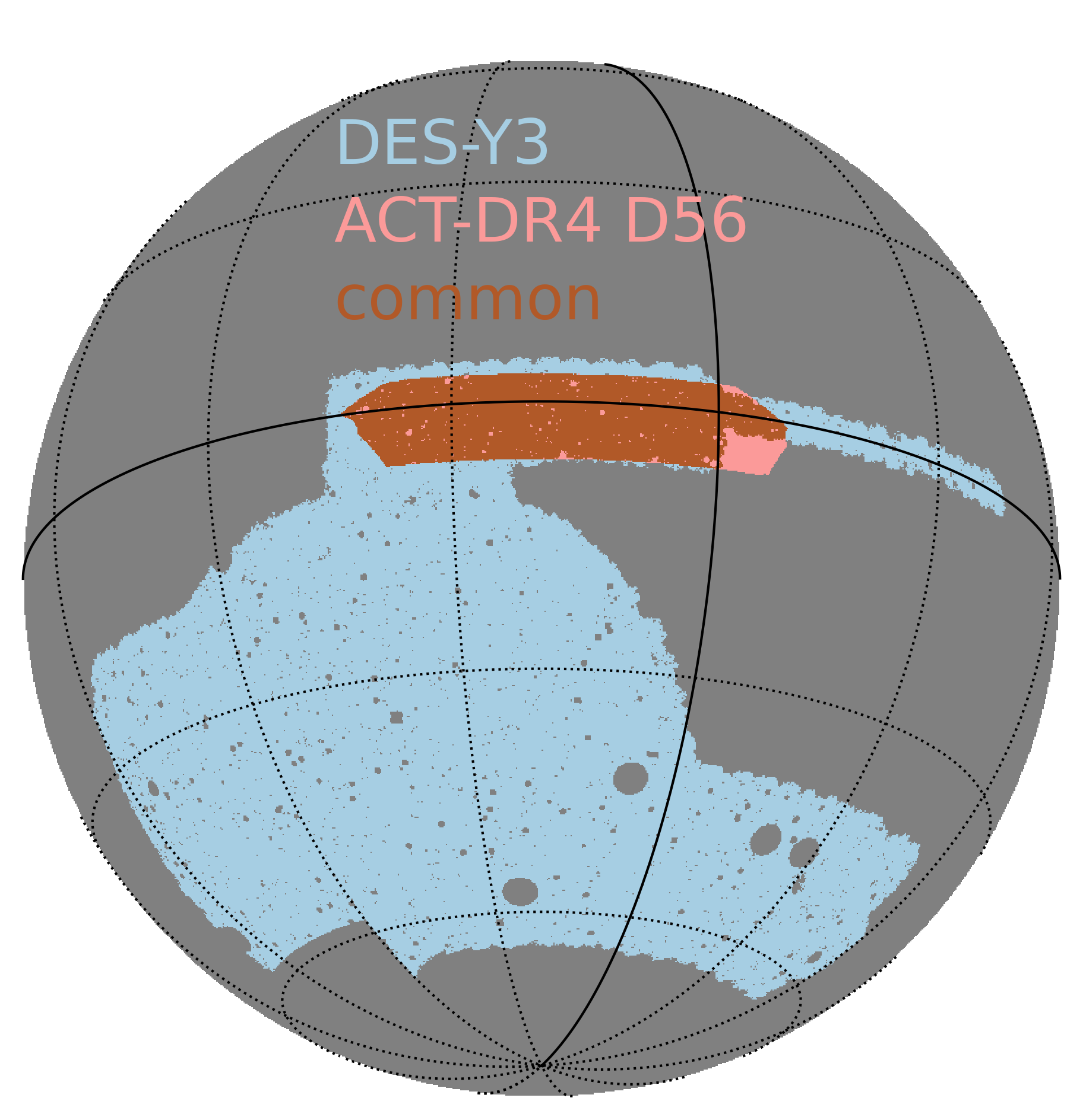}
 \caption{DES-Y3 and ACT-DR4 D56 footprints and their common footprint. The sky area common between them is around 450 ${\rm deg}^2$.} 
 \label{fig:ACTD56_DESY3_mask}
\end{figure}

\subsection{ACT CMB lensing data}\label{subsec:act_data}
We use the ACT-DR4 CMB lensing convergence maps from \cite{2021MNRAS.500.2250D}. These lensing maps are reconstructed using CMB temperature and polarization measurements by ACT in two frequency channels (98 and 150 GHz) during the 2014 and 2015 observing seasons \citep{2020JCAP...12..047A, 2021ApJS..255...11M}. The arcminute-resolution maps produced by the ACT Collaboration are described in \cite{2020JCAP...12..045C, 2020JCAP...12..047A, 2020PhRvD.102b3534M}. ACT-DR4 consists of lensing maps in two sky regions, Deep-56 (D56) and BOSS-North (BN), with respective sky areas 456 deg$^2$ and 1633 deg$^2$ \citep{2021MNRAS.500.2250D}. We use the lensing map in the D56 region, which overlaps with the DES-Y3 footprint, as shown in \cref{fig:ACTD56_DESY3_mask}. 

\begin{figure}
    \centering
    \includegraphics[width=0.45\textwidth]{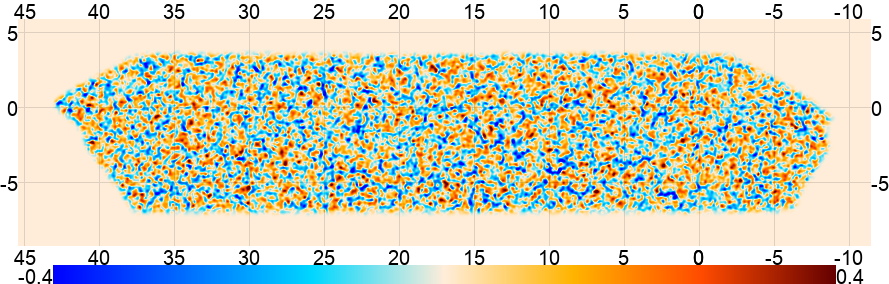}
    \caption{ACT-DR4 $\kcmb$ map reconstructed using ACT and \planck data in the D56 region. The map shown here is smoothed with a Gaussian kernel of 12 arcmin FWHM for visual purposes. The x-axis indicates right ascension, and the y-axis indicates declination. }
    \label{fig:act_planck_kappa}
\end{figure}

CMB lensing maps are obtained using the quadratic estimator \citep{2002ApJ...574..566H}. Signatures of extragalactic astrophysical processes present in the individual frequency maps, such as the Cosmic Infrared Background (CIB) and thermal Sunyaev-Zeldovich (tSZ) effect, lead to biases in the reconstructed convergence map \citep{2014JCAP...03..024O, 2014ApJ...786...13V}. These signals trace the large-scale structure and can lead to biases in the cross-correlation of $\kcmb$ with other large-scale structure probes, such as galaxy weak lensing. For the range of redshifts ($z \lesssim 1.0$) probed by ACT-DR4 and DES-Y3 \clkg, the biases due to tSZ are expected to be more prominent than those due to the CIB \citep{2019PhRvD..99b3508B}, which is sourced by galaxies spanning a broad range of redshift with the peak between $z \sim 1~\text{to}~2$ \citep{2015MNRAS.446.2696S}. ACT-DR4 provides two lensing maps: a \textit{tSZ-free} $\kcmb$ map where the contamination due to the tSZ effect is deprojected \citep{2018PhRvD..98b3534M}, and \textit{with-tSZ} $\kcmb$ map where the tSZ deprojection is not performed. We refer to results obtained using this latter map as `ACT-only'. The tSZ-free $\kcmb$ map uses \planck frequency maps along with the ACT data to perform the internal linear combination step required to deproject tSZ contamination and obtain the tSZ-free CMB map \citep{2018PhRvD..98b3534M, 2020PhRvD.102b3534M}. Hence, we refer to results derived from this map as `ACT+\emph{Planck}'. In the ACT-DR4 analysis, the CMB lensing maps are reconstructed using Fourier modes between $\ell^{\rm CMB}_{\rm min}$ and $\ell^{\rm CMB}_{\rm max}$. The lower multipole, $\ell^{\rm CMB}_{\rm min}$, is chosen to mitigate the effects of the atmospheric noise and the ACT mapmaker transfer function \citep{2021MNRAS.500.2250D}. $\ell^{\rm CMB}_{\rm max}$ is chosen to avoid contamination due to extragalactic foregrounds. The ACT-only convergence map is reconstructed with $\ell^{\rm CMB}_{\rm min} = 500$ and $\ell^{\rm CMB}_{\rm max} = 3000$. The tSZ-cleaned CMB map obtained using \planck frequency maps contains information on large angular scales, below $\ell < 500$. Hence, using ACT+\planck data and tSZ deprojection makes a wider range of CMB multipoles suitable for lensing reconstruction with $\ell^{\rm CMB}_{\rm min} = 100$ and $\ell^{\rm CMB}_{\rm max} = 3350$. In \cref{fig:act_planck_kappa}, we show the ACT+\planck \kcmb map over the D56 region. This map is smoothed using a Gaussian kernel of 12 arcmin FWHM for visual purposes only. We use the $\kcmb$ map without any additional smoothing in the analysis.

ACT lensing reconstruction is performed using a lensing analysis mask applied to the individual frequency maps or CMB maps. While computing the angular power spectrum, we use the square of this mask as the mask implicit in the reconstructed $\kcmb$ map. We also use 511 lensing reconstruction simulations made available by \cite{2021MNRAS.500.2250D} to obtain the lensing reconstruction noise.

The ACT $\kcmb$ maps are in the equirectangular plate carree (CAR) projection. We use the \pixell package to convert the maps from CAR projection to the \healpix pixelization at resolution \texttt{Nside} = 2048 \citep{2005ApJ...622..759G}.\footnote{\url{https://github.com/simonsobs/pixell}}

\subsection{DES-Y3 galaxy weak lensing data}\label{subsec:des_data}
DES is a photometric survey that carried out observations using the Dark Energy Camera \citep{2015AJ....150..150F} on Cerro Tololo Inter-American Observatory (CTIO) Blanco 4-meter Telescope in Chile. We use the weak lensing source galaxies catalogue of the DES-Y3 data. The catalogue is derived from the DES-Y3 GOLD data products \citep{2021ApJS..254...24S}. The shape measurement of source galaxies is performed using the \metacalibration algorithm \citep{Huff:2017qxu, 2017ApJ...841...24S} and is discussed in \cite*{2021MNRAS.504.4312G}. After various selection cuts are applied to reduce systematic biases, the catalogue contains the shape measurement ($e_1, e_2$) of $\sim 1 \times 10^8$ galaxies. It spans an effective (unmasked) area of 4143 deg$^2$ with effective number density $\neff = 5.59 ~ \rm{gal}/\rm{arcmin}^{2}$.

\begin{table}
	\centering
	\caption{Summary of source galaxy catalogue. $z^{\rm{PZ}}_1 - z^{\rm{PZ}}_2$ is the range of photometric redshifts of the given tomographic bin \citep*{2021MNRAS.505.4249M}, $\neff$ is the effective number density of source galaxies in units of $\rm{gal}/\rm{arcmin}^{2}$, and $\sigma_{\rm e}$ is uncertainty in the measurement of one component of the shape \citep*{2022PhRvD.105b3514A}. $\bar{R}_1$ and $\bar{R}_2$ are the average \metacalibration responses for two galaxy ellipticity components \citep*{2021MNRAS.504.4312G}. }
	\label{tab:desy3_ng_sigma_e}
	\begin{tabular}{lcccr} 
		\hline
            \hline
		Redshift Bin & Bin-1 & Bin-2 & Bin-3 & Bin-4\\
            \hline
		$z^{\rm{PZ}}_1 - z^{\rm{PZ}}_2$ & 0.0-0.36 & 0.36-0.63 & 0.63-0.87 & 0.87-2.0\\
		\hline
		$\neff$ & 1.476 & 1.479 & 1.484 & 1.461\\
		\hline
		$\sigma_{\rm e}$ &0.243 & 0.262 & 0.259 & 0.310 \\
		\hline
		$\bar{R}_1$ & 0.767 & 0.726 & 0.701 & 0.629 \\
		\hline
		$\bar{R}_2$ & 0.769 & 0.727 & 0.702 & 0.630 \\
		\hline  
            \hline
	\end{tabular}
\end{table}

The galaxies in the source catalogue are distributed in four tomographic redshift bins shown in \cref{fig:dndz_wz}. Photometric redshifts of these galaxies are estimated using the SOMPZ algorithm \citep*{2021MNRAS.505.4249M} using deep observations and additional colour bands from the DES deep fields \citep*{2022MNRAS.509.3547H}. The effective number of sources and the uncertainty in one component of the ellipticity measurement ($\sigma_{\rm e}$) for each redshift bin are shown in Table \ref{tab:desy3_ng_sigma_e}.

The catalogue provides the inverse variance weight ($w$) for the shape measurement of each galaxy. When computing the angular power spectrum, we use these weights to form the mask to be applied to the shear maps. We use the sum-of-weights scheme discussed in \cite{2021JCAP...03..067N} to prepare this mask. We discuss this procedure in \cref{subsec:map_making}.

\subsubsection{Blinding}\label{subsubsec:des_blinding}
We use catalogue-level blinding to guard ourselves against experimenter bias which may drive our analysis towards known values of cosmological parameters from existing experiments. We transform the shape catalogue in the same way as in the DES-Y3 analysis \citep*{2021MNRAS.504.4312G}. This blinding method involves changing ellipticity values with the transformation:
\begin{align}
    |\eta| &\equiv 2 \rm{arctan}|\bm{e}|  \nonumber\\
           &\rightarrow f|\eta| \nonumber
\end{align}
where $f$ is an \textit{unknown factor} between 0.9 and 1.1 \citep*{2021MNRAS.504.4312G}, which we keep the same for all four tomographic bins. This transformation limits the ellipticity values within unity and re-scales the estimated shear. Note that DES-Y3 analysis uses two-stage blinding; the first stage is at the catalogue level, and the second is at the level of summary statistics. In this work, we only perform catalogue-level blinding. 

We performed all of our null and validation tests and an initial round of internal collaboration review of the manuscript with the blinding factor still included. During this stage, we did not plot or compare the data bandpowers with the theoretical \clkg. We plotted the figures showing parameter inference from the blinded data without the axis values. After we finalized the analysis pipeline, we removed the unblinding factor and updated the manuscript accordingly to discuss the results.

\section{Method}\label{sec:method}
In this work, we infer the cosmological, astrophysical and observational systematic parameters using the angular cross-power spectrum between the CMB lensing convergence and the tomographic galaxy weak lensing fields. In this section, we discuss the analysis methodology.

\subsection{Simulations}\label{subsec:sims}
We use simulations of CMB convergence \kcmb and galaxy shape $\bm{\gamma}$ with realistic noise to validate the analysis pipeline and obtain the covariance matrices for $\clkg$. Weak lensing convergence and shear are not expected to be exact Gaussian random fields. A lognormal distribution provides a good approximation of weak lensing convergence and shear fields \citep{2011A&A...536A..85H}. Generating lognormal simulations is computationally cheap compared to N-body simulations and/or ray tracing. The feasibility of the lognormal simulations for the covariance matrices of the power spectrum is discussed in \cite*{2018PhRvD..98b3508F}.

We simulate \kcmb and $\bm{\gamma}$ signal maps as correlated lognormal random fields with zero mean using the publicly available code package \flask \citep{2016MNRAS.459.3693X}. We generate full sky, correlated signal maps of both convergence and shear at \nside = 2048, corresponding to 1.7 arcmin pixel resolution. To generate signal-only map realizations, inputs to \flask are 
(1) the theory angular power spectra describing the auto and cross spectra of convergence field ($\kappa_{\rm C/g}$) for the CMB and the source galaxies ($C^{\kappa_{\rm C/g} \kappa_{\rm C/g}}_{\ell}$), (2) the galaxy source redshift distribution $n(z)$, and (3) the lognormal shift parameter which determines the skewness of the lognormal distribution for a given variance. The auto and cross power spectra are computed using \ccl with the \halofit matter power spectrum. $n(z)$ is the DES-Y3 source galaxy redshift distribution, which is also used as input to \ccl while computing $C^{\kappa_{\rm C/g} \kappa_{\rm C/g}}_{\ell}$. We use the same lognormal shift parameter values as used in \cite*{2021MNRAS.508.3125F} and \cite*{2023PhRvD.107b3529O}. These are 0.00453, 0.00885, 0.01918, and 0.03287 for four DES-Y3 source redshift bins and 2.7 for the CMB. We then apply the ACT-D56 mask to the convergence fields and the DES-Y3 mask to the shear field to obtain signal-only maps over the respective survey footprints. The DES-Y3 mask used at this stage is a binary mask with a pixel value equal to zero if the pixel does not contain any source galaxy and a value of one otherwise. 

We use the following procedure to obtain the \kcmb and $\bm{\gamma}$ maps with noise that has the correlated signal part. In the simulations, a particular realization of the reconstructed \kcmb is generally obtained by reconstructing the lensing convergence from a simulated CMB map that has been lensed by a given \kcmb signal realization. However, in this work, we do not perform such an end-to-end \kcmb reconstruction with our lognormal signal-only \kcmb maps. Instead, we use existing ACT-DR4 \kcmb signal realizations and reconstruction simulations. The signal in these simulations is not correlated with our large scale structure simulations, so they cannot be used directly. Instead, we subtract the signal realization from these reconstructed \kcmb maps to obtain a realization of \kcmb reconstruction noise. We then add these resultant noise maps to our lognormal \kcmb signal-only map generated using \flask. For these \flask simulations, we have also generated \kcmb signal maps which are correctly correlated with the $\bm{\gamma}$ signal. We generate simulations of the noise in the shear (the uncertainty caused by the intrinsic galaxy shape) using the random rotation of galaxy ellipticities in the DES-Y3 shear catalogue: $(e_1 + i e_2) \rightarrow \exp{(2i \phi)} (e_1 + i e_2)$, where $\phi$ is a uniform random number in the range $[0, 2 \pi)$. A shear noise map is obtained using this catalogue where the galaxy shapes are rotated. We add these shear noise maps to the shear signal-only maps to obtain shear maps with realistic noise.

\subsection{Shear map making}\label{subsec:map_making}
\begin{figure}
    \centering
    \includegraphics[scale=0.70]{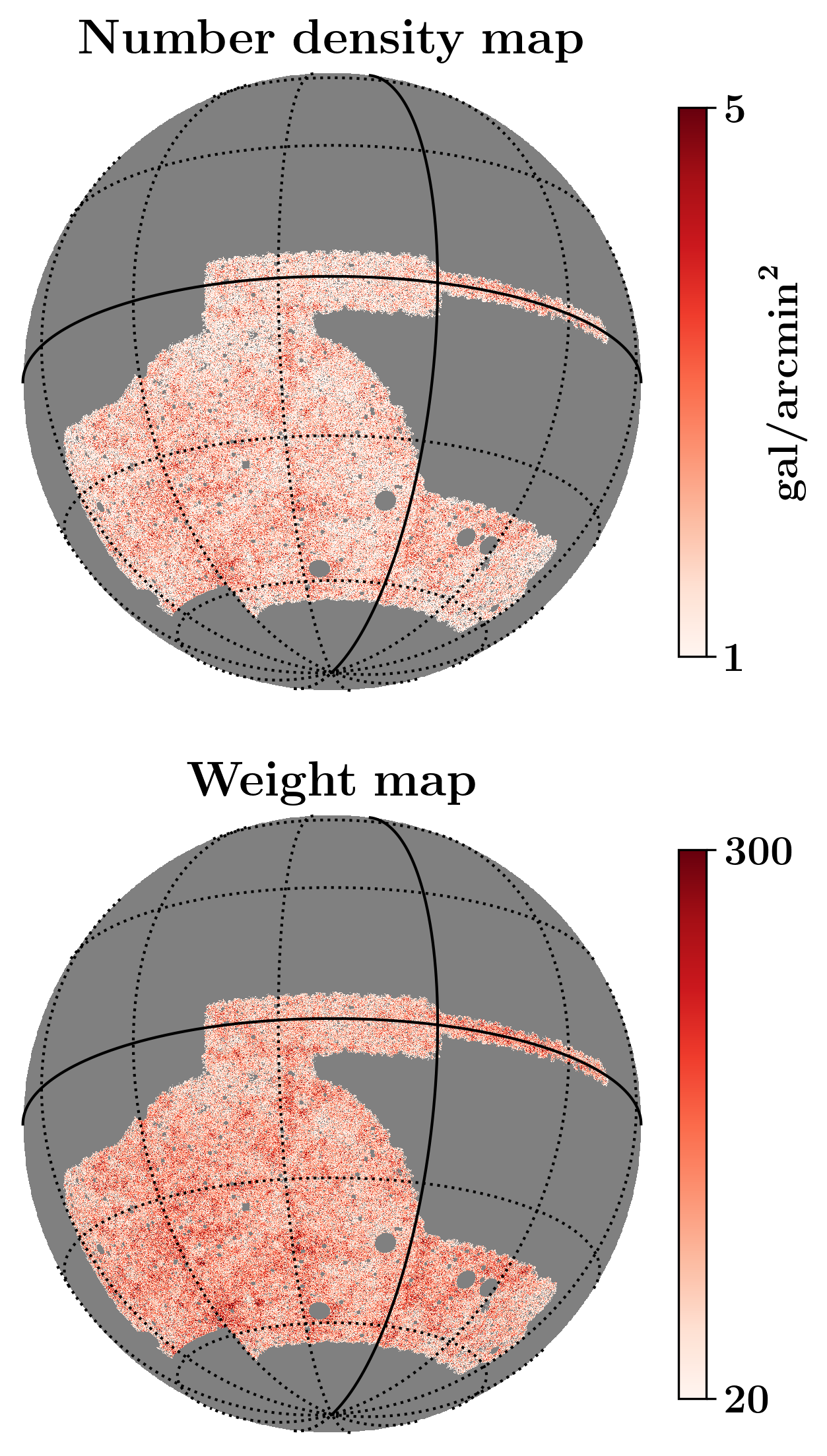}
    \caption{The source galaxy number density map in $\rm{gal/arcmin^2}$ unit (\emph{top}) and the weight map (\emph{bottom}) for the galaxies in DES-Y3 Bin-4. For the weight map, the value in each pixel is the summation of the inverse variance weights of all the galaxies that fall within that pixel. The weight map is used as the shear field mask without apodization.}
    \label{fig:des_maps_1}
\end{figure}
\begin{figure}
    \centering
    \includegraphics[scale=0.60]{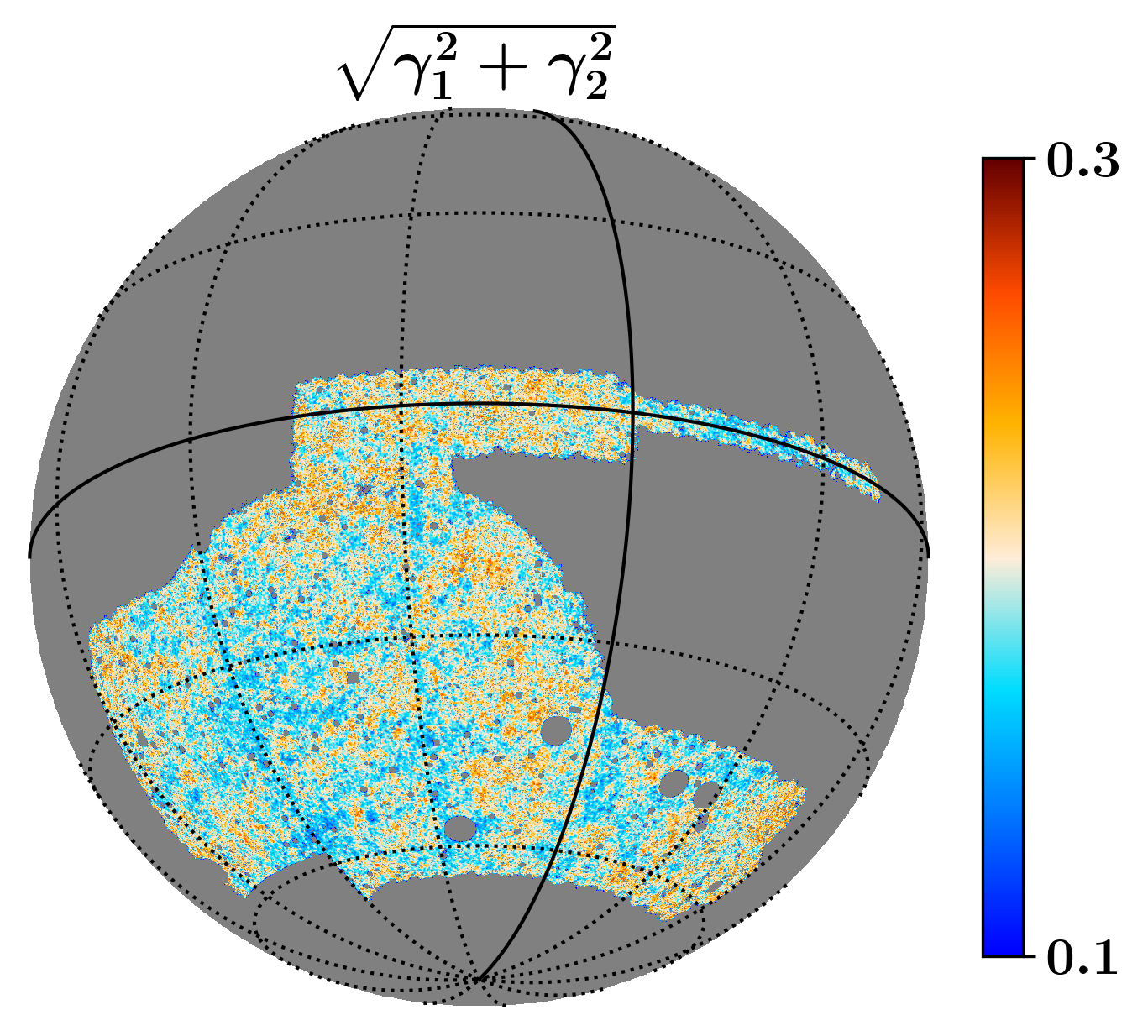}
    \caption{Map of the magnitude of shear ($\sqrt{\gamma^2_1 + \gamma^2_2}$) for the tomographic Bin-4, where the value of shear in each pixel is estimated using \cref{eq:gamma_map}. The map is smoothed with a Gaussian kernel of 12 arcmin FWHM for visual purposes.}
    \label{fig:des_maps_2}
\end{figure}

We perform our analysis in harmonic space on the maps prepared in the \healpix pixelization. 
Along with the shape measurements, the DES-Y3 shape catalogue contains weights and \metacalibration response $(R_1, R_2)$ for each galaxy. The $R_1$ and $R_2$ are the diagonals of the response matrix $\bm{R}$ discussed in \cref{sec:theory}. While estimating shear from the shape measurement, we do not 
use the response for each galaxy, but the average response as used in \cite*{2021MNRAS.504.4312G}. We first subtract the non-zero mean of each ellipticity component for each galaxy using the weighted average and then correct for the response using the following expression:
\begin{equation}\label{eq:sub_e_mean}
 \hat{e}_i = \frac{1}{\bar{R}} \Big(e_i - \frac{\sum_j w_j e_j}{\sum_j w_j} \Big),
\end{equation}
where the average response $\bar{R}$ for each ellipticity component of four tomographic bins is given in \cref{tab:desy3_ng_sigma_e} and the labels $i$ and $j$ run over all of the galaxies in a given tomographic bin. The above subtraction is carried out for each tomographic bin separately. The shear map for a given bin is obtained from these mean subtracted and response-corrected galaxy shapes. The shear estimate for a given pixel $p$ is the inverse variance weighted average of galaxy ellipticities
\begin{equation}\label{eq:gamma_map}
 \boldsymbol{\gamma}(n_p) = \frac{\sum_{i \in p } w_i \boldsymbol{\hat{e}}_i }{\sum_{i \in p } w_i},
\end{equation}
where the summation is over all the galaxies that fall within the area of pixel $p$. 

To obtain the mask to be used with the shear maps, we use the sum-of-weights scheme, where we form the map from the inverse variance weights ($w_i$) given in the DES-Y3 catalogue \citep{2021JCAP...03..067N},
\begin{equation}\label{eq:SoW_mask}
    W(n_p) = \sum_{i \in p} w_i.
\end{equation}
We show the representative shear mask in \cref{fig:des_maps_1}, along with the source galaxy number density map. These maps indicate the inhomogeneity in the galaxy count and their weights. \cref{fig:des_maps_2} show the maps of the shear component obtained using the procedure discussed in this section.

\subsection{Power spectrum bandpowers and covariance matrix}\label{subsec:pseudo_cl}
While computing the angular power spectrum with the partial sky map, one needs to consider the correlations between the spherical harmonic coefficients induced by the mask. This problem is addressed by the pseudo-$C_{\ell}$ formalism, such as in the \texttt{MASTER} algorithm \citep{2002ApJ...567....2H}. The algorithm deconvolves the effect of the mask and provides an estimate of the power spectrum ($\hat{C}_{\bm{q}}$) binned over a certain range of multipoles $\ell \in \bm{q}$. The ensemble average of $\hat{C}_{\bm{q}}$ is equal to the weighted average of the underlying angular power spectrum of the full sky map over a range of multipoles. The bandpower window function specifies the range of multipoles and the multipole weights. To compute these power spectrum bandpowers on the partial sky maps, we use the \texttt{MASTER} algorithm and its application for spin-2 fields \citep{2011MNRAS.412...65H}, as implemented in \namaster \citep{2019MNRAS.484.4127A}.

In \namaster, we specify separate masks for the \kcmb and $\bm{\gamma}$ fields. For \kcmb, we use the ACT-DR4 analysis mask for the D56 region. We convert this mask from CAR pixelization to \healpix pixelization at \nside = 2048 resolution using the \texttt{reproject} module of the \pixell package. The original mask in CAR projection is apodized. We do not introduce any extra apodization in the mask after reprojecting to \healpix. This analysis mask is applied to CMB maps used in the quadratic estimator while reconstructing \kcmb; hence, the reconstructed \kcmb map has the mask implicit in it. We use the square of the analysis mask as the mask implicit in reconstructed \kcmb.\footnote{This information is specified in \namaster using \texttt{masked\_on\_input = True} keyword argument.} For the shear field, we use the sum of inverse variance weights mask as expressed in \cref{eq:SoW_mask} and depicted in \cref{fig:des_maps_1}. This procedure is equivalent to dividing by the variance of the shear estimate in \cref{eq:gamma_map}. Compared to the \kcmb mask, the shear mask is highly non-uniform, as evident from \cref{fig:des_maps_1}. We do not apodize this mask because apodization would lead to losing a substantial sky fraction. Using the simulations described above, we verify that our masking choices do not affect the recovered data vector. In the remaining section, we discuss the computation of pseudo-$C_{\ell}$ and validation of the simulations at the power spectrum level.

\begin{figure*}
 \centering
 \includegraphics[width = 0.9\textwidth]{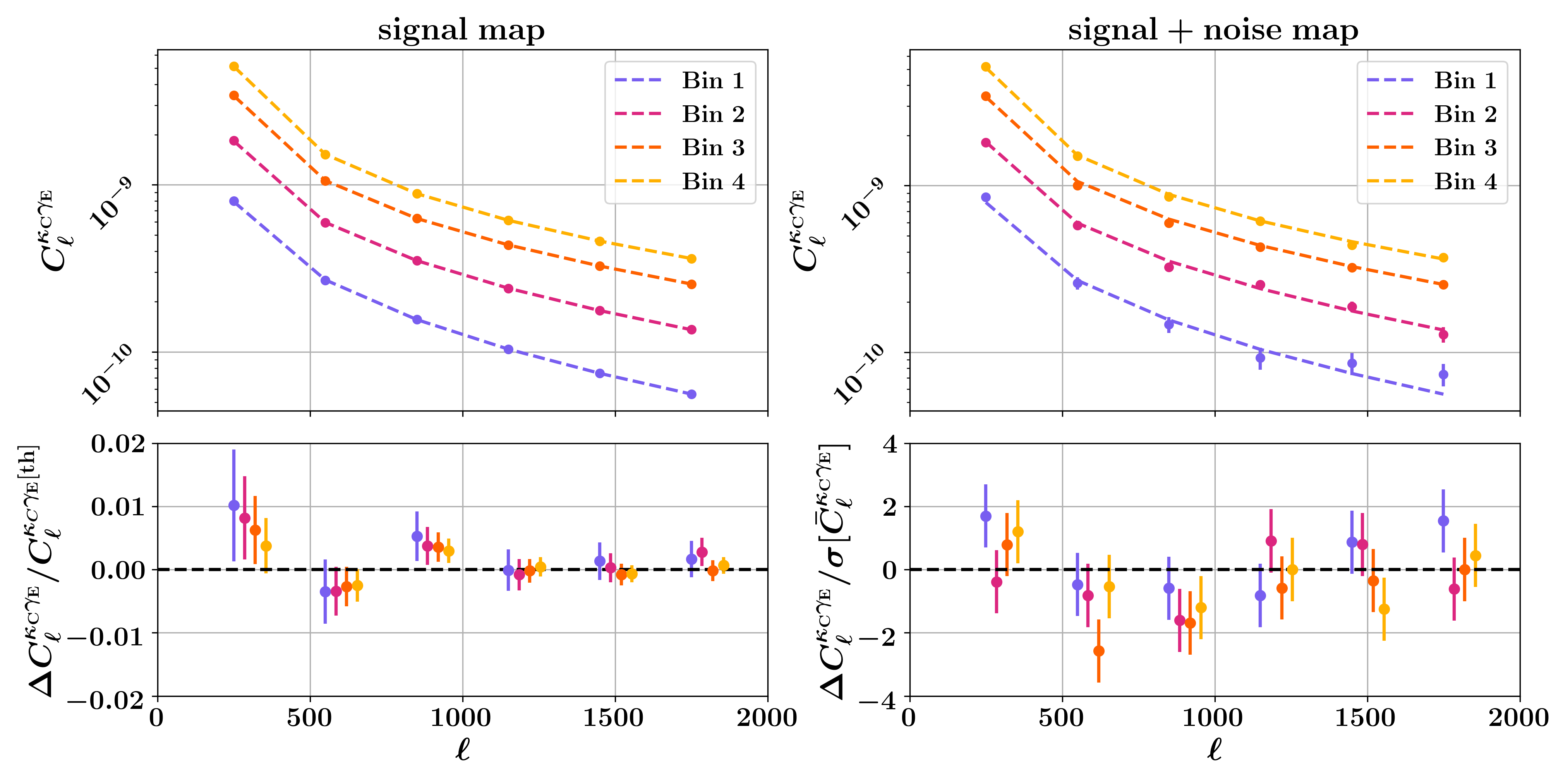}
 \caption{Comparison of mean of \clkg (\barclkg) from signal only (\emph{left panel}) and signal + noise (\emph{right panel}) simulations with the theory \clkg, enabling us to compare our pipeline against expectations. 
 \emph{Left top panel}: Comparison between input theory power spectra (dashed line) and the mean of the power spectrum of 511 signal-only simulated maps (points). \emph{Left bottom panel}: The relative difference between the two. The error bars represent the uncertainty on the mean of 511 simulations. 
 \emph{Right top panel}: Comparison between input theory power spectra and the mean of the power spectrum of 511 simulated maps with signal and noise. \emph{Right bottom panel}: The difference between the two in units of the uncertainty on the mean of \clkg. The error bars are the standard deviation of the mean, i.e. $\sigma[\barclkg] = \sigma[\clkg]/\sqrt{\rm{N}_{\rm sims}}$. }
 \label{fig:compare_simCls}
\end{figure*}

In the reconstructed \kcmb map, lower multipoles are affected by the mean-field bias caused by statistical anisotropy due to non-lensing effects, such as the analysis mask, inhomogeneous noise and other non-idealities in the data. For ACT D56 \kcmb, multipoles below $\ell \approx 50$ are affected by the mean-field \citep{2021MNRAS.500.2250D}. Hence, we neglect the first bandpower of \clkg computed in the range $\ell = 0 - 100$. This analysis uses the \clkg computed at multipoles above $\lmin = 100$. The distribution of matter by the baryonic processes also affects the weak lensing angular power spectrum at small scales. To accurately model these scales, one needs to consider the effect of baryons in the modelling. In this work, we do not consider the modelling of the baryons and choose $\lmax = 1900$ so that the effect of baryons on \clkg is negligible at the given statistical uncertainty. To assess the effect of baryons, we use the halo model with baryon modelling considered in \hmcode \citep{2015MNRAS.454.1958M} as implemented in \camb \citep{2000ApJ...538..473L,2012JCAP...04..027H}. Modelling of baryons in \hmcode is done through two parameters: halo concentration parameter $A_{\rm HM}$ (\texttt{HMCode\_A\_baryon}) and the halo profile parameter $\eta$ (\texttt{HMCode\_eta\_baryon}). We compute theory \clkg for the fiducial cosmological parameters and over the range of values of $A_{\rm HM}$ and $\eta$. For $A_{\rm HM}$, we consider the range $A_{\rm HM} = 2 ~ \rm{to} ~ 4.5$ and $\eta$ is determined by the empirical relation $\eta=1.03 - 0.11 A_{\rm HM}$ \citep{2015MNRAS.454.1958M}. We compare theory \clkg with the uncertainty on \clkg with ACT-DR4 and DES-Y3. We find that, over the range of baryon parameters considered here, the relative effect of baryons on \clkg for $\ell > 1000$ can be up to 20\%. At the redshift where $\wcmb(z) \wgalaxy(z)$ has the peak, $\lmax = 1900$ corresponds to the comoving wavenumber of $k_{\rm max} \equiv \lmax/\chi(z) = 0.96, 0.72, 0.53, 0.44 ~ \rm{Mpc}^{-1}$ for four DES-Y3 tomographic redshift bins, respectively. The matter perturbations at these scales are non-linear and sensitive to baryonic processes \citep{2015MNRAS.454.1958M}. However, the effect of baryons is still well within two per cent of the statistical uncertainty on \clkg up to $\ell = 1900$.
Moreover, for the given noise level, the expected SNR of \clkg is saturated beyond $\ell \approx 1900$. Hence, we choose $\ell = 1900$ as the optimal choice for $\lmax$ in this analysis. We choose the multipole bin width $\Delta \ell = 300$ with uniform weights for the power spectrum binning.

We compare the pseudo-$C_{\ell}$ computed from simulated maps with the input theory power spectrum. In the left column of \cref{fig:compare_simCls}, we compare the \flask signal-only simulation bandpowers computed over the survey footprint and the input theory \clkg. We compare the mean of the pseudo-$C_{\ell}$ from 511 simulations with the binned input theory power spectrum. We perform the binning of the theory \clkg while properly taking into account the effect of bandpower window as discussed in Section 2.1.3 of \cite{2019MNRAS.484.4127A}. As shown in the left panel of \cref{fig:compare_simCls}, we see no significant bias between pseudo-$C_{\ell}$ computed from simulated maps and the input \clkg and conclude that our simulated signal maps are consistent with the appropriate cosmological signal. This also verifies that the mode decoupling by \namaster for the given masks gives an unbiased power spectrum estimate. We then compute the pseudo-$C_{\ell}$ of the 511 simulations with signal and noise. In the right column of \cref{fig:compare_simCls}, we compare the mean of these 511 bandpowers with the input theory. Here also, we do not see a significant bias and find that the bandpowers computed from the noisy simulations are consistent with the input theory. This validates the power spectrum computation part of the analysis pipeline.

We use these 511 simulation bandpowers to obtain the covariance matrix for the \clkg data vector. We expect this covariance matrix to accurately capture features of real data relevant to the power spectrum analysis. These include the non-Gaussianity of the signal modelled as the lognormal field, the inhomogeneous and non-Gaussian nature of \kcmb reconstruction noise, the inhomogeneous nature of shear noise arising from variations in the number count and the inverse variance weights. Each simulation bandpower realization is obtained using \namaster with the same mask treatment as applied to the data and hence captures the effect of using the partial sky. 

\begin{figure}
 \centering
 \includegraphics[width = 0.45\textwidth]{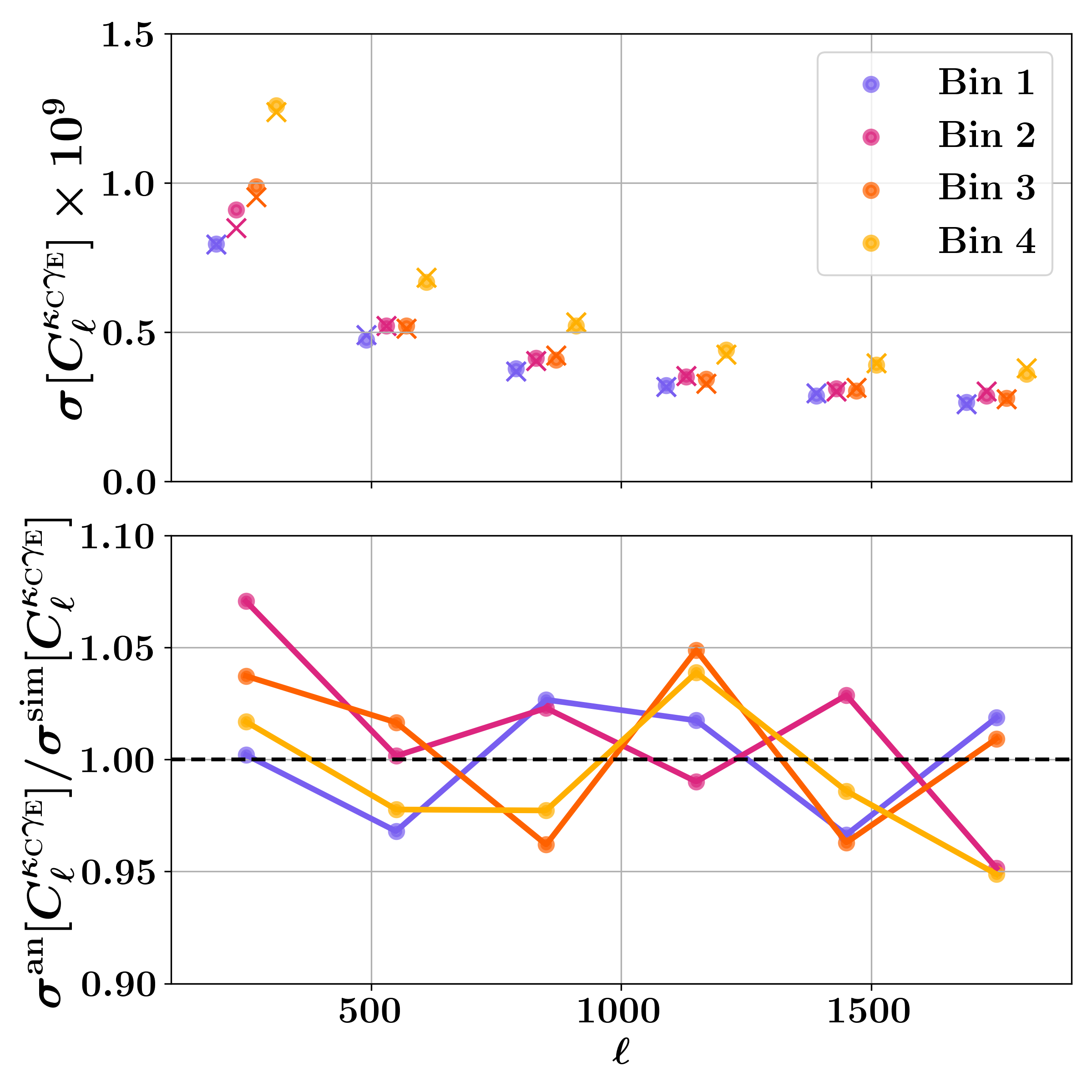}
 \caption{\emph{Top:} Square root of the diagonal of the analytical (dot) and simulation (cross) covariance matrices for four tomographic bins. 
 \emph{Bottom:} The ratio of the diagonal of two covariance matrices, indicating agreement between the two within $\pm 5 \%$.}
 \label{fig:cov_mat_diag}
\end{figure}

\begin{figure}
 \centering
 \includegraphics[width = 0.45\textwidth]{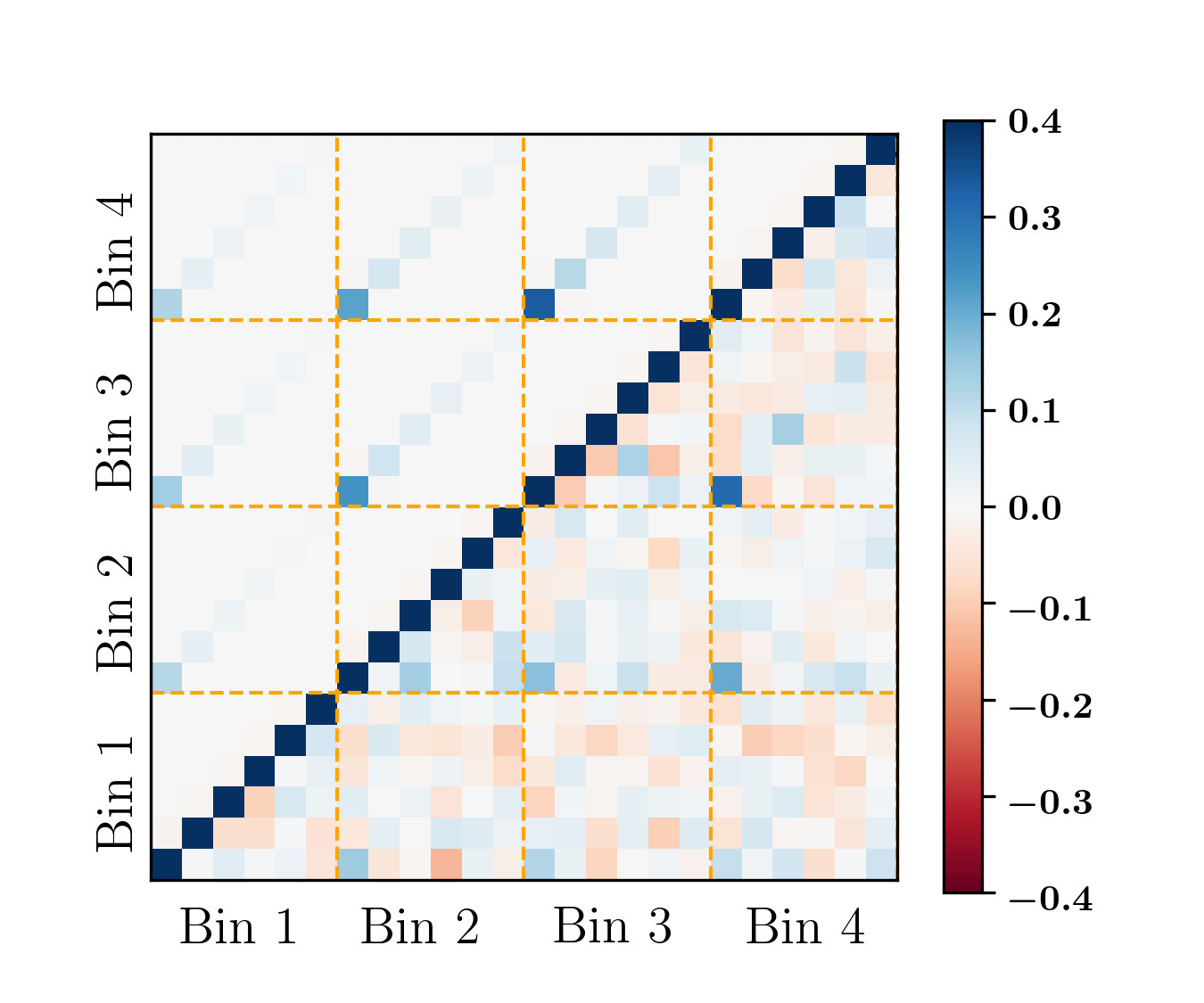}
 \caption{Correlation matrix obtained from the covariance matrix over the multipole range of $\ell = 100 ~ \text{to} ~ 1900$ with $\Delta \ell = 300$. The upper triangle shows the elements of the correlation matrix obtained from the analytical covariance matrix, and the lower triangle shows that from the simulation covariance matrix. Note that the colour scale is saturated at $\pm 0.4$ to clearly show the fluctuations in the off-diagonal terms.}
 \label{fig:corr_matrix}
\end{figure}

We also construct a theoretical covariance matrix for the pseudo-$C_\ell$ using \namaster. This covariance matrix takes into account the effect of the mask, the Gaussian contribution based on the auto and cross theory $C_{\ell}$ of \kcmb and \gE, and the noise power spectrum $N_{\ell}$ of the respective field. For the \kcmb noise power spectrum, we use the mean of the noise power spectra obtained from 511 maps of the \kcmb noise simulation. We obtain the shear noise power spectrum, $N^{\gamma \gamma}_{\ell}$, using the following analytical expression \citep{2021JCAP...03..067N}:
\begin{equation}\label{eq:an_nlgg}
    N^{\gamma \gamma}_{\ell} = A \frac{\sum_i w^2_i \sigma^2_{e,i} }{(\sum_i w_i)^2},
\end{equation}
where $A$ is the sky area and $\sigma^2_{e,i} = (e^2_{i,1} + e^2_{i,2})/2$. The summation in the above equation is carried out only for the galaxies within the common region between DES-Y3 and ACT D56 region. We find that using $N^{\gamma \gamma}_{\ell} = \sigma^2_{e}/\neff$, with the $\sigma^2_{e}$ and $\neff$ values given in \cref{tab:desy3_ng_sigma_e}, leads to relatively lower $N^{\gamma \gamma}_{\ell}$ than the one obtained using \cref{eq:an_nlgg} evaluated for galaxies only over the ACT D56 region. This is because $\sigma^2_{e}$ and $\neff$ given in \cref{tab:desy3_ng_sigma_e} are obtained from all the galaxies within the respective tomographic redshift bin. In contrast, for cross-correlation, we only need $\sigma^2_{e}$ and $\neff$ for the region that overlaps with the ACT D56 region. In \cref{fig:cov_mat_diag}, we compare the diagonal of two covariance matrices. We find good agreement between the two covariance matrix estimates.
In \cref{fig:corr_matrix}, we show the correlation matrix obtained from the simulation covariance matrix as well as the Gaussian covariance matrix. With the choice of $\Delta \ell = 300$, we see no significant correlation between nearby bandpowers. Some of the matter lensing the source galaxies in two different redshift bins is the same. This leads to a correlation between the \clkg corresponding to these redshift bins. The Gaussian covariance part in \cref{fig:corr_matrix} clearly shows the non-zero off-diagonal terms that arise due to these correlations, and as expected, the correlations between the two highest redshift bins, Bin 3 and Bin 4, are relatively larger. We include this inter-redshift bin correlation in our simulations, which is considered in the parameter inference. 

\section{Likelihood and inference}
\label{sec:inference}
To evaluate the likelihood for the \clkg bandpowers, we make use of the Simons Observatory Likelihoods and Theories \soliket framework.\footnote{\url{https://github.com/simonsobs/SOLikeT/}} \soliket is a unified framework for analysing cosmological data from CMB and LSS experiments being developed for the Simons Observatory \citep[SO,][]{2019JCAP...02..056A}. Here we use the \texttt{KappaGammaLikelihood} module to compute the theory \clkg bandpowers at a given set of parameters $\boldsymbol{\theta}$. This computation uses \camb \citep{2000ApJ...538..473L,2012JCAP...04..027H} matter power spectra with Limber integrals evaluated by \ccl \citep{2019ApJS..242....2C}. The \texttt{KappaGammaLikelihood} module has been verified to reproduce published results.\footnote{\url{https://github.com/simonsobs/SOLikeT/pull/58##issuecomment-1213989444} where the measurement of \cite{2015PhRvD..91f2001H} is reproduced.}
\subsection{Cosmological model and parameters}
\label{subsec:cosmological_model}
\begin{figure*}
    \centering
    \includegraphics[width=\textwidth]{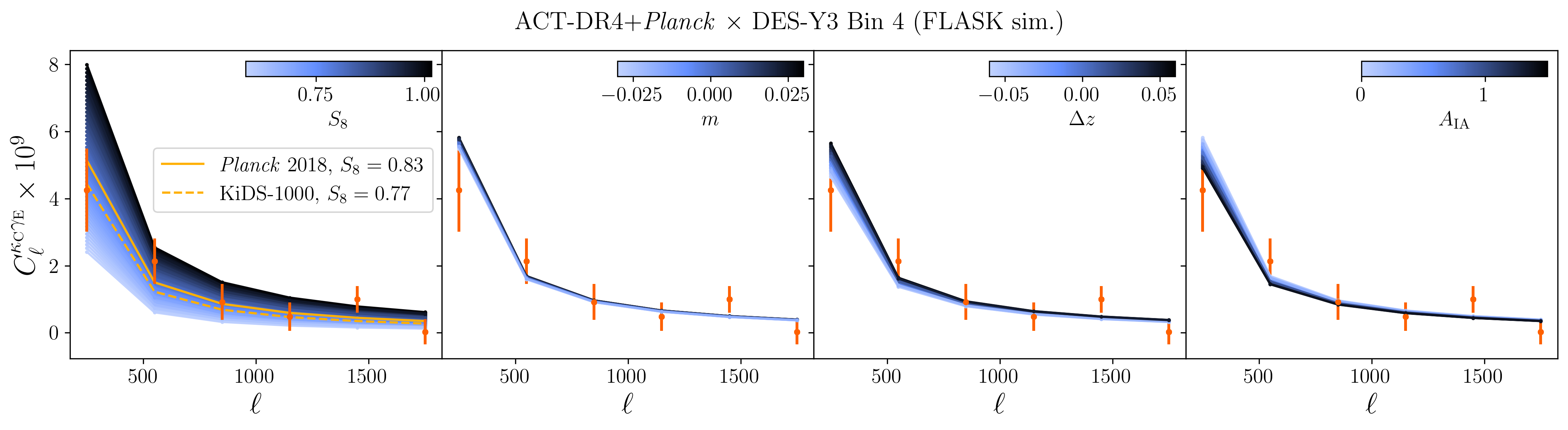}
    \caption{Illustrative changes in predicted data vectors as the cosmological and nuisance parameters are individually varied, compared to the simulated data vector described in \cref{subsec:sims} (to be concise, we only show tomographic Bin 4 as this is the highest SNR bin). The left-most panel also shows cosmologies with $S_8$ as found by \planck 2018 primary CMB \citep{2020A&A...641A...6P} and the KiDS-1000+BOSS+2dFLenS galaxy clustering and weak lensing survey \citep{2021A&A...646A.140H} as examples of the range of values present in the current literature.}
    \label{fig:models_onepanel}
\end{figure*}

We consider a cosmology with fiducial parameters as given by the \cite{2020A&A...641A...6P} \planck ``base-$\Lambda$CDM'' \texttt{TT,TE,EE+lowE+lensing} model, with values as described in \cref{tab:params}. Parameters are held at these fixed values for simulations, while parameter inference runs are initialised centred around these values, then sampled within the prior ranges shown and marginalised over to give results on other parameters. Where no prior is shown, the values are kept fixed throughout. Our main results are the posterior of the parameters $\Omega_{\rm m}$, $\sigma_8$ and $S_8 \equiv \sigma_8 \left( \Omega_{\rm m} / 0.3  \right)^{0.5}$, where $S_8$ is the standard parameter optimally constrained by galaxy lensing, in contrast to $S_8^{\rm CMBL} \equiv \sigma_8 \left( \Omega_{\rm m} / 0.3  \right)^{0.25}$ which is optimally constrained by CMB lensing alone. The leftmost panel of \cref{fig:models_onepanel} shows a simulation of our data vector (as described in \cref{subsec:sims}) plotted against predictions from cosmologies with different relevant $S_8$ values, giving an idea of the constraining power of the data.
\subsection{Nuisance model and parameters}
\label{subsec:nuisance_model}
Systematic uncertainties on galaxy cosmic shear power spectra are frequently dealt with by marginalising over simple parameterised models. Here, we consider models of three cases, following the choices in the baseline DES-Y3 analyses:
\begin{itemize}
    \item \textbf{Multiplicative shear bias}: The process of measuring weak lensing shear from noisy images can induce biases on the inferred power spectrum \citep[see, e.g.][]{2022MNRAS.509.3371M}. For current experiments, including DES-Y3, it has been shown to be adequate to model these using a single multiplicative parameter per tomographic bin $m_i$ \citep{2006MNRAS.368.1323H, 2006MNRAS.366..101H, 2020OJAp....3E..14K}, which modifies the power spectra as:
    \begin{equation}
        {C}^{\kappa_{\rm C} \gamma_{\rm E}, i }_{\ell} \rightarrow (1 + m_i) {C}^{\kappa_{\rm C} \gamma_{\rm E}, i }_{\ell}.
    \end{equation}
    In the second from the left panel of \cref{fig:models_onepanel} we show the effect of varying the $m$ nuisance parameter on the \clkg spectra alongside our simulated measurements for tomographic bin 4.
    \item \textbf{Source redshift distribution calibration}: Galaxy shear power spectra are highly sensitive to the redshift distribution function $n(z)$ within each tomographic bin of the sources used. Where the samples are selected using photometric information, as is the case in DES-Y3, the estimated $n(z)$ may have significant uncertainties in both the overall mean redshift and detailed shape. Though more sophisticated parameterisations of these uncertainties exist and are expected to be important for near-future experiments, it has been shown that for the weak lensing source galaxies in DES-Y3, it is adequate only to consider the uncertainty on the mean of the $n(z)$ within each tomographic bin \cite*[e.g.][]{2022MNRAS.511.2170C}. We include four additional nuisance parameters $\Delta z_{i}$ for a shift in the mean of each tomographic bin: at each likelihood evaluation step, we shift the distribution in each tomographic bin $i$ according to:
    \begin{equation}
        n_i(z) \rightarrow n_i(z + \Delta z_{i}).
    \end{equation}
    The second from the right panel of \cref{fig:models_onepanel} shows the effect of varying the $z$ nuisance parameter on the \clkg spectra alongside our simulated measurements for tomographic bin 4.
    \item \textbf{Galaxy Intrinsic Alignments}: The use of galaxy images as a proxy for gravitational shear relies on the assumption that intrinsic galaxy shapes are randomly oriented, which is not the case in reality. Physically close pairs of galaxies will tend to align their major axes towards overdensities local to them in positively correlated `intrinsic-intrinsic' (II) alignments. Negative `shear-intrinsic' (GI) correlations are also created when distant galaxies are tangentially sheared by lensing from foreground overdensities, which more nearby galaxies are gravitationally aligned towards. The power spectrum of the contaminating Intrinsic Alignments (IA) can be physically modelled in a number of ways \citep[see][and references therein]{2023MNRAS.524.2195S}. Here, we adopt the Nonlinear Linear Alignment (NLA) model \citep{2007MNRAS.381.1197H, 2007NJPh....9..444B} which makes the simplifying assumption that IAs are from E-mode GI alignments only, neglecting the intrinsic-intrinsic B-mode term which is also possible to consider in the \clkg observable. The NLA model treats the GI alignment power spectrum as a simple scaling of the matter power spectrum with a redshift evolution. We infer the two parameters $A_{\rm IA}$ and $\eta_{\rm{IA}}$ across all tomographic bins, corresponding to a substitution in the galaxy lensing kernel:
    \begin{equation}
    \label{eq:ia_nla}
        \wgalaxy(z) \rightarrow \wgalaxy(z) -A_{\rm IA} C_{1} \rho_{\rm cr} \frac{\Omega_{\rm m}}{G(z)} n(z)
         \Big( \frac{1 + z}{1 + z_0} \Big)^{\eta_{\rm{IA}}}.
    \end{equation}
    Here, $z_0$ is pivot redshift fixed to $0.62$ as in \cite*{2022PhRvD.105b3515S}, $G(z)$ is the linear growth factor and $C_{1} = 5\times 10^{14} \, M^{-1}_{\odot} h^{-2} \mathrm{Mpc}^3$ is the normalisation constant. The rightmost panel of \cref{fig:models_onepanel} shows the effect of varying $A_{\rm IA}$ on the theory spectra for tomographic bin 4. Whilst the more sophisticated Tidal Alignment and Tidal Torquing (TATT) model was adopted as fiducial for the DES-Y3 3x2pt analysis of \cite{2022PhRvD.105b3520A}, when considering only the shear part of the data, \cite*{2022PhRvD.105b3515S} find a mild preference for the simpler NLA model (row three of Table III in that work) which we therefore choose to adopt for reasons of both model and implementation simplicity.
\end{itemize}

\subsection{Likelihood computation}
\label{subsec:likelihood}
We compute a simple Gaussian likelihood ($\mathcal{L}$) between our data vector bandpowers and binned theory vector at a given set of cosmological and nuisance parameters $\boldsymbol{\theta}$ using the covariance matrix, $\mathbb{C}$, calculated in \cref{subsec:pseudo_cl}:
\begin{equation}
    -2\ln{\mathcal{L}}=\sum_{\ell\ell^\prime}\big[\hat{C}^{\kappa_{\rm{C}} \gamma_{\rm{E} } }_{\ell} -{C}^{\kappa_{\rm{C}} \gamma_{\rm{E} } }_{\ell}(\boldsymbol{\theta})\big]{\mathbb{C}}^{-1}_{\ell\ell^\prime}\big[\hat{C}^{\kappa_{\rm {C} }\gamma_{\rm {E} } }_{\ell^\prime} - {C}^{\kappa_{\rm {C} } \gamma_{\rm {E} }}_{\ell^\prime}(\boldsymbol{\theta})\big],
 \end{equation}
where $\hat{C}^{\kappa_{\rm{C}} \gamma_{\rm{E} } }_{\ell}$ is the data vector and ${C}^{\kappa_{\rm{C}} \gamma_{\rm{E} } }_{\ell}$ is the model power spectrum. The posterior probability for the parameters is then proportional to the likelihood multiplied by the priors ($\Pi$): $P(\boldsymbol{\theta} | \hat{C}^{\kappa_{\rm{C}} \gamma_{\rm{E} } }_{\ell}) \propto \mathcal{L}(\hat{C}^{\kappa_{\rm{C}} \gamma_{\rm{E} } }_{\ell} | \boldsymbol{\theta}) \Pi(\boldsymbol{\theta})$. The choices of prior distributions are detailed in \cref{subsec:priors}.

\subsubsection{Hartlap correction}
\label{subsubsec:hartlap}
Because the fiducial covariance matrix is estimated from a finite number of simulations, the inverse covariance matrix used in the likelihood computation is known to be a biased estimate of the true inverse covariance matrix \citep{Anderson2003, 2007A&A...464..399H}. To account for this, we apply the well-known Hartlap correction to the inverse covariance matrix:
\begin{equation}
    \mathbb{C}^{-1} \rightarrow \alpha \mathbb{C}^{-1}; \alpha \equiv \frac{N_{\rm sims} - N_{\rm data} - 2}{N_{\rm sims} - 1},
\end{equation}
where $N_{\rm sims}$ is the number of simulations and $N_{\rm data}$ is the length of the data vector. 
We use the corrected covariance matrix for computing our likelihood. For our 511 simulations and 24 data points, the size of the Hartlap correction is $\alpha = 0.951$. The choice of $\Delta \ell = 300$ reduces the total number of data points in the data vector, which is optimal compared to $\Delta \ell < 300$ because, for a given number of simulations, fewer data points minimize the impact of the Hartlap correction.

\subsection{Prior choice}
\label{subsec:priors}
In \cref{tab:params}, we show the set of cosmological and nuisance parameters varied in our Monte Carlo chains. Fiducial cosmological parameters (the values at which simulations are performed and that in inference sampling runs are used to initialise the chains) are chosen to coincide with those of the \planck Collaboration's ``base-$\Lambda$CDM'' \texttt{TT,TE,EE+lowE+lensing} model from \cite{2020A&A...641A...6P} and priors are wide enough to capture all reasonable cosmologies at the time of writing. Whilst our \clkg observable depends only weakly on the Hubble expansion parameter $H_0$, we found in initial runs based on a simulated data vector that when this parameter was kept fixed, a sharp boundary appeared in the two-dimensional $(\sigma_8, \Omega_{\rm m})$ plane, with the lower right section of the `banana' shape being cut off. This did not affect the posterior on $S_8$ but did lead to an artificial bi-modality in the one dimensional $\Omega_{\rm m}$ constraint. Allowing $H_0$ to vary removed this effect and is in line with the prior treatments of CMB lensing and cross-correlation data vectors \citep[e.g.][]{2023PhRvD.107b3530C, ACT:2023kun}.

\subsection{Posterior sampling}
\label{subsec:sampling}
We sample from the posterior using the Markov Chain Monte Carlo Metropolis sampler distributed with \cobaya \citep{2002PhRvD..66j3511L, 2013PhRvD..87j3529L,2019ascl.soft10019T,2021JCAP...05..057T}. We first run a chain in our fiducial parameterisation and with the simulated data vector to convergence without defined scales for the mixed Gaussian-exponential proposal distribution used for taking steps\footnote{As described in \url{https://cobaya.readthedocs.io/en/latest/sampler_mcmc.html\#covariance-matrix-of-the-proposal-pdf}. }. We subsequently use the proposal covariance matrix learned during this chain to speed up convergence for all subsequent chains. We regard chains as converged when the Gelman-Rubin criteria reach a value $R - 1 < 0.01$ and the first $30\%$ of chains are removed as burn-in. For all of our chains, this results in a number of effective samples in the range of 1500-2000.

\begin{table}
	\centering	
	\caption{	
        The parameters and priors used in the model specification within a $\Lambda$CDM cosmology. Fiducial values are used for simulations and initialisation of inference chains. Priors are either Uniform $\mathcal{U} [\mathrm{min}, \mathrm{max}]$ or Gaussian $\mathcal{N}(\mu,\sigma)$. Unlisted other cosmological parameters and model choices are fixed to their default values in \camb v1.3.5 \citep{antony_lewis_2022_6137931}}
	\vspace{-0.2cm}
	\begin{tabular}{ccc}
		\hline
            \hline
		Parameter & Fiducial & Prior \\\hline
		\multicolumn{3}{l}{\textbf{Cosmology Sampled}}  \\
		$\Omega_{\rm c} h^2$ & 0.120  & $\mathcal{U}[0.05, 0.99]$ \\ 
		$\log(A_\mathrm{s}10^{10})$ & 3.042 &  $\mathcal{U}[1.6, 4.0]$  \\ 
		$H_0$  & 67.36 & $\mathcal{U}[40, 100]$   \\
		\hline

            \multicolumn{3}{l}{\textbf{Cosmology Fixed} } 	 \\
            $\Omega_{\rm b} h^2$ & 0.0224 & - \\
            $n_s$ & 0.9649 & - \\
            $\sum m_{\nu} \, [\mathrm{eV}]$ & 0.06 & - \\
            \hline
  
  		\multicolumn{3}{l}{\textbf{Galaxy Intrinsic Alignment} } 	 \\
		$A_{\rm IA}$ & 0.35 & $\mathcal{N}(0.35, 0.65)$ \\
            $\eta_{\rm IA}$ & 1.66 & $\mathcal{N}(1.66, 4)$ \\
            \hline
		
		\multicolumn{3}{l}{\textbf{Galaxy redshift calibration} }\\
		$\Delta z_1 $ & 0.0 & $\mathcal{N}(0.0, 0.018)$\\
            $\Delta z_2 $ & 0.0 & $\mathcal{N}(0.0, 0.015)$\\
            $\Delta z_3 $ & 0.0 & $\mathcal{N}(0.0, 0.011)$\\
            $\Delta z_4 $ & 0.0 & $\mathcal{N}(0.0, 0.017)$\\
            \hline
		
		\multicolumn{3}{l}
            {\textbf{Galaxy shear calibration}}\\
            $m_1 $ & $-0.006$ & $\mathcal{N}(-0.006, 0.009)$\\
            $m_2 $ & $-0.020$ & $\mathcal{N}(-0.020, 0.008)$\\
            $m_3 $ & $-0.024$ & $\mathcal{N}(-0.024, 0.008)$\\
            $m_4 $ & $-0.037$ & $\mathcal{N}(-0.037, 0.008)$\\

            \hline
		\hline
	\end{tabular}
	\label{tab:params}

\end{table}

\section{Validation of data and method}\label{sec:validation}
We validate the data vector using two null tests and check for systematic contamination from Galactic dust and stars. Before applying it to the data, we validated the parameter inference methodology using simulations. This includes checking for the absence of bias in the inferred parameters, robustness to the choice of the covariance matrix, and robustness to the effect of the intrinsic galaxy alignment modelling.
\subsection{Data vector null tests}\label{subsec:data_null_tests}
We check the data for some non-idealities that may be present. We compute the $\chi^2$ of the statistic under consideration, for which we set PTE = 0.05 as a threshold for considering the test failed. Our unblinding decision was based on the $\chi^2$ and PTEs computed with bandpowers of four redshift bins considered together. We consider a null test to be passed if the PTE exceeds this threshold.\footnote{Unblinding of the data vector was performed assuming a one-sided PTE threshold, with PTE < 0.05 indicating a failed null test.
However, as demonstrated in this section, the data vectors for the full dataset pass the null test even when considering two-sided PTE distributions.} 

\begin{figure}
    \centering
    \includegraphics[width=0.45\textwidth]{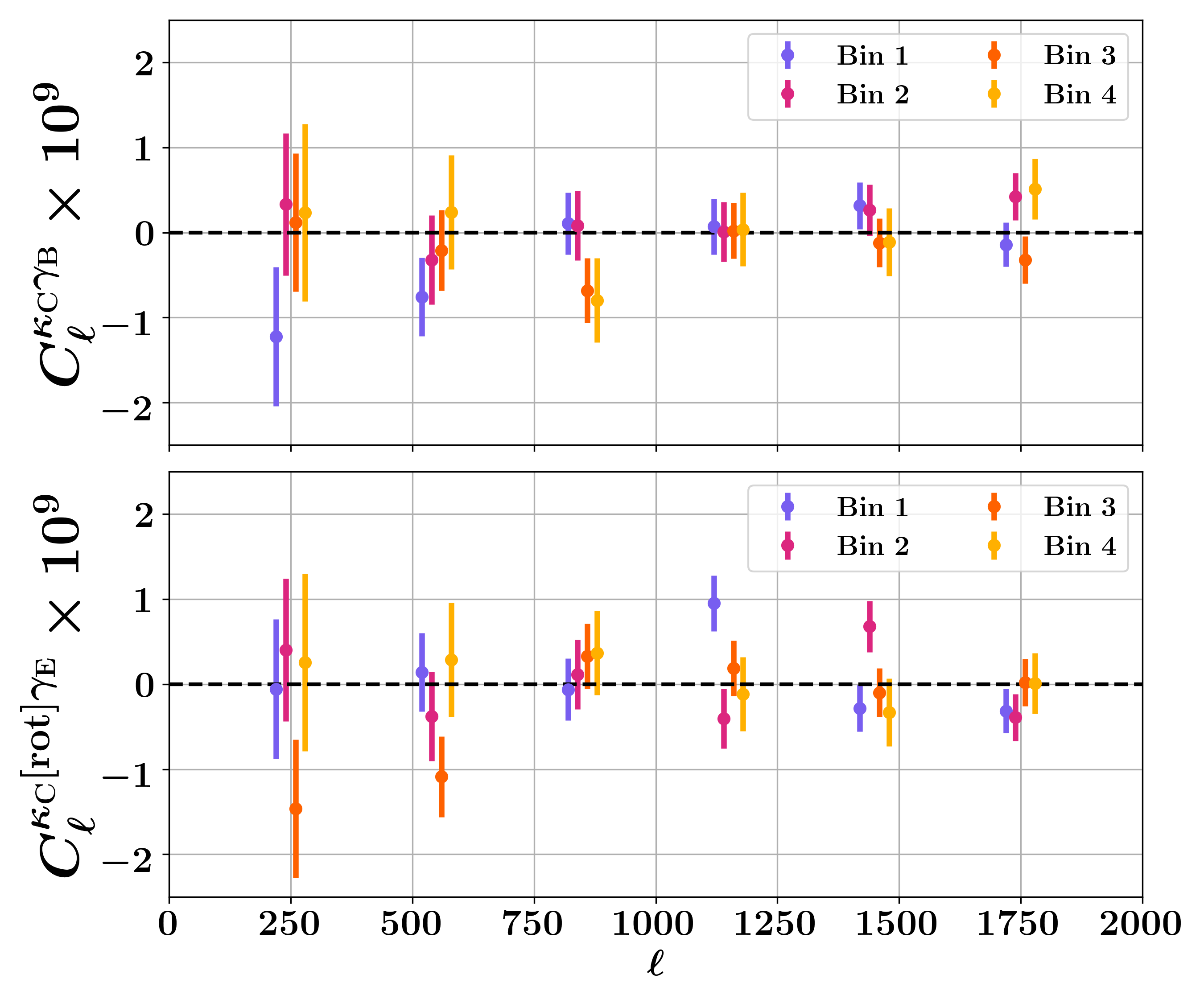}
    \caption{\emph{Top:} The power spectrum between \kcmb and the B-mode of the shear ($C^{\kappa_{\rm C} \gamma_{\rm B} }_{\ell}$). \emph{Bottom:} The correlation between \kcmb and the E-mode of the shear map ($C^{\kappa_{\rm C} [\rm rot] \gamma_{\rm E} }_{\ell}$) obtained from the catalogue in which DES-Y3 catalogue ellipticities are randomly rotated. The error bar indicates $1 \sigma$ uncertainty of the statistic. We find both sets of bandpowers to be consistent with zero.}
    \label{fig:null_test_AP}
\end{figure}

At linear order and under the Born approximation, weak lensing of galaxies by large-scale structure is not expected to give rise to B-modes in the shear map. Therefore, we do not expect any significant B-mode signal of cosmological origin in the shear map. Moreover, obtaining E and B modes from the partial sky shear maps can cause mixing between E and B modes. In the absence of such spurious B-modes, the correlation of the B-modes in the shear maps with the \kcmb is expected to be consistent with zero. In \cref{fig:null_test_AP}, we show the $C^{\kappa_{\rm C} \gamma_{\rm B} }_{\ell}$ bandpowers for four redshift bins with their error bars. With the blinded data vector, we find the four data vectors together are consistent with zero with PTE = 0.81, indicating the absence of spurious B-modes in the shear data. The PTEs for the individual redshift bin bandpowers are 0.47, 0.75, 0.57, and 0.59, respectively.

We also correlate the \kcmb map with the shear map obtained from the DES-Y3 catalogue, where ellipticities are randomly rotated. The random rotation is expected to wash out any cosmological signal in the shear maps and, indeed, is how we obtain the shear noise for the mock shear simulations, as discussed in \cref{subsec:sims}. Hence, the correlation of these maps with the \kcmb map tests for any non-cosmological features in shear maps that may correlate with the \kcmb map. From \cref{fig:null_test_AP}, with the blinded data vector, we find this correlation is also consistent with zero with PTE = 0.19. The PTE values for individual redshift bin bandpowers are 0.07, 0.20, 0.19, and 0.96. Note that we do not regard the 0.96 as a failure, as these per-bin PTE numbers were not part of our unblinding criteria. For the larger set of PTEs generated by including per-bin calculations, it is more likely that a failure appears by chance from sampling this part of the Uniform distribution expected for the PTE values. These tests provide an important check of the analysis pipeline. 

We obtain the covariance matrices for both tests using 511 simulations. For the B-mode null test, the covariance matrix is obtained using $C_{\ell}^{\kappa_{\rm C} \gamma_{\rm B}}$ bandpowers computed using 511 simulations. For the rotation null test, we compute \clkg bandpowers; hence the covariance matrix is the same as used for the signal \clkg bandpower.

\subsection{Diagnostic tests using survey property maps}\label{subsec:survey_prop}
\begin{figure}
    \centering
    \includegraphics[width=0.45\textwidth]{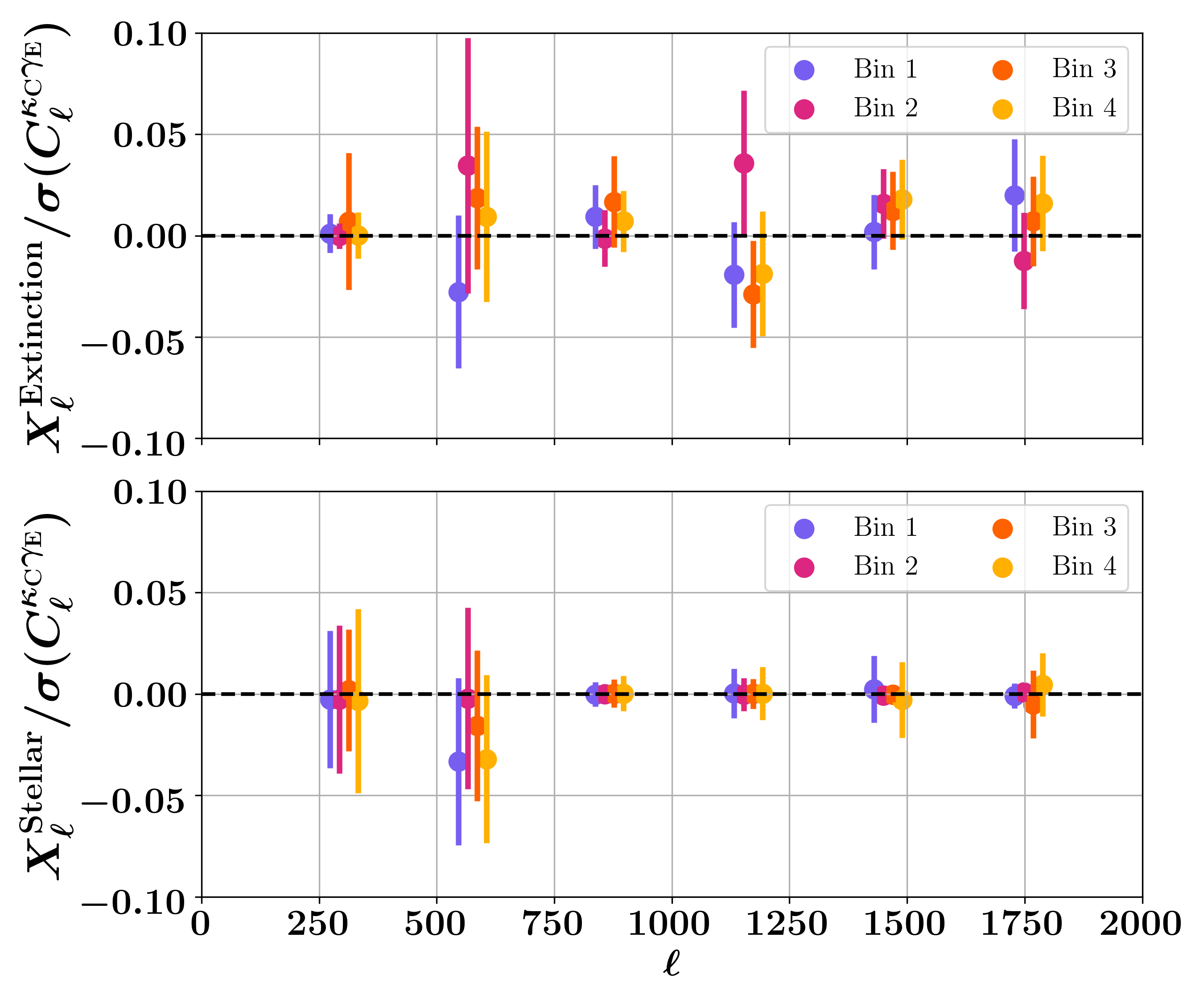}
    \caption{Survey property correlation statistics $X_{\ell}$ in units of error bar of \clkg, for the dust extinction map (top panel) and the stellar density (bottom panel). The error bars are obtained using $X_{\ell}$ computed over 28 jackknife samples of the data. For both survey properties, $X^{S}_{\ell}$ is less than $5\%$ of the error bar on \clkg and consistent with zero.}
    \label{fig:X_survey_prop}
\end{figure}

Any systematic effect or contamination ($S$) that simultaneously affects both the observables, \kcmb reconstructed from the observed CMB and $\bm{\gamma}$ estimated using galaxy shape measurements, can lead to a bias in the measurement of \clkg. For example, the Galactic dust can affect CMB lensing reconstruction through its presence in the CMB map, and the extinction by dust can affect the measurement of galaxy properties. The following statistic captures the amplitude of contamination to \clkg coming from a given survey property $S$ \citep{2019PhRvD.100d3501O, 2023PhRvD.107b3530C}:
\begin{equation}
    X^{S}_{\ell} = \frac{C^{\kappa_{\rm C} S}_{\ell} C^{\gamma_{\rm E} S}_{\ell}}{C^{S S}_{\ell}}.
\end{equation}
In this work, we consider two survey properties: dust extinction, where $S$ is the map of E(B-V) reddening \citep{1998ApJ...500..525S} and stellar density, where $S$ is the map of stellar density \citep{2021ApJS..254...24S, 2021ApJS..255...20A, 2023MNRAS.tmp.2370S}. In \cref{fig:X_survey_prop}, we show $X^{S}_{\ell}$ for both survey properties with its error bar. We obtained the error bar using the delete-one patch jackknife method, where we used 28 jackknife samples of the data and the survey property maps. For both survey properties, the effect on \clkg is within a few per cent of the error on \clkg and hence is of less concern. For dust extinction $X_{\ell}$, we obtain the PTE for four bins as 0.895, 0.698, 0.769, and 0.659, indicating no significant detection of dust contamination in cross-correlation. For the stellar density, the PTE values are 0.993, 0.999, 0.999, and 0.992. Whilst these PTE values are high and close to one, they indicate an over-consistency with zero according to the estimated error bars.  The jackknife method overestimates the error bar in general \citep{2009MNRAS.396...19N, 2021MNRAS.505.5833F}. Therefor, we do not regard high PTE as problematic in the case of a diagnostic test.

\subsection{Model validation}\label{subsec:model_tests}
\begin{figure}
    \centering
    \includegraphics[width=0.475\textwidth]{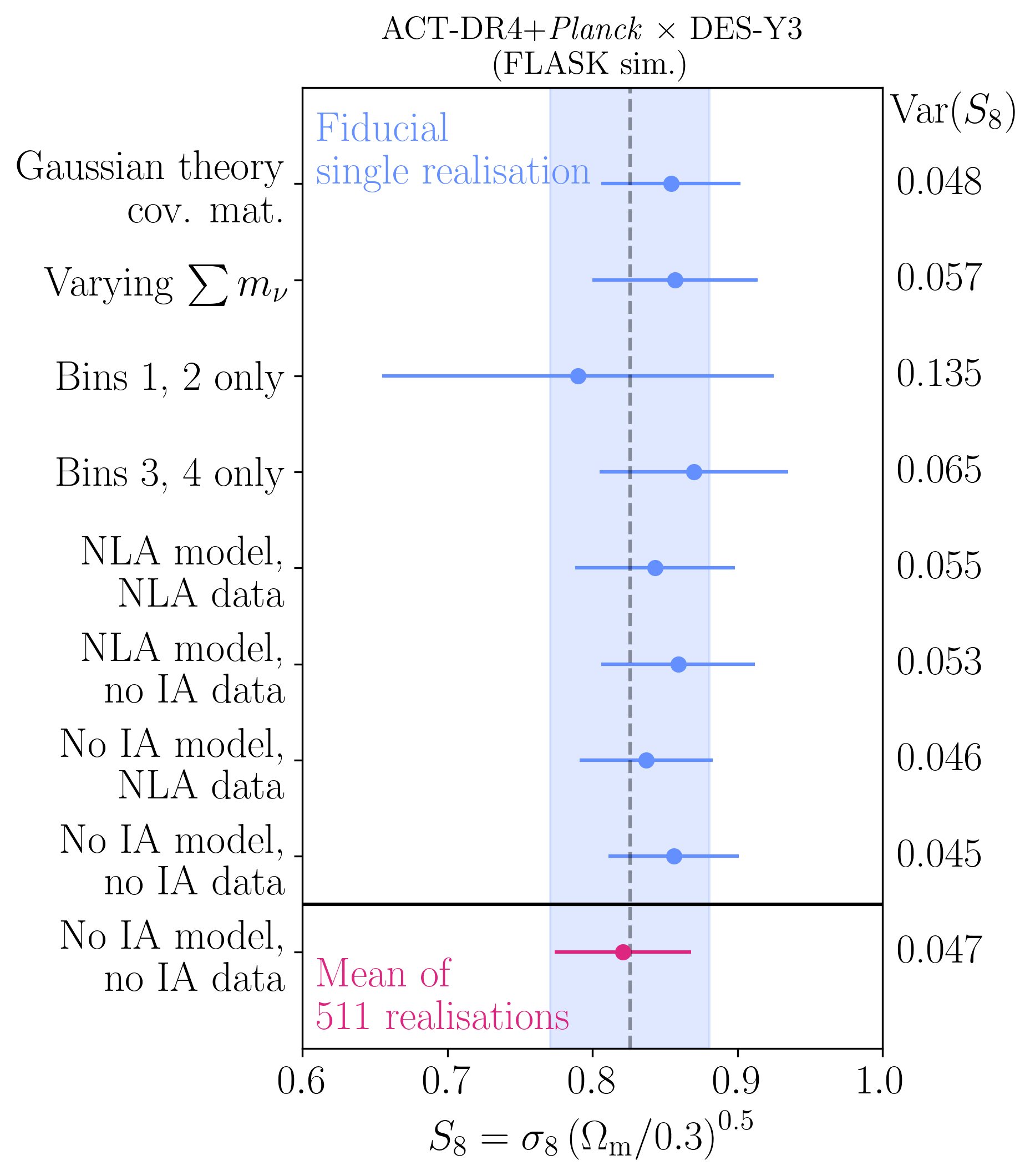}
    \caption{The stability of the recovery of the $S_8$ parameter from our simulated data vector as we change the model used for the inference. The dashed vertical line represents the true value of the input to the simulation, and the shaded band is the error bar in the fiducial model setup. Note that the slightly (but not significantly) high value recovered in the upper rows is consistent with what is expected for this single realisation (see \cref{fig:alens_model_recovery}). The final row shows the result for one model using a data vector which is the mean of 511 realisations.}
    \label{fig:model_stability}
\end{figure}

\subsubsection{IA model robustness}
\label{subsubsec:ia_robustness}
In order to assess the robustness of our inference to the model chosen for intrinsic galaxy alignments, we create two simulated data vectors, one without any Intrinsic Alignment (IA) signal and one in which the observed angular power spectrum $C_\ell$s have additional power added from a Non-linear Linear Alignment (NLA) model as described in \cref{eq:ia_nla} with the fiducial parameter values $\lbrace A_{\rm IA}, \eta_{\rm IA}\rbrace = \lbrace 0.35, 1.66 \rbrace$ (corresponding to the mean posterior values from DES-Y3 shear-only analysis in Table III of \citealt*{2022PhRvD.105b3515S}) as shown in \cref{tab:params}. In \cref{fig:model_stability}, we show the stability of our measurement of $S_8$ to the choice of IA model, with no significant parameter shifts observed when considering mismatched choices of model and data (e.g. when the NLA model is used on a data vector with no IA signal and vice-versa). We chose to include the full NLA model parameterisation for our fiducial inference runs on the data.

\begin{figure}
\centering
\includegraphics[width = 0.49\textwidth]{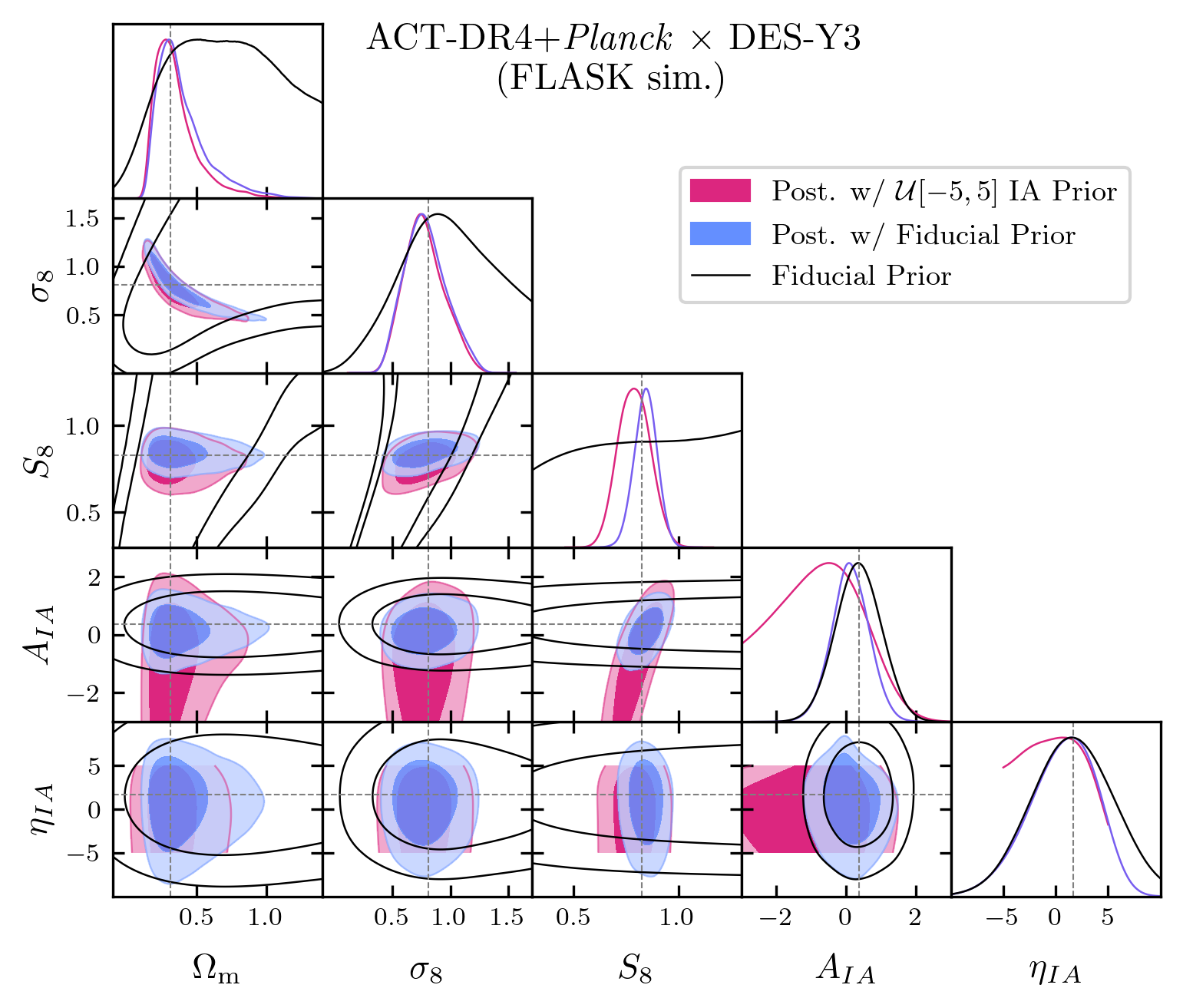}
\caption{Posteriors showing recovery of the input model from data simulations under the fiducial prior (blue) and an alternate prior choice on galaxy intrinsic alignment (IA) parameters (red), that is less localized in $A_{\rm IA}$ direction. The fiducial prior (black) is informed by the DES-Y1 cosmic shear analyses (which use data independent from those used here) and expectations for the NLA model. The wide uninformative prior causes the mild correlation between $A_{\rm IA}$ and $S_8$ to drag the posterior on the latter to lower values. See text for further discussion.}
\label{fig:wideia_prior}
\end{figure}

Within this parameterization, we choose to include an informative prior on $A_{\rm IA} \sim \mathcal{N}(0.35, 0.65)$ and $\eta_{\rm IA} \sim \mathcal{N}(1.66, 4)$, with prior widths a factor four wider than the posterior on NLA IA parameters found in DES-Y1 data from the 3x2pt analysis of \cite{2018PhRvD..98d3526A}. Note that the DES-Y1 data are in a sky region not included in our analysis so this prior can be regarded as independent. We refer to this prior as our `fiducial prior' and it is shown throughout as black unfilled contours. In order to assess the impact of this choice, we also run an inference chain on the simulated data vector with broad priors $\mathcal{U}(-5,5)$ on both parameters, matching the \emph{priors} used in the DES-Y3 analyses. Note that the latter is a broader prior in the sense that it is less localized in the $A_{\rm IA}$ direction compared to the fiducial prior. The results can be seen in \cref{fig:wideia_prior}. Here, it can be seen that whilst the 1D posteriors on \Omegam and $\sigma_8$ are not affected, a mild degeneracy between $A_{\rm IA}$ and $S_8$ causes a widening of the posterior on $S_8$ and shift of the peak to lower values. For positive values of $A_{\rm IA}$, the constraint is relatively unaffected by the choice of prior, and we can indeed see some constraining power of the data appearing due to the similar upper limits from both the wide and the informative prior. For negative values of $A_{\rm IA}$, it can be seen that the lower limit is dominated by the prior in the fiducial case, with the posterior extending significantly further for the wide prior. This lack of constraining power causes a `projection effect', which lowers the inferred value of $S_8$. However, the galaxy formation physics represented by the NLA IA model is expected to result in $A_{\rm IA} > 0$ for red galaxies, and this has been observationally shown to be the case (with KiDS+GAMA \citealt{2019A&A...624A..30J} find $A^{\rm Red}_{\rm IA} = 3.18^{+0.47}_{-0.46}$ and with DES-Y1 \citealt{2019MNRAS.489.5453S} find $A^{\rm Red}_{\rm IA} = 2.38^{+0.32}_{-0.31}$). For blue galaxies in the NLA model, $A_{\rm IA} < 0$ is possible, but observations have so far been consistent with zero and inconsistent with large negative values (\citealt{2019A&A...624A..30J} find $A^{\rm Blue}_{\rm IA} = 0.21^{+0.37}_{-0.36}$ and \citealt{2019MNRAS.489.5453S} find $A^{\rm Blue}_{\rm IA} = 0.05^{+0.10}_{-0.09}$). For weak lensing samples such as DES-Y3, which contain a mixture of red and blue galaxies, with red fraction $f_{\rm Red} \sim 20\%$ this means $A_{\rm IA} = A^{\rm Red}_{\rm IA}f_{\rm Red} + A^{\rm Blue}_{\rm IA}(1 - f_{\rm Red})$ is even more constrained to be positive. 
Motivated by these considerations, we keep the informative prior on IA parameters for the fiducial analysis. 
\subsubsection{Neutrino model robustness}
\label{subsubsec:neutrino_robustness}
We chose as our baseline model three neutrinos of degenerate mass, consistent with the model choice of the \cite{2022PhRvD.105b3520A}, with $\sum m_{\nu} = 0.06\,$eV in our fiducial analysis. We find that differences between our \clkg observable are at most $\sim0.5\%$ when comparing three degenerate neutrinos to the normal hierarchy case and at most $\sim2\%$ when comparing to a single massive neutrino scenario.

The relevant row of \cref{fig:model_stability} shows the stability of our $S_8$ measurement to the marginalisation over the sum of neutrino mass, with a uniform prior $\sum m_{\nu} \, [\mathrm{eV}] \sim \mathcal{U}[0.0, 1.0]$, in this model.

\subsection{Covariance matrix validation}\label{subsec:covmat_validation}
In addition to the covariance matrix computation with \flask simulations, as detailed in \cref{subsec:pseudo_cl}, we also construct an analytical Gaussian covariance matrix for the pseudo-$C_\ell$ estimator. The covariance matrix estimated using simulation bandpowers is expected to model non-Gaussian contributions more correctly than the theoretical case but suffers from realisation noise effects since it is an average over the finite number of simulations. The results of parameter constraints obtained using the analytical covariance matrix are shown in the relevant row of \cref{fig:model_stability}. As can be seen, both methods give consistent posteriors on our simulated data vector. Therefore, we choose to continue with the simulation-based covariance matrix. This choice is somewhat arbitrary, but on the understanding that as the constraining power of the data improves with future data releases, the effects modelled in the simulations will become more significant. 

\subsection{Recovery of input model from mock data}
\label{subsec:input_recovery}
\begin{figure}
    \centering
    \includegraphics[width=0.4\textwidth]{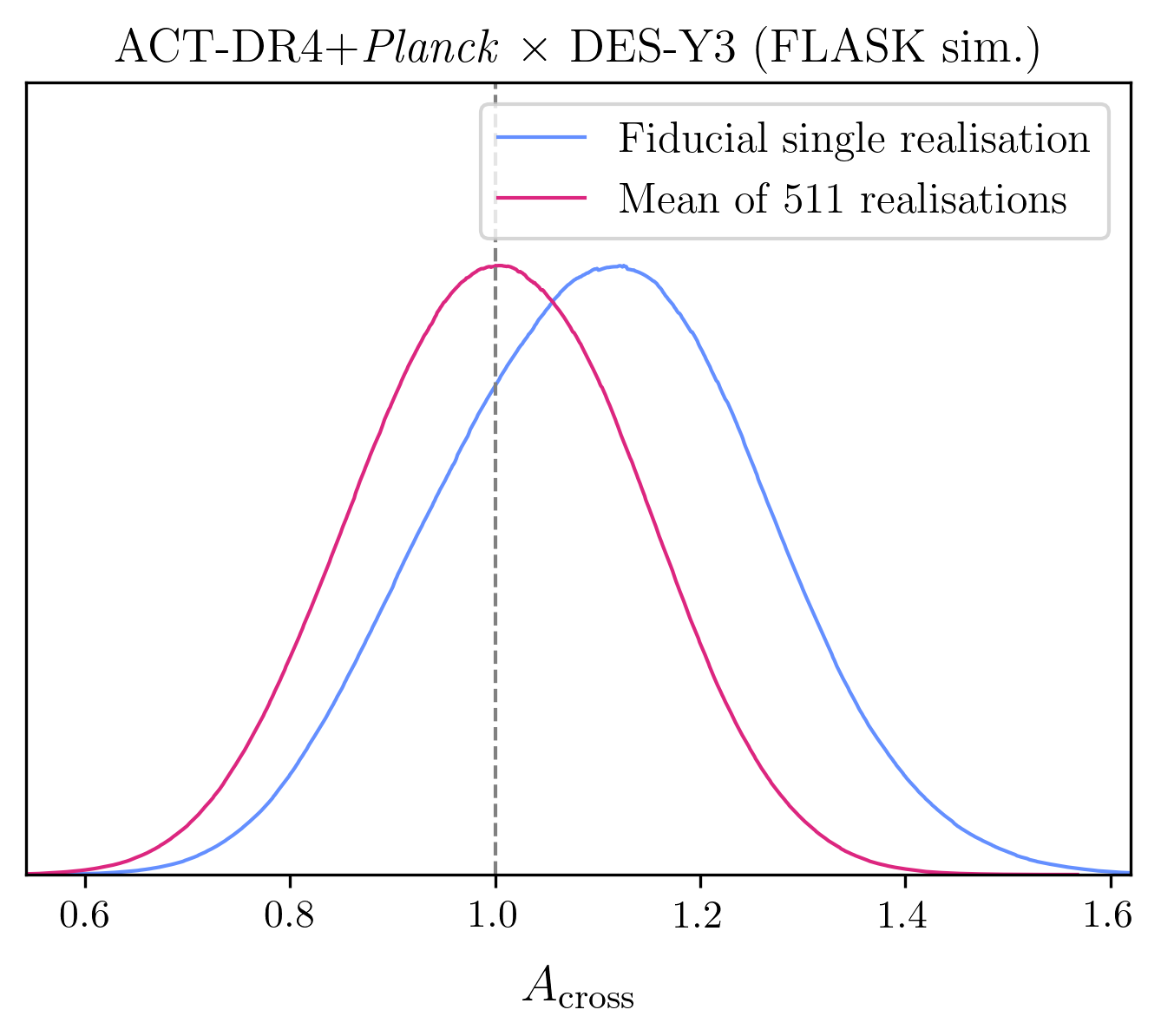}
    \caption{Recovery of the $A_{\rm cross}$ parameter from our simulated data vector. All cosmological and nuisance parameters are held at their true input values, and $A_{\rm cross}$ is an overall amplitude parameter for the \clkg spectra, which has the value one at the true input model.}
    \label{fig:alens_model_recovery}
\end{figure}
In order to verify that our inference pipeline is capable of making an unbiased recovery of cosmological parameters, we run it on a data vector recovered from one of the 511 \flask simulations described in \cref{subsec:sims}. The results of this analysis are shown in \cref{fig:wideia_prior} for cosmological and IA parameters. In \cref{fig:model_recovery_cosm,fig:model_recovery_nuisance}, we show the posterior and prior, including those for observational nuisance parameters, but zoomed out to show the full shapes of the prior. The priors are specified in \cref{tab:params} and are shown as unfilled contours in the space of the inferred parameters, indicating the level of information gained from the data (or lack of it in the case of the prior-dominated nuisance parameters). As can be seen, all inferred parameters are recovered with biases smaller than the $68\%$ credible interval, as may be expected from realisation noise. In addition to this full parameter inference on a single simulation, we have also inferred only the $A_{\rm cross}$ parameter for this simulation \emph{and} a data vector which is the mean of the 511 simulations. $A_{\rm cross}$ is a phenomenological parameter which modifies the overall amplitude of the lensing spectra with respect to that predicted by a model with a fixed set of cosmological parameters (here, the true input parameters to the simulation):
\begin{equation}
    {C^{\kappa_{\rm C} \gamma_{\rm E} }_{\ell}}_{\rm obs} = A_{\rm cross} {C^{\kappa_{\rm C} \gamma_{\rm E} }_{\ell}}_{\rm true}
\end{equation}
As shown in \cref{fig:alens_model_recovery}, this gives a mean value and $68\%$ credible interval of $A_{\rm cross} = 1.00\pm 0.13$ for the mean data vector and $A_{\rm cross} = 1.10\pm 0.15$ for the fiducial realisation, confirming the finding that this single realisation has a random fluctuation towards higher clustering amplitude $S_8$, as seen in the full inference in \cref{fig:model_stability}.

\subsection{Internal consistency}
\label{subsec:internal_consistency}
\begin{figure}
    \centering
    \includegraphics[width=0.45\textwidth]{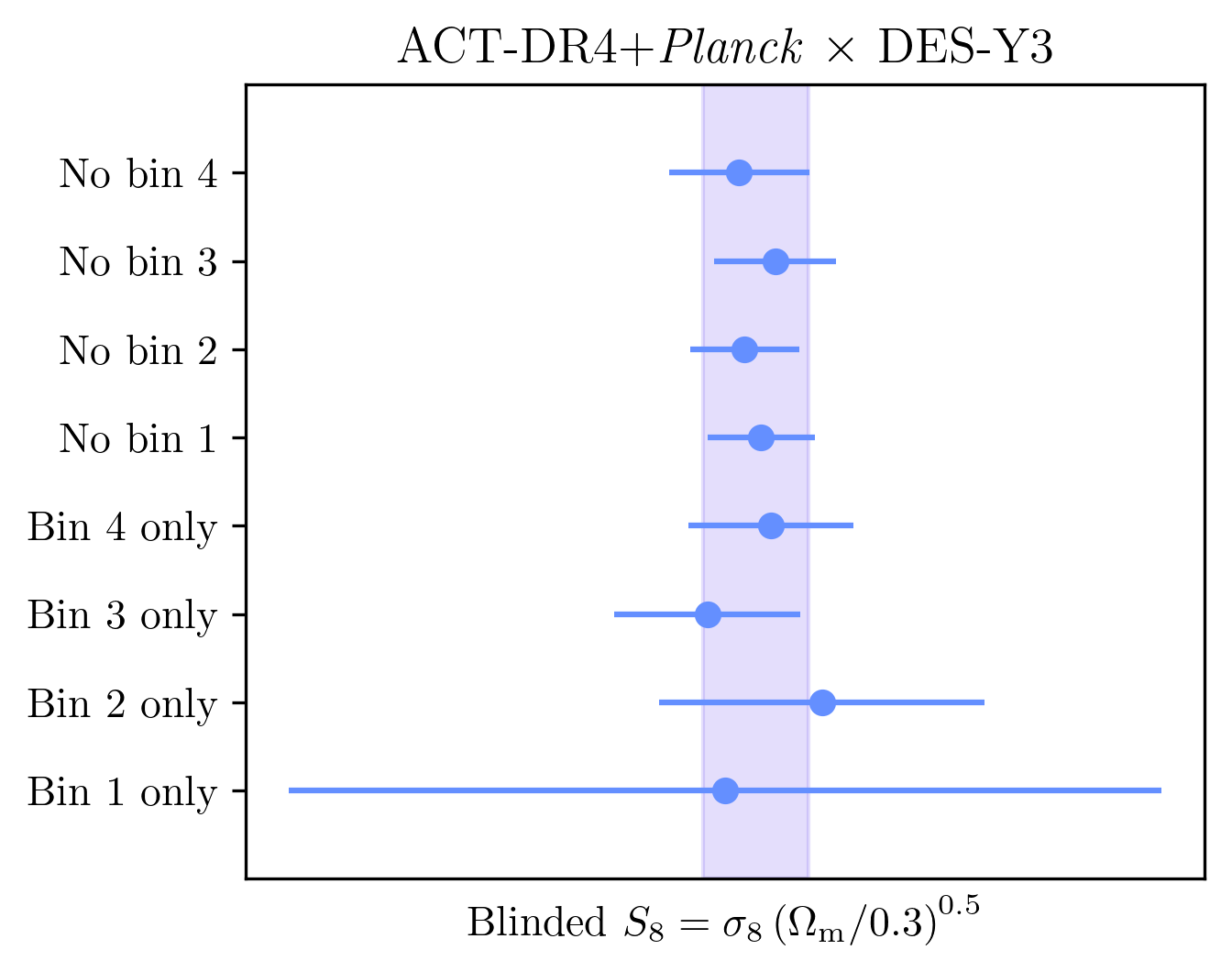}
    \caption{Stability of our 1D marginalised measurement of $S_8$ when using different parts of the full fiducial data vector, with the fiducial result shown as the shaded band. In each row, we either remove a single DES tomographic bin or use the data from only one.}
    \label{fig:split_stability}
\end{figure}
In order to test the robustness of our result, we perform the inference on a number of different splittings of the full data set. These involve i) leaving out data from individual tomographic bins in turn and ii) only using data from an individual tomographic bin in turn.
The results of these runs are shown in \cref{fig:split_stability}. These checks were performed \emph{before} the blinding factor applied to the DES shear catalogue (see \cref{subsubsec:des_blinding} for details) was removed in order to act as an extra confirmation of the adequacy of the analysis pipeline. As can be seen, each data split is consistent with all others within the expected scatter, and the behaviour of the error bars matches physical expectations, with progressively larger amounts of cosmic shear lensing signal contained in higher redshift tomographic bins.

\section{Results}
\label{sec:results}
\begin{figure*}\label{fig:data_cls}
    \centering
    \includegraphics[width=\textwidth]{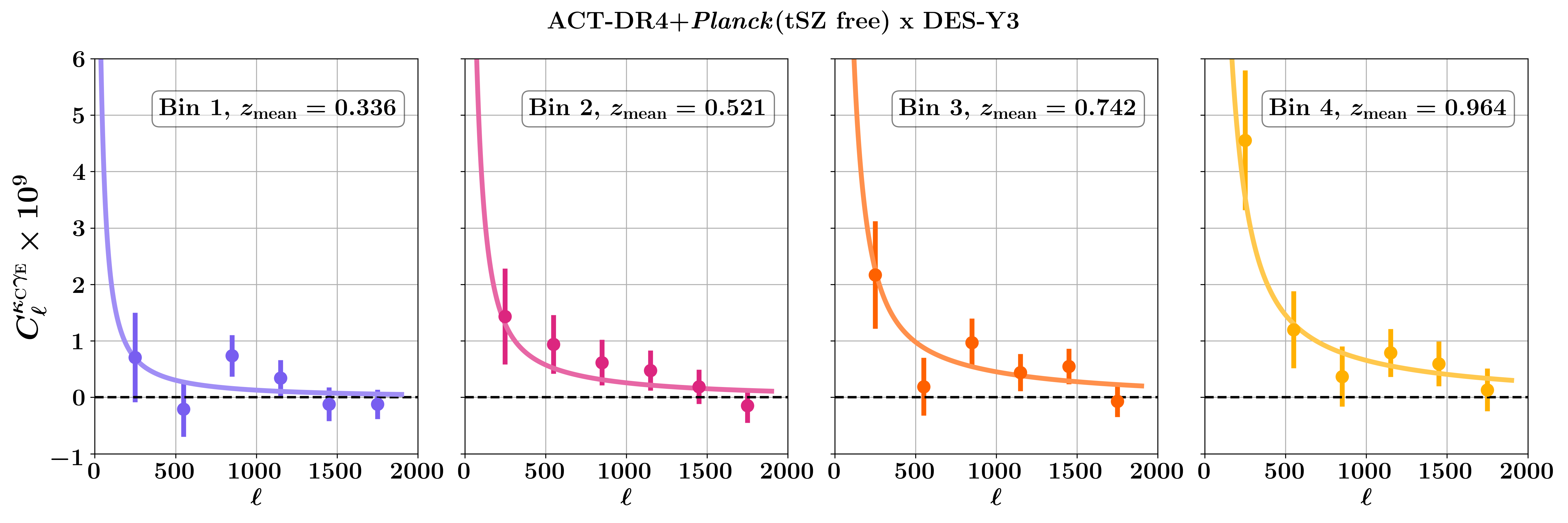}
    \caption{The points with error bar show \clkg bandpowers with ACT+\planck tSZ-free \kcmb and DES-Y3 shear, with four redshift bins. Error bars are the square root of the diagonal of the simulation covariance matrix, and $z_{\rm mean}$ is the mean redshift of the source galaxy distribution taken from \citep{2022PhRvD.105b3520A}. The curves show the best-fit theory \clkg corresponding to best-fit parameters in \cref{tab:results_parameters}.}
\end{figure*}

With the above work demonstrating that we have a data vector passing null tests for systematics contamination and that we have a working unbiased measurement pipeline, we now present the results using our fiducial ACT-DR4+\emph{Planck}-tSZ deprojected data vector. We unblind the data vector by obtaining the numerical value of the blinding factor $f$ (as described in \cref{subsubsec:des_blinding}) and applying the inverse of the blinding transformation to the catalogue shear values. We then re-make our maps and two-point data vectors and proceed. \cref{fig:data_cls} shows the data for four tomographic bins along with the best-fit theory model \clkg. As a validation, we also compute the full set of results for \clkg with the ACT-DR4-only \kcmb, and these are presented in \cref{app:actonly}.

\subsection{Lensing amplitude $A_{\rm cross}$ }
\label{subsec:alens_results}
We first consider the case in which we fix all other parameters to our fiducial values as in \cref{tab:params} and vary only the normalisation ($A_{\rm cross}$) of the observed \clkg spectrum relative to the prediction from this model:
\begin{equation}
{C^{\kappa_{\rm C} \gamma_{\rm E} }_{\ell}}_{\rm obs} = A_{\rm cross} {C^{\kappa_{\rm C} \gamma_{\rm E} }_{\ell}}_{\rm Planck}
\end{equation}
Under a uniform prior of $A_{\rm cross} \sim \mathcal{U}[0.0, 2.0]$ we find a measurement of $A_{\rm cross} = 0.84^{+0.16}_{-0.13}$ indicating agreement with the \planck result but mildly favouring a lower amplitude. This compares to previous determinations of $A_{\rm cross}$ using \clkg data: from POLARBEAR (polarisation lensing)$\times$HSC of $1.70\pm 0.48$ \citep{2019ApJ...882...62N}; from \planck\unskip$\times$HSC of $0.81\pm0.25$ \citep{2020ApJ...904..182M} and ACT+\planck\unskip$\times$KiDS-1000 of $0.69\pm0.14$ \citep{2021A&A...649A.146R}. Other previous measurements of $A_{\rm cross}$ from \clkg have used earlier \planck data releases for the baseline cosmology so are not directly comparable: from ACT$\times$CS82 \citep{2015PhRvD..91f2001H}; from \planck\unskip$\times$CFHTLenS \citep{2015PhRvD..92f3517L}; from SPT+\planck\unskip$\times$DES-Science Verification \citep{2016MNRAS.459...21K}; from \planck\unskip$\times$RCSLenS and CFHTLenS \citep{2016MNRAS.460..434H}; from \planck\unskip$\times$KiDS-450 \citep{2017MNRAS.471.1619H}; from \planck\unskip$\times$SDSS \citep{2017MNRAS.464.2120S} and from SPT+\planck\unskip$\times$DES-Y1 \citep{2019PhRvD.100d3501O}.

\subsection{Matter clustering $S_8$ and other parameters}
\label{subsec:s8_results}
\begin{figure*}
    \centering
    \includegraphics[width=0.8\textwidth]{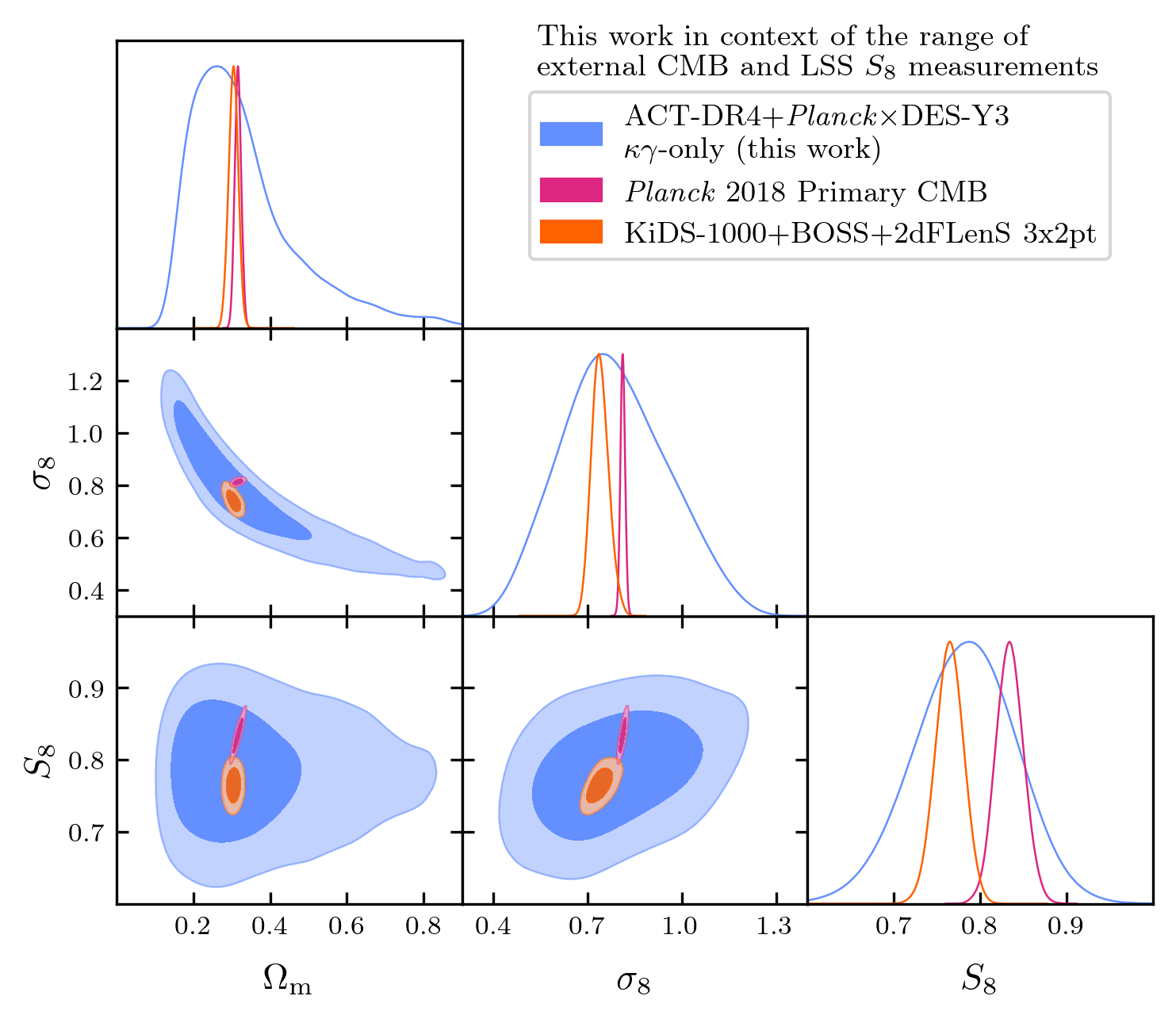}
    \caption{The inferred cosmological parameters $\sigma_8$, $\Omega_{\rm m}$ and $S_8 = \sigma_8 \left( \Omega_{\rm m} / 0.3  \right)^{0.5}$ from our fiducial DES-Y3$\times$ACT-DR4+\planck-tSZ deprojected data vector. To give context for our result, we also show results from other experiments with which we minimally share any data and cover the prominent range of $S_8$ values available in the literature. These are from the CMB at high redshift (\planck primary CMB) and LSS at lower redshifts (KiDS-1000+BOSS+2dFLenS 3x2pt). The full plot of this posterior, including nuisance parameters, is shown in \cref{fig:results_cosmology_full}.}
    \label{fig:result_cosmology}
\end{figure*}

\begin{figure}
    \centering
    \includegraphics[width=0.475\textwidth]{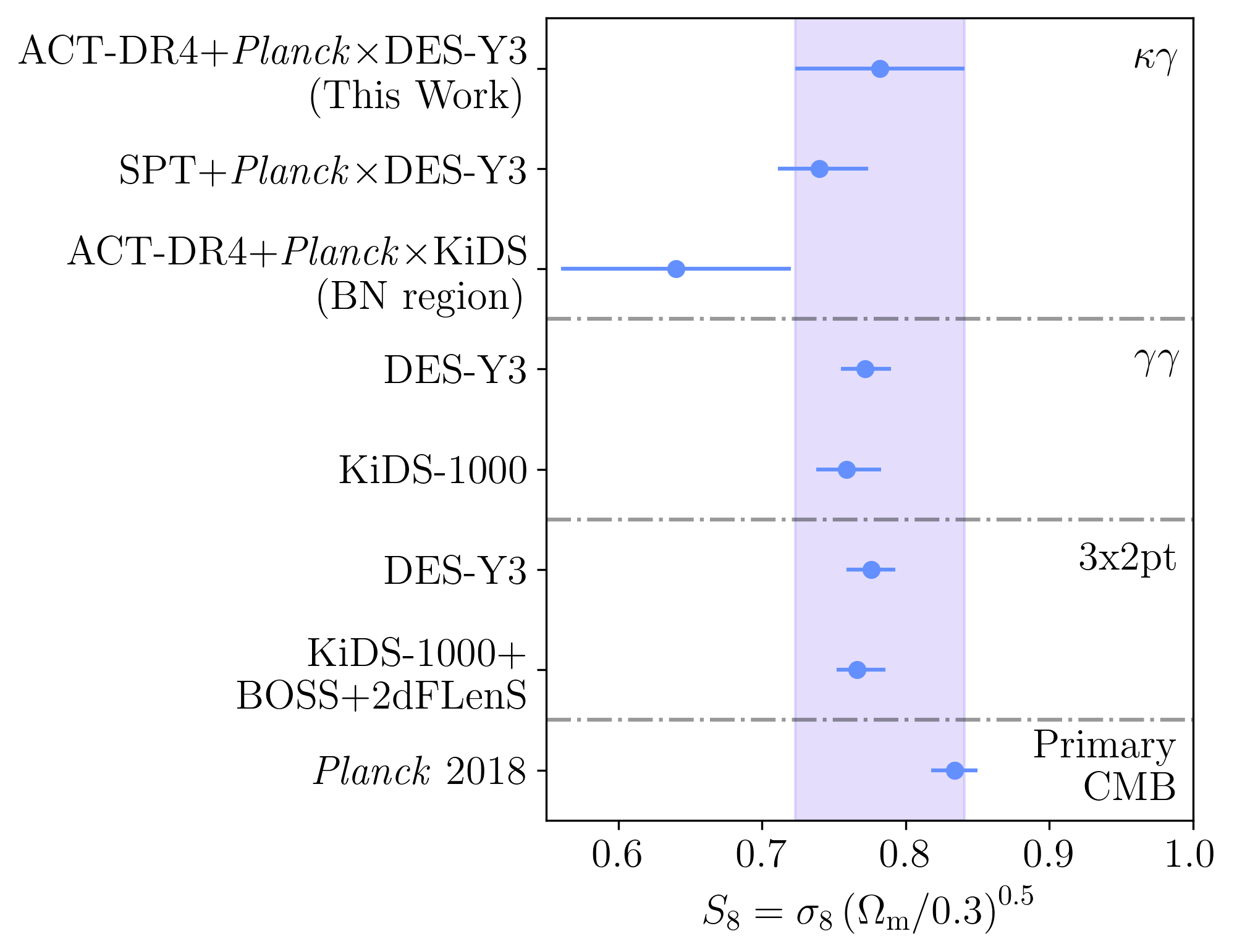}
    \caption{The measured cosmological parameter $S_8$ from our fiducial DES-Y3$\times$ACT-DR4+\planck-tSZ deprojected data vector alongside a number of other measurements of the same parameters. Note that the ACT-DR4+\planck$\times$KiDS measurement uses the \kcmb of the BN region of ACT-DR4 data \citep{2021A&A...649A.146R}.}
    \label{fig:s8_compilation}
\end{figure}

In \cref{fig:result_cosmology}, we show the inferred posterior on cosmological parameters from the fiducial ACT-DR4+\planck$\times~$DES-Y3 data vector. The ten galaxy weak lensing nuisance parameters (galaxy intrinsic alignment, redshift calibration, and shear calibration) are also varied but are prior-dominated and omitted in the plot, as is the $H_0$ parameter (see \cref{app:posterior_zooms} for the plot including them). We also show the constraints on the cosmological parameters from other experiments. We chose these experiments as being external data sets using very different techniques to measure the same cosmological parameters: the \planck 2018 primary CMB result from \cite{2020A&A...641A...6P}, and the KiDS-1000 3x2pt result from \cite{2021A&A...646A.140H}. Though of lower constraining power, our result contains fully independent data and probes a different set of redshift and physical scales to the two other experiments (see \cref{fig:dndz_wz}). Additionally, we show in \cref{fig:s8_compilation} our 1D marginalised measurement of the $S_8$ parameter alone against a further catalogue of other measurements (SPT+\emph{Planck}$\times$DES-Y3 \citealt{2023PhRvD.107b3530C}; ACT-DR4+\planck\unskip$\times$KiDS \citealt{2021A&A...649A.146R}; DES-Y3 shear only \citealt{2022PhRvD.105b3514A}, \citealt*{2022PhRvD.105b3515S}; KiDS-100 shear only \citealt{2021A&A...645A.104A}; DES-Y3 3x2pt \citealt{2022PhRvD.105b3520A}; KiDS-1000 3x2pt \citealt{2021A&A...646A.140H}; \planck 2018 Primary CMB \citealt{2020A&A...641A...6P}). Summary statistics from our inference for the full set of parameters are shown in \cref{tab:results_parameters}. The marginalised mean values and 1D $68\%$ credible regions for the matter density parameter and the amplitude of the fluctuations in the matter distribution are:
\begin{itemize}
\centering
\item[] $\Omegam = 0.338^{+0.05}_{-0.17}$;
\item [] $\sigma_8 = 0.79^{+0.16}_{-0.19}$ ;
\item [] $S_8 = 0.782\pm 0.059$.
\end{itemize}
This inference is drawn when informative priors are used on the nuisance parameters. Constraints on the density of the matter $\Omegam$, even if not most precise, are still a significant improvement over the assumed prior on $\Omegam$. The value of $S_8$ inferred in this analysis is consistent with that inferred from the \planck \texttt{TT,TE,EE+lowE} CMB measurements $S_8=0.834 \pm 0.016$ \citep{2020A&A...641A...6P} with the difference of $0.85 \sigma$, when adding the statistical uncertainties in quadrature to obtain the uncertainty on the difference. It is also consistent with the $S_8 = 0.766^{+0.020}_{-0.014}$ inferred using KiDS-1000 3x2pt analysis \citep{2021A&A...646A.140H}, with around $0.3 \sigma$ difference. A companion study performs the cross-correlation analysis of ACT-DR4 D56 \kcmb and DES-Y3 MAGLIM galaxies finding $S_8 = 0.75^{+0.04}_{-0.05}$ \citep{2023arXiv230617268M}, which differs only by $0.4 \sigma$ with the $S_8$ inferred in this work.

\begin{table}
	\centering	
	\caption{1D Marginalised posterior mean and $68\%$ credible interval for the parameters sampled during our main analysis.}
	\vspace{-0.2cm}
	\begin{tabular}{ccc}
		\hline
            \hline
		Parameter & Prior & Posterior \\\hline
		\multicolumn{3}{l}{\textbf{Cosmology}}  \\
		$\Omega_{\rm c} h^2$ & $\mathcal{U}[0.05, 0.99]$ & $0.161^{+0.042}_{-0.073}$\\ 
		$\log(A_\mathrm{s}10^{10})$ &  $\mathcal{U}[1.6, 4.0]$ & --- \\ 
		$H_0$  & $\mathcal{U}[40, 100]$ & --- \\
            $\sigma_8$  & --- & $0.79^{+0.16}_{-0.19}$ \\
            $\Omega_{\rm m }$  & --- & $0.338^{+0.05}_{-0.17}$ \\
            $S_8 = \sigma_8 \left( \Omega_{\rm m} / 0.3  \right)^{0.5}$  & ---  & $0.782\pm 0.059$ \\
		\hline
  
  		\multicolumn{3}{l}{\textbf{Galaxy Intrinsic Alignment} } 	 \\
		$A_{\rm AIA}$ & $\mathcal{N}(0.35, 0.65)$ & $0.31\pm 0.57$\\
            $\eta_{\rm AIA}$ & $\mathcal{N}(1.66, 4)$ & $-1.0^{+3.8}_{-3.1}$ \\
            \hline
		
		\multicolumn{3}{l}{\textbf{Galaxy redshift calibration} }\\
		$\Delta z_1 $ & $\mathcal{N}(0.0, 0.018)$ & $0.001\pm 0.018$\\
            $\Delta z_2 $ & $\mathcal{N}(0.0, 0.015)$ & $0.001\pm 0.015$\\
            $\Delta z_3 $ & $\mathcal{N}(0.0, 0.011)$ & $-0.001\pm 0.011$\\
            $\Delta z_4 $ &  $\mathcal{N}(0.0, 0.017)$ & $0.000\pm 0.017$\\
            \hline
		
		\multicolumn{3}{l}
            {\textbf{Galaxy shear calibration}}\\
            $m_1 $ & $\mathcal{N}(-0.006, 0.009)$ & $-0.0062\pm 0.0089$\\
            $m_2 $ & $\mathcal{N}(-0.020, 0.008)$ & $-0.0198\pm 0.0080$\\
            $m_3 $ & $\mathcal{N}(-0.024, 0.008)$ & $-0.0240\pm 0.0080$\\
            $m_4 $ & $\mathcal{N}(-0.037, 0.008)$ & $-0.0370\pm 0.0080$\\

            \hline
		\hline
	\end{tabular}
	\label{tab:results_parameters}

\end{table}

\subsection{$S_8$ at different redshifts}
\label{subsec:s8_z_results}
\begin{figure}
    \centering
    \includegraphics[width=0.475\textwidth]{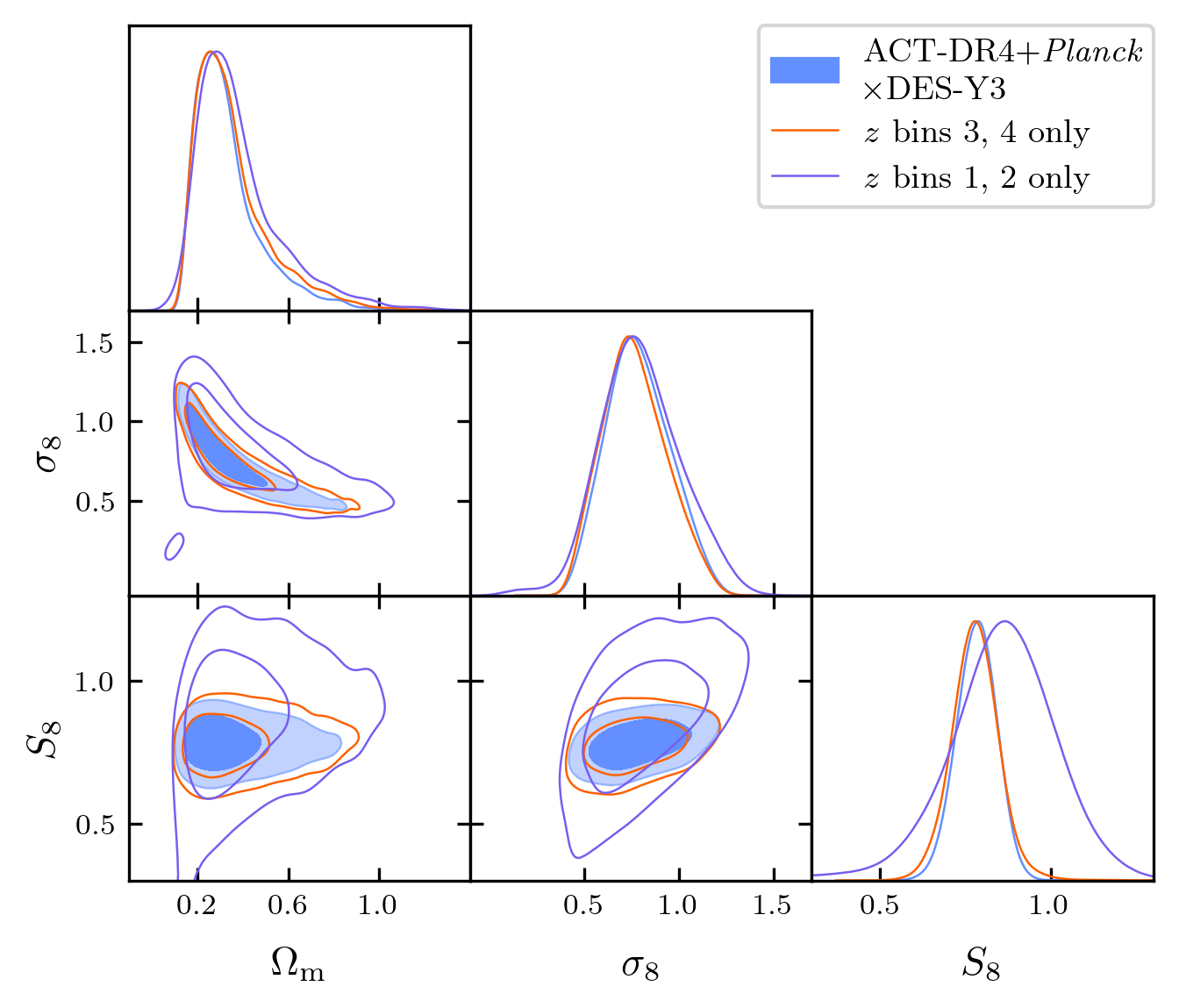}
    \caption{Measurement of the cosmological parameters in two subsets of the data, one covering galaxy redshifts $0 < z \leq 0.63$ and with the resulting \clkg kernel peaking below $z = 0.5$ and the other covering redshifts $ 0.63 < z < 2.0$ and with the resulting \clkg kernel peaking above $z = 0.5$. We find both subsets of the inferred parameters to be consistent.}
    \label{fig:results_zdep}
\end{figure}

As discussed in \cref{sec:intro}, across the multiple measurements of $S_8$ from various observables, it has been noted that higher redshift probes often favour a higher value \citep[e.g. the primary CMB in][]{2020A&A...641A...6P, 2020JCAP...12..047A, 2021PhRvD.104b2003D}, whilst lower redshift ones favour a lower value \citep[e.g. galaxy weak lensing in][]{2021A&A...646A.140H,2022PhRvD.105b3520A,More:2023knf,Miyatake:2023njf,Sugiyama:2023fzm}. In light of this, we split our data vector into two different sub-sets and constrain the $S_8$ parameter independently in each one. One subset contains only the spectra made with DES-Y3 tomographic bins 1 and 2 (covering redshifts $0 < z \leq 0.63$ and with the resulting \clkg kernel peaking below $z = 0.5$), and the other contains only tomographic bins 3 and 4 (covering redshifts $ 0.63 < z < 2.0$ and with the resulting \clkg kernel peaking above $z = 0.5$). In \cref{fig:results_zdep}, we show the two posteriors on cosmological parameters, along with the one from our fiducial analysis with all four tomographic bins. For the sample at lower redshift (bins 1 and 2), we obtained $\Omegam = 0.385^{+0.073}_{-0.22}$ and $S_8 = 0.85^{+0.17}_{-0.13}$, $\sigma_8 = 0.80^{+0.19}_{-0.23}$. Consistently, for the sample at higher redshift (bins 3 and 4), we found $\Omegam = 0.357^{+0.052}_{-0.20}$, $S_8 = 0.779 \pm 0.073$, $\sigma_8 = 0.77^{+0.15}_{-0.19}$. Our analysis reveals that the constraining power is significantly stronger at higher redshifts, primarily due to the better overlap with the CMB lensing kernel. This suggests that the dominant contribution to the overall constraining power when utilizing the entire sample stems from these bins.

\subsection{Weak lensing nuisance parameters}
\label{subsec:nuisance_results}
\begin{figure}
    \centering
    \includegraphics[width=0.475\textwidth]{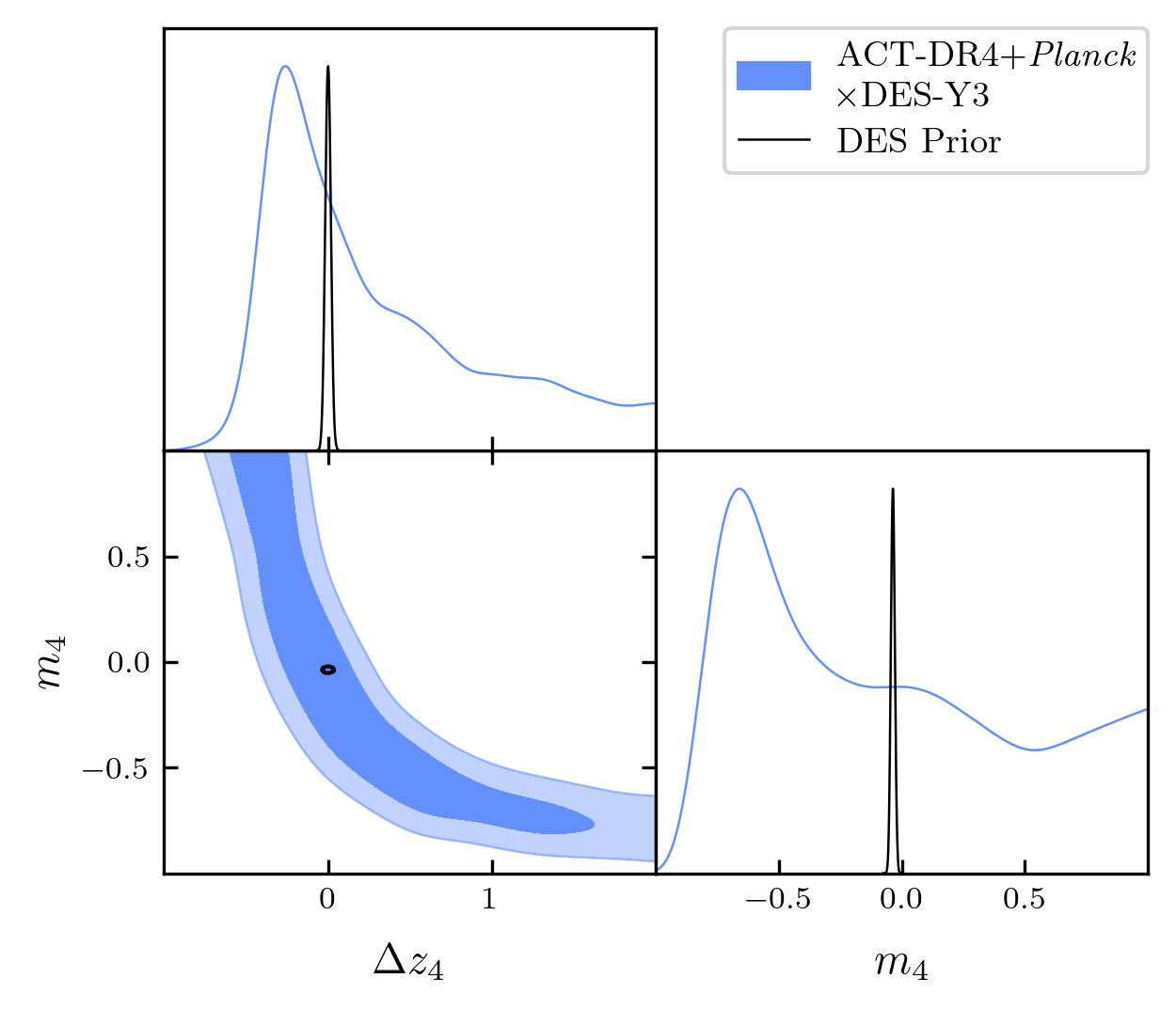}
    \caption{Constraints on the galaxy weak lensing nuisance parameters in DES-Y3 tomographic bin 4 ($0.87 < z < 2.0$). The DES-Y3 prior used in the main analysis is shown as unfilled contours and is consistent with the filled contours obtained with fixed cosmological parameters but broad priors on nuisance parameters.}
    \label{fig:results_nuisance}
\end{figure}

The priors on weak lensing galaxy redshift and shear calibration detailed in \cref{tab:params} and used in the above inference runs are derived from a series of simulations and deep training data implemented as part of the DES-Y3 analysis pipelines. They are therefore informative and dominate the posterior for the nuisance parameters (as seen in \cref{fig:model_recovery_nuisance}). It is interesting to use the CMB lensing from ACT-DR4 as an extra high redshift lensing bin to attempt to independently calibrate these nuisance parameters and validate the priors available from simulations. This has been previously advocated as a productive use of \clkg data sets \citep[e.g.][]{2013arXiv1311.2338D}. Though these simulation-derived priors are often given as uncorrelated, wider priors may result in degeneracies in 3x2pt analyses. In such a case, the \clkg observable may provide useful degeneracy breaking thanks to the differences in redshift and scale dependence. Here, we make use of only the highest redshift and highest signal tomographic bin (Bin 4), fix all other cosmological and nuisance parameters to their fiducial values and infer only the redshift and shear calibration parameters $(\delta z_4, m_4)$ with broad priors $\delta z_4 \sim \mathcal{U}[-1, 2]$ and $m_4 \sim \mathcal{U}[-1, 1]$. These priors are a factor 100 wider than the Gaussian priors applied in the main analysis and span the plausible range of possible calibration uncertainties. The inferred posterior is shown in \cref{fig:results_nuisance}. Though the constraining power of our data is far lower than that of the DES-Y3 prior, the posterior is consistent, meaning the informative prior passes as accurate within the terms of this low precision test.

\section{Conclusions}\label{sec:conclusions}
In this analysis, we measured \clkg, the angular power spectrum between the CMB weak lensing map of ACT-DR4 in the D56 region and the DES-Y3 cosmic shear catalogue consisting of around 100 million galaxies. The measurement is over the common sky area of around 450 deg$^2$ between the two surveys. To avoid one of the main extragalactic foreground biases which originate from the tSZ contamination in \kcmb, we used the tSZ-free \kcmb map obtained using ACT-DR4 and \emph{Planck} data for the baseline analysis \citep{2021MNRAS.500.2250D}. The analysis is carried out in harmonic space over the multipole range of $\ell = 100~\rm{to}~1900$. Over this range, we measure the cross-correlation at SNR = 7.1. As demonstrated in \cref{subsec:data_null_tests}, the measured data vector passes specific null tests, indicating the lack of significant detection of some of the non-idealities that generally affect \clkg measurements. We also tested for contamination due to stars and Galactic dust. We found their effect is negligible compared with the statistical uncertainty and saw no significant evidence of their contamination. We performed the initial analysis with the blinding procedure described in \cref{subsubsec:des_blinding}. After the data vector passed the null tests and we confirmed that the analysis pipeline recovered the unbiased input values from simulations, we unblinded the catalogue and the parameters inferred from the unblinded data vector.

We used this \clkg measurement to infer the matter density parameter ($\Omegam$) and the amplitude of fluctuations in the matter distribution ($S_8$). We inferred $\Omegam = 0.338^{+0.05}_{-0.17}$, and $S_8 = 0.782\pm 0.059$. Our main result is shown in Figure \ref{fig:result_cosmology}. These values were inferred using informative but well-motivated priors on the observational and astrophysical nuisance parameters, which were marginalized while inferring $\Omegam$ and $S_8$. We investigated the validity of the priors on galaxy intrinsic alignment parameters by significantly relaxing them and checking the consistency of the resulting posteriors. As depicted in \cref{fig:wideia_prior}, we found the posteriors with this broader IA prior are consistent with those with the fiducial IA priors but have relatively weak constraints on the cosmological parameters. We also assessed the consistency between the inference obtained using various subsets of the data, shown in \cref{fig:results_zdep}. 
Our results are statistically consistent with many recent cosmic shear studies, including those utilizing the DES-Y3 data alone (\citealt{2022MNRAS.515.1942D, 2022PhRvD.105b3520A}, \citealt*{2022PhRvD.105b3515S}, \citealt{2022PhRvD.105b3514A}) and cross-correlation with SPT and \planck CMB lensing \citep{2023PhRvD.107b3530C}. Furthermore, our results are in agreement with $S_8$ inferred using KiDS data \citep{2021A&A...646A.140H}, although slightly higher than the value inferred using the cross-correlation with ACT-DR4 BOSS North and \planck CMB lensing \citep{2021A&A...649A.146R}. However, we note that these differences fall within $\sim 1.4\sigma$, indicating a relatively minor deviation. We summarized this comparison in \cref{fig:s8_compilation}.

Measurements of the clustering of matter from combinations of CMB and optical weak lensing are rapidly growing in precision, with the highest-yet SNR achieved being that of SPT+\planck $\times$ DES-Y3 at $18 \sigma$ \citep{2023PhRvD.107b3530C}. We note that this analysis is carried out on a substantially larger sky area of 3920 deg$^2$ than the 450 deg$^2$ area of the ACT D56 region considered in this work. For a given sky area, comparatively higher SNR obtained in this work is owing to the lower \kcmb reconstruction noise of the ACT D56 observations. We find the $S_8$ inferred in these two studies in statistical agreement, as depicted in \cref{fig:s8_compilation}. Recently, the ACT Collaboration has completed an analysis of the reconstructed CMB lensing map using the DR6 sky area of 9400 deg$^2$, which overlaps with nearly the entire DES-Y3 survey footprint \citep{ACT:2023dou, ACT:2023kun, ACT:2023ubw}. These data will provide a great opportunity to continue the work done here by performing cross-correlation with various probes of large-scale structure, including galaxy lensing and galaxy density \citep{2023arXiv230617268M}.  Further on the horizon, correlations between the Simons Observatory \citep{2019JCAP...02..056A} and CMB-S4 \citep{CMB-S4:2016ple} lensing maps with shear data from the \emph{Euclid} satellite \citep{2018LRR....21....2A} and the Rubin Observatory Legacy Survey of Space and Time \citep{2019ApJ...873..111I} will be even more precise. These analyses, with their higher statistical precision, will need to be carried out with more careful theoretical modelling of astrophysical effects of baryons and galaxy intrinsic alignment along with the modelling of observational systematics \citep[see, e.g.][]{Kilo-DegreeSurvey:2023gfr} than required for the data analysed in this work.

\section*{Acknowledgements}
SS acknowledges support from the Beus Center for Cosmic Foundations. IH, EC, SG, and HJ acknowledge support from the European Research Council (ERC) under the European Union's Horizon 2020 research and innovation programme (Grant Agreement No. 849169).  CS acknowledges support from the Agencia Nacional de Investigaci\'on y Desarrollo (ANID) through FONDECYT grant no.\ 11191125 and BASAL project FB210003. KM acknowledges support from the National Research Foundation of South Africa. GSF acknowledges support through the Isaac Newton Studentship, the Helen Stone Scholarship at the University of Cambridge, and from the European Research Council (ERC) under the European Union’s Horizon 2020 research and innovation programme (Grant agreement No. 851274). KM acknowledges support from the National Research Foundation of South Africa. LBN acknowledges support from the Simons Foundation. JCH acknowledges support from NSF grant AST-2108536, NASA grants 21-ATP21-0129 and 22-ADAP22-0145, the Sloan Foundation, and the Simons Foundation. SKC acknowledges support from NSF award AST-2001866. KMH acknowledges NSF award number 1815887. OD acknowledges support from SNSF Eccellenza Professorial Fellowship (No. 186879).

We acknowledge the support of the Supercomputing Wales project, which is part-funded by the European Regional Development Fund (ERDF) via Welsh Government. We thank Agn\`es Fert\'e and Jessie Muir for help loading MultiNest chains.

Support for ACT was through the U.S.~National Science Foundation through awards AST-0408698, AST-0965625, and AST-1440226 for the ACT project, as well as awards PHY-0355328, PHY-0855887 and PHY-1214379. Funding was also provided by Princeton University, the University of Pennsylvania, and a Canada Foundation for Innovation (CFI) award to UBC. ACT operated in the Parque Astron\'omico Atacama in northern Chile under the auspices of the Agencia Nacional de Investigaci\'on y Desarrollo (ANID). The development of multichroic detectors and lenses was supported by NASA grants NNX13AE56G and NNX14AB58G. Detector research at NIST was supported by the NIST Innovations in Measurement Science program.  Computing for ACT was performed using the Princeton Research Computing resources at Princeton University, the National Energy Research Scientific Computing Center (NERSC), and the Niagara supercomputer at the SciNet HPC Consortium.

Funding for the DES Projects has been provided by the U.S. Department of Energy, the U.S. National Science Foundation, the Ministry of Science and Education of Spain, 
the Science and Technology Facilities Council of the United Kingdom, the Higher Education Funding Council for England, the National Center for Supercomputing 
Applications at the University of Illinois at Urbana-Champaign, the Kavli Institute of Cosmological Physics at the University of Chicago, 
the Center for Cosmology and Astro-Particle Physics at the Ohio State University,
the Mitchell Institute for Fundamental Physics and Astronomy at Texas A\&M University, Financiadora de Estudos e Projetos, 
Funda{\c c}{\~a}o Carlos Chagas Filho de Amparo {\`a} Pesquisa do Estado do Rio de Janeiro, Conselho Nacional de Desenvolvimento Cient{\'i}fico e Tecnol{\'o}gico and 
the Minist{\'e}rio da Ci{\^e}ncia, Tecnologia e Inova{\c c}{\~a}o, the Deutsche Forschungsgemeinschaft and the Collaborating Institutions in the Dark Energy Survey. 

The Collaborating Institutions are Argonne National Laboratory, the University of California at Santa Cruz, the University of Cambridge, Centro de Investigaciones Energ{\'e}ticas, 
Medioambientales y Tecnol{\'o}gicas-Madrid, the University of Chicago, University College London, the DES-Brazil Consortium, the University of Edinburgh, 
the Eidgen{\"o}ssische Technische Hochschule (ETH) Z{\"u}rich, 
Fermi National Accelerator Laboratory, the University of Illinois at Urbana-Champaign, the Institut de Ci{\`e}ncies de l'Espai (IEEC/CSIC), 
the Institut de F{\'i}sica d'Altes Energies, Lawrence Berkeley National Laboratory, the Ludwig-Maximilians Universit{\"a}t M{\"u}nchen and the associated Excellence Cluster Universe, 
the University of Michigan, NSF's NOIRLab, the University of Nottingham, The Ohio State University, the University of Pennsylvania, the University of Portsmouth, 
SLAC National Accelerator Laboratory, Stanford University, the University of Sussex, Texas A\&M University, and the OzDES Membership Consortium.

Based in part on observations at Cerro Tololo Inter-American Observatory at NSF's NOIRLab (NOIRLab Prop. ID 2012B-0001; PI: J. Frieman), which is managed by the Association of Universities for Research in Astronomy (AURA) under a cooperative agreement with the National Science Foundation.

The DES data management system is supported by the National Science Foundation under Grant Numbers AST-1138766 and AST-1536171.
The DES participants from Spanish institutions are partially supported by MICINN under grants ESP2017-89838, PGC2018-094773, PGC2018-102021, SEV-2016-0588, SEV-2016-0597, and MDM-2015-0509, some of which include ERDF funds from the European Union. IFAE is partially funded by the CERCA program of the Generalitat de Catalunya.
Research leading to these results has received funding from the European Research
Council under the European Union's Seventh Framework Program (FP7/2007-2013) including ERC grant agreements 240672, 291329, and 306478.
We  acknowledge support from the Brazilian Instituto Nacional de Ci\^encia
e Tecnologia (INCT) do e-Universo (CNPq grant 465376/2014-2).

This manuscript has been authored by Fermi Research Alliance, LLC under Contract No. DE-AC02-07CH11359 with the U.S. Department of Energy, Office of Science, Office of High Energy Physics.


\section*{Data Availability} 
The data underlying this article are available in a GitHub repository at \url{https://github.com/itrharrison/actdr4kappa-x-desy3gamma-data}, in the Zenodo repository referenced there (the DOI generated will be added here on publication), and will be made available on the LAMBDA data service \url{https://lambda.gsfc.nasa.gov/} upon publication.



\bibliographystyle{mnras_2author}
\bibliography{references_kappaXgamma} 
\section*{Author Affiliations}
$^{1}$ School of Earth and Space Exploration, Arizona State University, Tempe, AZ 85287, USA\\
$^{2}$ School of Physics and Astronomy, Cardiff University, CF24 3AA, UK\\
$^{3}$ Fermi National Accelerator Laboratory, P. O. Box 500, Batavia, IL 60510, USA\\
$^{4}$ Kavli Institute for Cosmological Physics, University of Chicago, Chicago, IL 60637, USA\\
$^{5}$ Cerro Tololo Inter-American Observatory, NSF's National Optical-Infrared Astronomy Research Laboratory, Casilla 603, La Serena, Chile\\
$^{6}$ Laborat\'orio Interinstitucional de e-Astronomia - LIneA, Rua Gal. Jos\'e Cristino 77, Rio de Janeiro, RJ - 20921-400, Brazil\\
$^{7}$ Department of Physics, University of Michigan, Ann Arbor, MI 48109, USA\\
$^{8}$ Institute of Astronomy, University of Cambridge, Madingley Road, Cambridge CB3 0HA, UK\\
$^{9}$ Kavli Institute for Cosmology, University of Cambridge, Madingley Road, Cambridge CB3 0HA, UK\\
$^{10}$ Department of Physics and Astronomy, University of Southern California, Los Angeles, CA 90089, USA\\
$^{11}$ Institute of Cosmology and Gravitation, University of Portsmouth, Portsmouth, PO1 3FX, UK\\
$^{12}$ Department of Astronomy, Cornell University, Ithaca, NY 14853, USA\\
$^{13}$ Argonne National Laboratory, 9700 South Cass Avenue, Lemont, IL 60439, USA\\
$^{14}$ Department of Physics and Astronomy, University of Pennsylvania, Philadelphia, PA 19104, USA\\
$^{15}$ CNRS, UMR 7095, Institut d'Astrophysique de Paris, F-75014, Paris, France\\
$^{16}$ Sorbonne Universit\'es, UPMC Univ Paris 06, UMR 7095, Institut d'Astrophysique de Paris, F-75014, Paris, France\\
$^{17}$ Department of Physics, Northeastern University, Boston, MA 02115, USA\\
$^{18}$ Canadian Institute for Theoretical Astrophysics, University of Toronto, 60 St. George Street, Toronto, ON, M5S 3H8, Canada\\
$^{19}$ Department of Physics \& Astronomy, University College London, Gower Street, London, WC1E 6BT, UK\\
$^{20}$ Kavli Institute for Particle Astrophysics \& Cosmology, P. O. Box 2450, Stanford University, Stanford, CA 94305, USA\\
$^{21}$ SLAC National Accelerator Laboratory, Menlo Park, CA 94025, USA\\
$^{22}$ Instituto de Astrofisica de Canarias, E-38205 La Laguna, Tenerife, Spain\\
$^{23}$ Universidad de La Laguna, Dpto. Astrof\'isica, E-38206 La Laguna, Tenerife, Spain\\
$^{24}$ Institut de F\'{\i}sica d'Altes Energies (IFAE), The Barcelona Institute of Science and Technology, Campus UAB, 08193 Bellaterra (Barcelona) Spain\\
$^{25}$ Physics Department, William Jewell College, Liberty, MO, 64068\\
$^{26}$ Department of Astronomy and Astrophysics, University of Chicago, Chicago, IL 60637, USA\\
$^{27}$ Department of Physics, Duke University Durham, NC 27708, USA\\
$^{28}$ NASA Goddard Space Flight Center, 8800 Greenbelt Rd, Greenbelt, MD 20771, USA\\
$^{29}$ Department of Physics, Cornell University, Ithaca, NY 14853, USA\\
$^{30}$ Hamburger Sternwarte, Universit\"{a}t Hamburg, Gojenbergsweg 112, 21029 Hamburg, Germany\\
$^{31}$ Universit\'{e} de Gen\`{e}ve, D\'{e}partement de Physique Th\'{e}orique et CAP, 24 quai Ernest-Ansermet, CH-1211 Gen\`{e}ve 4, Switzerland\\
$^{32}$ School of Mathematics and Physics, University of Queensland, Brisbane, QLD 4072, Australia\\
$^{33}$ Department of Physics, IIT Hyderabad, Kandi, Telangana 502285, India\\
$^{34}$ Universit\'e Grenoble Alpes, CNRS, LPSC-IN2P3, 38000 Grenoble, France\\
$^{35}$ Department of Physics and Astronomy, University of Waterloo, 200 University Ave W, Waterloo, ON N2L 3G1, Canada\\
$^{36}$ Department of Applied Mathematics and Theoretical Physics, University of Cambridge, Cambridge CB3 0WA, UK\\
$^{37}$ Lawrence Berkeley National Laboratory, 1 Cyclotron Road, Berkeley, CA 94720, USA\\
$^{38}$ Berkeley Center for Cosmological Physics, Department of Physics, University of California, Berkeley, CA 94720, USA\\
$^{39}$ Institute of Theoretical Astrophysics, University of Oslo. P.O. Box 1029 Blindern, NO-0315 Oslo, Norway\\
$^{40}$ University Observatory, Faculty of Physics, Ludwig-Maximilians-Universit\"at, Scheinerstr. 1, 81679 Munich, Germany\\
$^{41}$ Center for Astrophysical Surveys, National Center for Supercomputing Applications, 1205 West Clark St., Urbana, IL 61801, USA\\
$^{42}$ Department of Astronomy, University of Illinois at Urbana-Champaign, 1002 W. Green Street, Urbana, IL 61801, USA\\
$^{43}$ Department of Physics, Columbia University, 538 West 120th Street, New York, NY, USA 10027\\
$^{44}$ Santa Cruz Institute for Particle Physics, Santa Cruz, CA 95064, USA\\
$^{45}$ Center for Cosmology and Astro-Particle Physics, The Ohio State University, Columbus, OH 43210, USA\\
$^{46}$ Department of Physics, The Ohio State University, Columbus, OH 43210, USA\\
$^{47}$ Department of Physics, Florida State University, Tallahassee, FL 32306, USA\\
$^{48}$ Center for Astrophysics $\vert$ Harvard \& Smithsonian, 60 Garden Street, Cambridge, MA 02138, USA\\
$^{49}$ Centre for Radio Astronomy Techniques and Technologies, Department of Physics and Electronics, Rhodes University, P.O. Box 94, Makhanda 6140, South Africa\\
$^{50}$ South African Radio Astronomy Observatory, 2 Fir Street, Observatory 7925, South Africa\\
$^{51}$ Jet Propulsion Laboratory, California Institute of Technology, 4800 Oak Grove Dr., Pasadena, CA 91109, USA\\
$^{52}$ Departamento de F\'isica Matem\'atica, Instituto de F\'isica, Universidade de S\~ao Paulo, CP 66318, S\~ao Paulo, SP, 05314-970, Brazil\\
$^{53}$ George P. and Cynthia Woods Mitchell Institute for Fundamental Physics and Astronomy, and Department of Physics and Astronomy, Texas A\&M University, College Station, TX 77843, USA\\
$^{54}$ Centro de Investigaciones Energ\'eticas, Medioambientales y Tecnol\'ogicas (CIEMAT), Madrid, Spain\\
$^{55}$ Instituci\'o Catalana de Recerca i Estudis Avan\c{c}ats, E-08010 Barcelona, Spain\\
$^{56}$ Max Planck Institute for Extraterrestrial Physics, Giessenbachstrasse, 85748 Garching, Germany\\
$^{57}$ Astrophysics Research Centre, University of KwaZulu-Natal, Westville Campus, Durban 4041, South Africa\\
$^{58}$ School of Mathematics, Statistics \& Computer Science, University of KwaZulu-Natal, Westville Campus, Durban 4041, South Africa\\
$^{59}$ Department of Physics, Stanford University, 382 Via Pueblo Mall, Stanford, CA 94305, USA\\
$^{60}$ Instituto de F\'isica Gleb Wataghin, Universidade Estadual de Campinas, 13083-859, Campinas, SP, Brazil\\
$^{61}$ Department of Physics, Yale University, 217 Prospect St., New Haven, CT 06511\\
$^{62}$ Department of Physics and Astronomy, Haverford College, Haverford, PA, USA 19041\\
$^{63}$ Observat\'orio Nacional, Rua Gal. Jos\'e Cristino 77, Rio de Janeiro, RJ - 20921-400, Brazil\\
$^{64}$ Institute for Astronomy, University of Edinburgh, Edinburgh EH9 3HJ, UK\\
$^{65}$ Department of Physics and Astronomy, Stony Brook University, Stony Brook, NY 11794, USA\\
$^{66}$ Brookhaven National Laboratory, Bldg 510, Upton, NY 11973, USA\\
$^{67}$ Instituto de F\'isica, Pontificia Universidad Cat\'olica de Valpara\'iso, Casilla 4059, Valpara\'iso, Chile\\
$^{68}$ School of Physics and Astronomy, University of Southampton, Southampton, SO17 1BJ, UK\\
$^{69}$ Computer Science and Mathematics Division, Oak Ridge National Laboratory, Oak Ridge, TN 37831\\
$^{70}$ Institut de Recherche en Astrophysique et Plan\'etologie (IRAP), Universit\'e de Toulouse, CNRS, UPS, CNES, 14 Av. Edouard Belin, 31400 Toulouse, France\\
$^{71}$ Instituto de Astrof\'isica and Centro de Astro-Ingenie\'ia, Facultad de F\'isica, Pontificia Universidad Cat\'olica de Chile, Av. Vicu\~na Mackenna 4860, 7820436 Macul, Santiago, Chile\\



\appendix

\section{Results with ACT-only Data Vector}\label{app:actonly}
\begin{figure}
    \centering
    \includegraphics[width=0.45\textwidth]{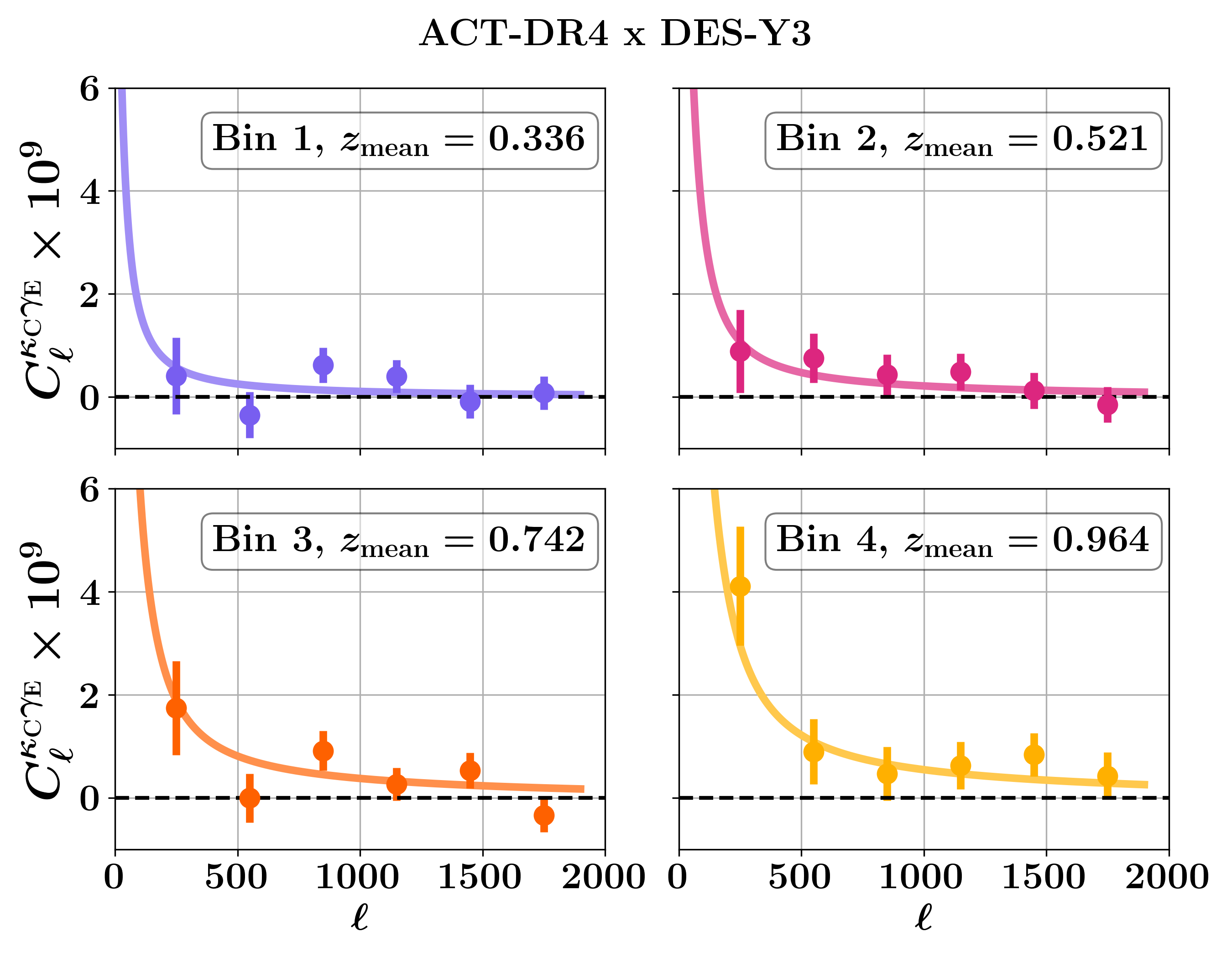}
    \caption{Similar to \cref{fig:data_cls}, but for ACT-only \kcmb. The points with the error bar show \clkg bandpowers with ACT-DR4 \kcmb and DES-Y3 shear for four DES-Y3 redshift bins. Error bars are the square root of the diagonal of the covariance matrix. $z_{\rm mean}$ is the mean redshift of the source galaxy distribution. The curves show the best-fit theory \clkg.}
    \label{fig:act_only_data_cls}
\end{figure}

\begin{figure}
    \centering
    \includegraphics[width=0.45\textwidth]{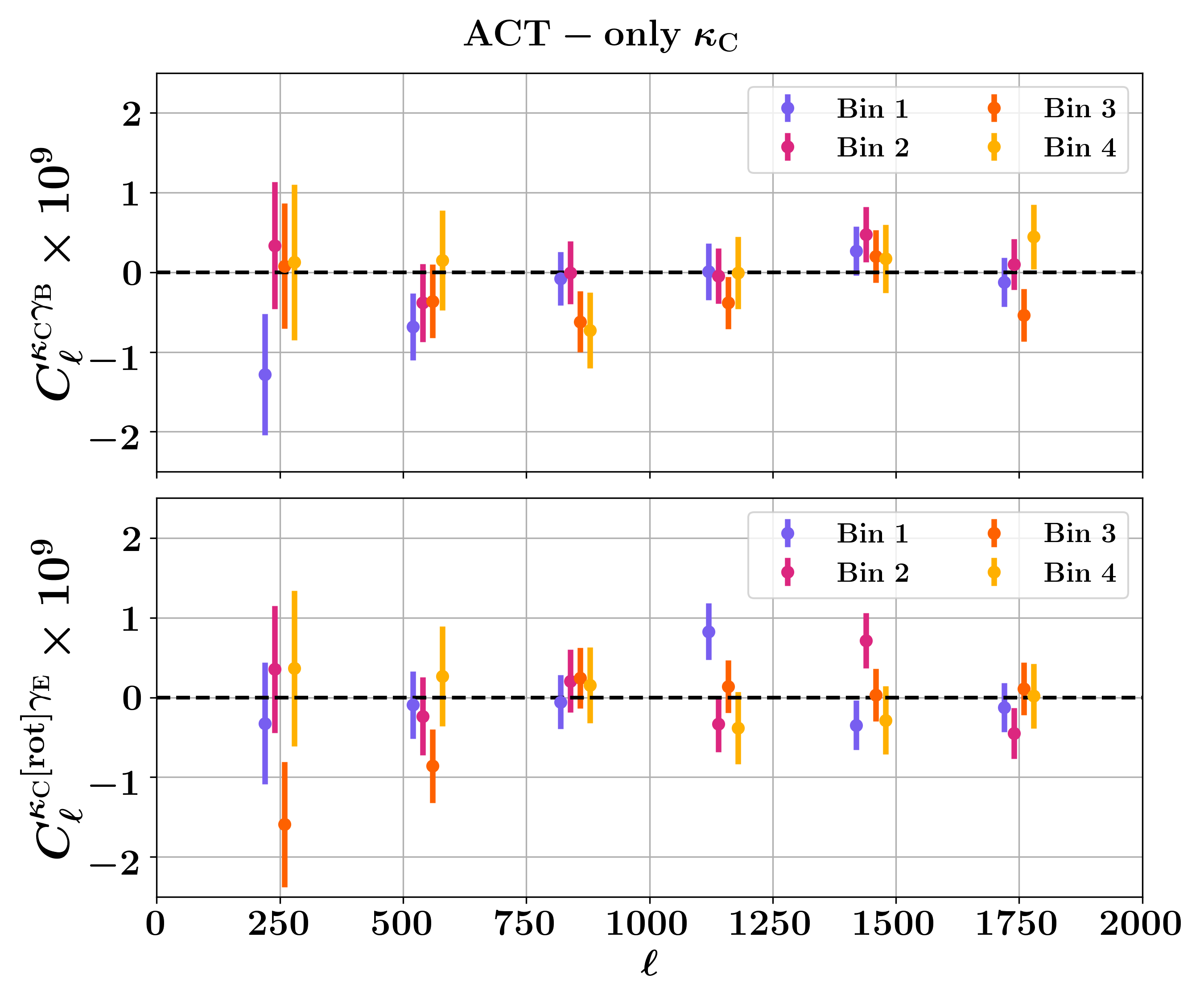}
    \caption{Similar to \cref{fig:null_test_AP}, but for ACT-only data. \emph{Top:} The power spectrum between \kcmb and the B-mode of the shear ($C^{\kappa_{\rm C} \gamma_{\rm B} }_{\ell}$), which is expected to be consistent with zero. \emph{Bottom:} The correlation between \kcmb and the E-mode of the shear map ($C^{\kappa_{\rm C} [\rm rot] \gamma_{\rm E} }_{\ell}$) obtained from the catalogue in which ellipticities are randomly rotated.}
    \label{fig:act_only_null_test}
\end{figure}

We also perform all of our analysis with the \clkg data vector containing only ACT-DR4 data (i.e., no \planck data and no tSZ deprojection). \cref{fig:act_only_data_cls} shows the ACT-only data vector and the best fit \clkg. With ACT-only \kcmb, we measure \clkg at the SNR of 6.6, which is slightly less than the ACT + \planck SNR = 7.1. ACT-only \clkg passes the null tests similar to the ACT+\planck \clkg with  PTE = $0.67$ for the B-mode null test and PTE = $0.43$ for the rotation null test. The null test bandpowers are depicted in \cref{fig:act_only_null_test}.

\subsection{Comparison of ACT-only and ACT+\planck data vector}\label{subsec:compare_A_AP}
\begin{figure}
    \centering
    \includegraphics[width=0.45\textwidth]{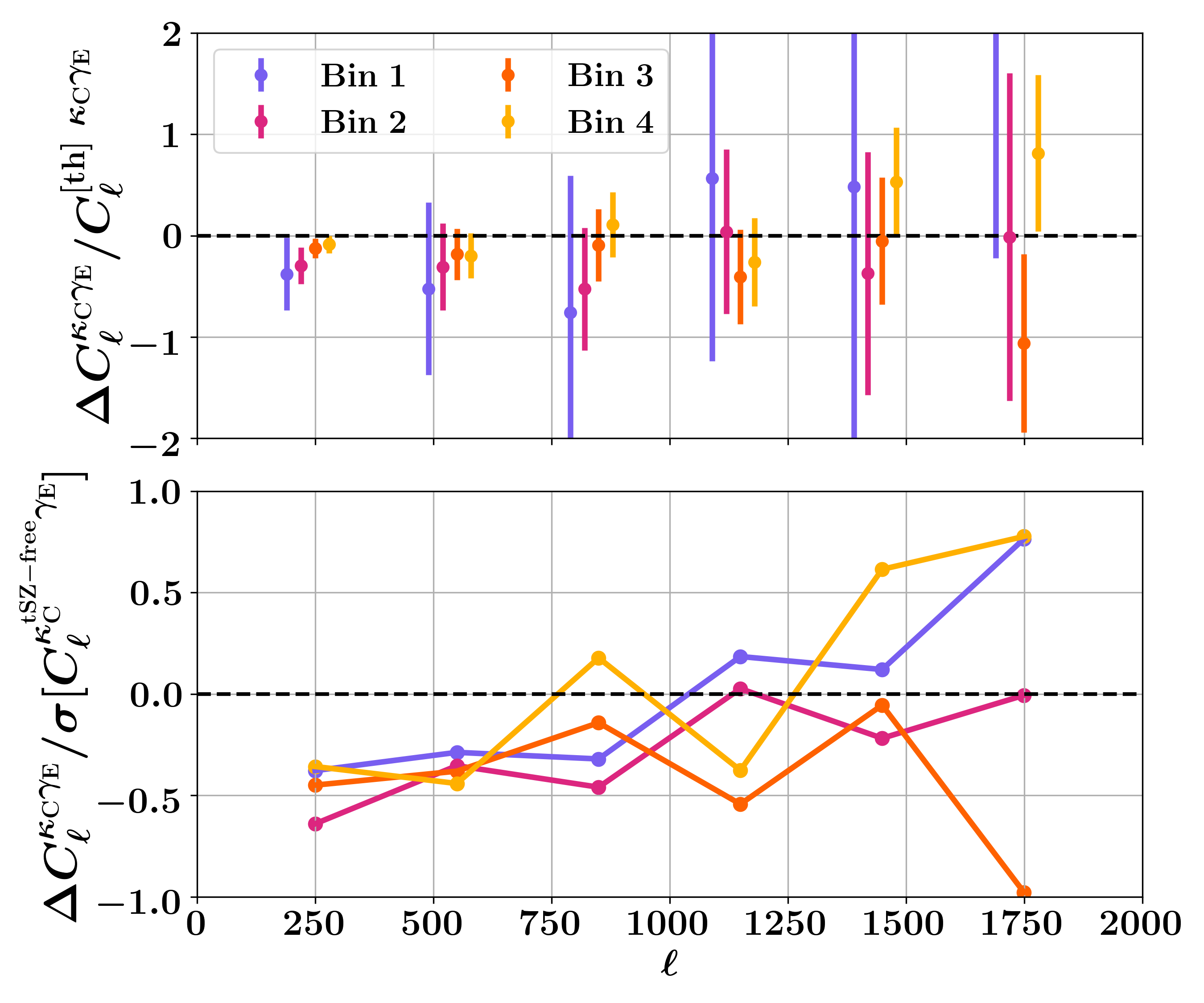}
    \caption{Comparison of \clkg [ACT-only, with-tSZ] and \clkg [ACT+\planck, tSZ-free]. \emph{Top:} The difference between two ($\Delta C^{\kappa_{\rm C} \gamma_{\rm E} }_{\ell} \equiv C^{\rm with-tSZ}_{\ell} - C^{\rm tSZ-free}_{\ell}$) relative to the fiducial theory \clkg. The error bars depict $\sigma(\Delta \clkg)/C^{{\rm [th]} \kcmb \gE}_{\ell}$, where $\sigma(\Delta \clkg)$ is uncertainty on $\Delta \clkg$. \emph{Bottom:} The same difference as in the top panel, $\Delta \clkg$, but now in units of error bar on tSZ-free $\clkg$.}
    \label{fig:compare_ACTonly_AP_data}
\end{figure}

In this section, we compare \clkg data vector obtained using ACT-only \kcmb and ACT+\planck \kcmb. An astrophysical systematic that is not modelled and is correlated with the reconstructed \kcmb and galaxy shear can lead to a bias in the measurement of \clkg. In addition to the lensing, observed CMB maps contain the imprints of other secondary anisotropies and foregrounds, such as the Sunyaev-Zel'dovich effect and emissions from galaxies, such as the CIB. If these effects are sufficiently non-Gaussian over the scales of the CMB temperature anisotropies used to reconstruct \kcmb, they can bias the lensing estimate. If, in addition, these secondary effects and extra-galactic foregrounds share redshift overlap with a tracer of large-scale structure, the cross-correlation of the reconstructed \kcmb with these tracers is biased \citep{2014JCAP...03..024O, 2014ApJ...786...13V}. For correlations with cosmic shear, the effect of the tSZ bias is shown to be more severe at lower redshift bins \citep{2019PhRvD..99b3508B}.
In \kcmb reconstruction with ACT 98 and 150 GHz maps, tSZ contamination is mitigated by finding and masking massive clusters in the temperature maps. The clusters detected above signal-to-noise ratio = 5 are masked in frequency maps, and the masked regions are inpainted. However, this procedure does not remove all of the tSZ bias in the reconstructed \kcmb. Hence, along with the ACT 98 and 150 GHz maps, \cite{2021MNRAS.500.2250D} use \planck frequency maps from 30 GHz to 545 GHz to deproject tSZ and reconstruct \kcmb where the effect of tSZ is nulled. In \cref{fig:compare_ACTonly_AP_data}, we show the difference between the \clkg bandpowers for \kcmb with ACT-only data and the ACT + \planck tSZ-free (which is our choice for the baseline analysis). We find that the four bandpowers are consistent with the hypothesis of no difference, with the PTEs for four redshift bin bandpowers being 0.86, 0.72, 0.68, and 0.65, respectively. 

\subsection{Analysis results}
\begin{figure}
    \centering
    \includegraphics[width=0.475\textwidth]{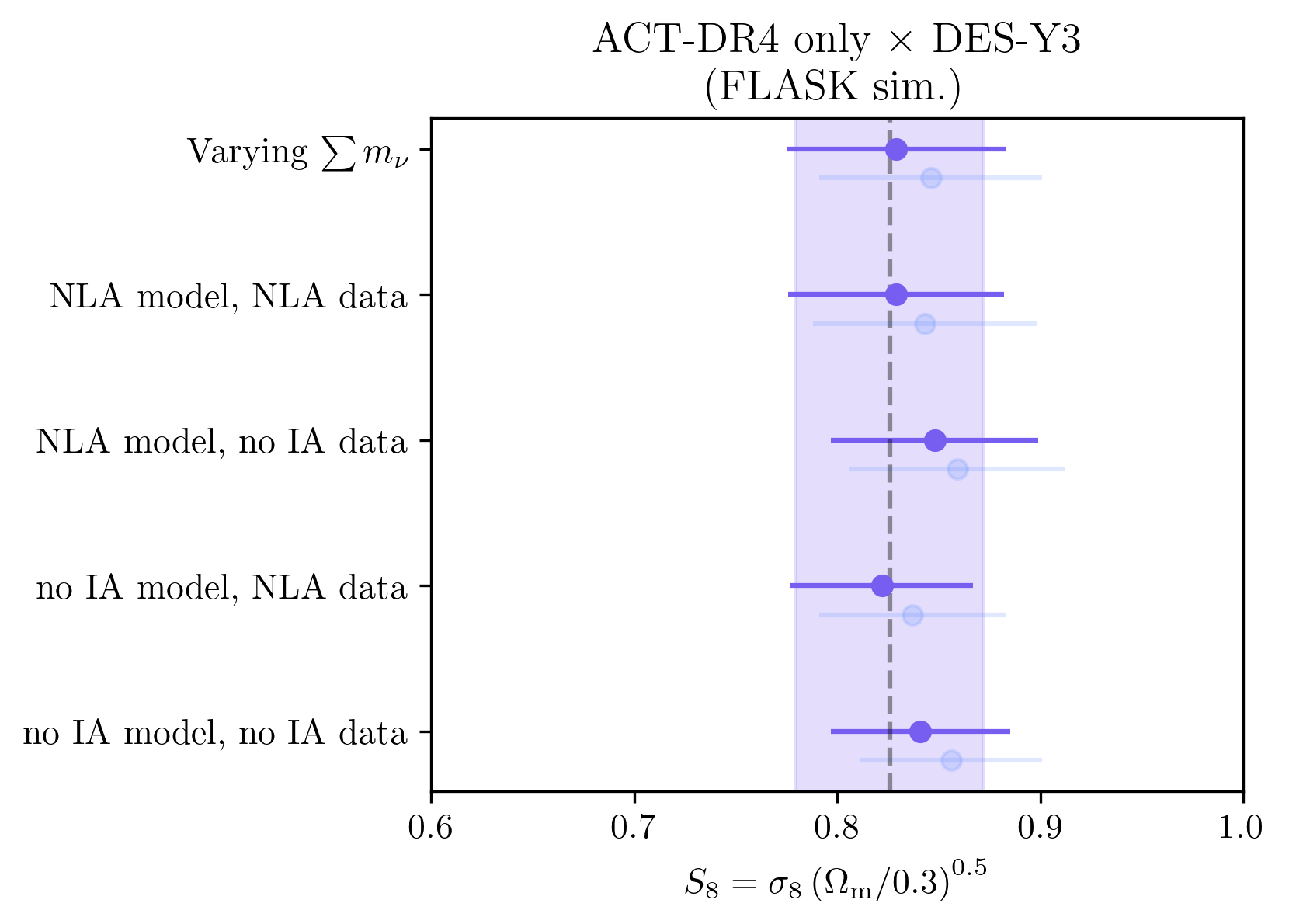}
    \caption{The stability of the recovery of the $S_8$ parameter from our simulated ACT-only data vector as we change the model used for the inference. Results from the fiducial ACT+\planck simulated data vector are shown as the faded points (and ACT-only results are not shown for all the model variations). The dashed vertical line represents the true value input to the simulation, and the shaded band is the error bar in the fiducial model setup.}
    \label{fig:model_stability_actonly}
\end{figure}

\begin{figure}
    \centering
    \includegraphics[width=0.475\textwidth]{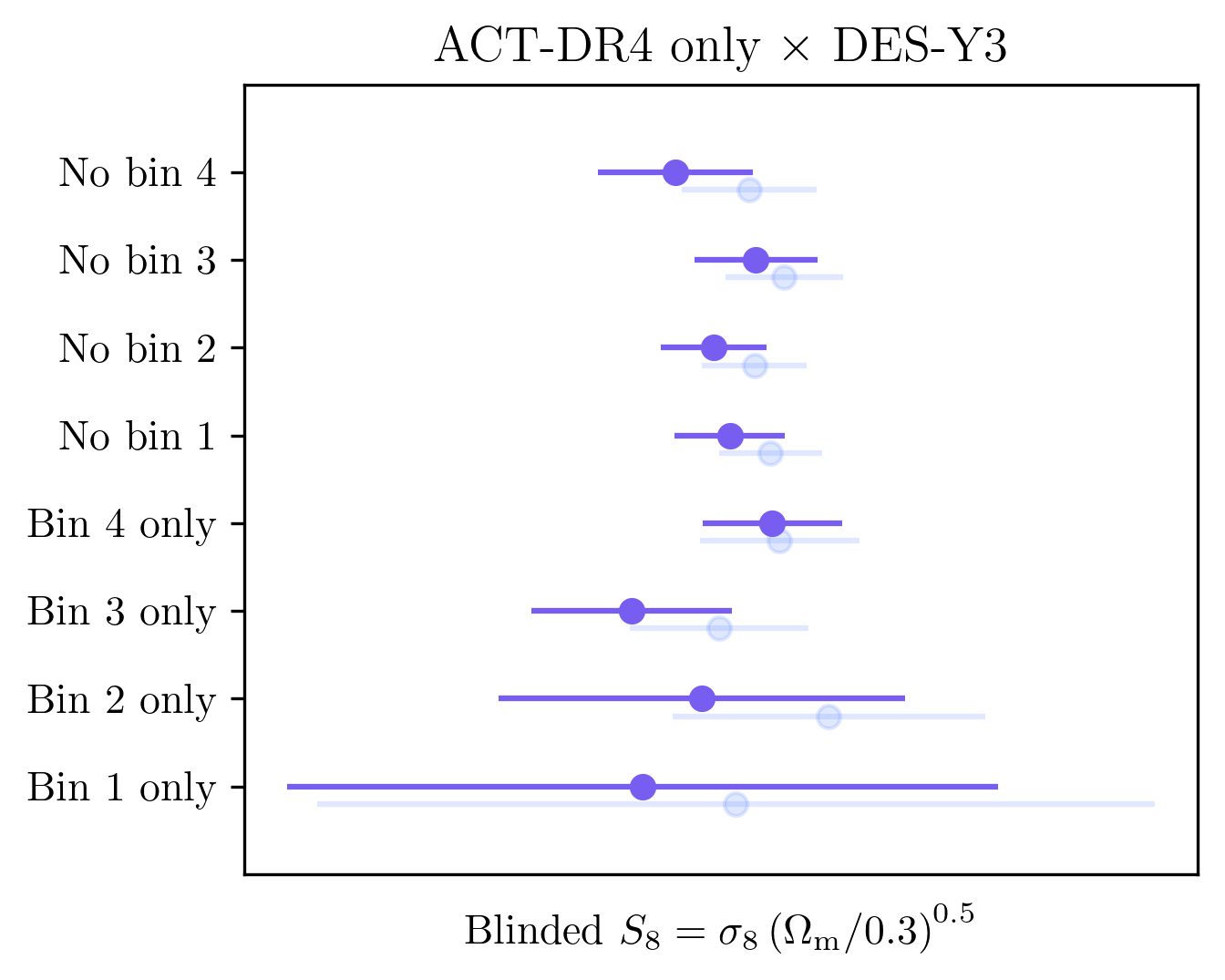}
    \caption{Stability of our 1D marginalised measurement of $S_8$ when using different parts of the full ACT only data vector (results from the fiducial ACT+\planck data vector are shown as points with lighter shade). In each row, we either remove a single DES-Y3 tomographic bin or use the data from only one of the four DES-Y3 bins.}
    \label{fig:split_stability_actonly}
\end{figure}

\begin{figure}
    \centering
    \includegraphics[width=0.45\textwidth]{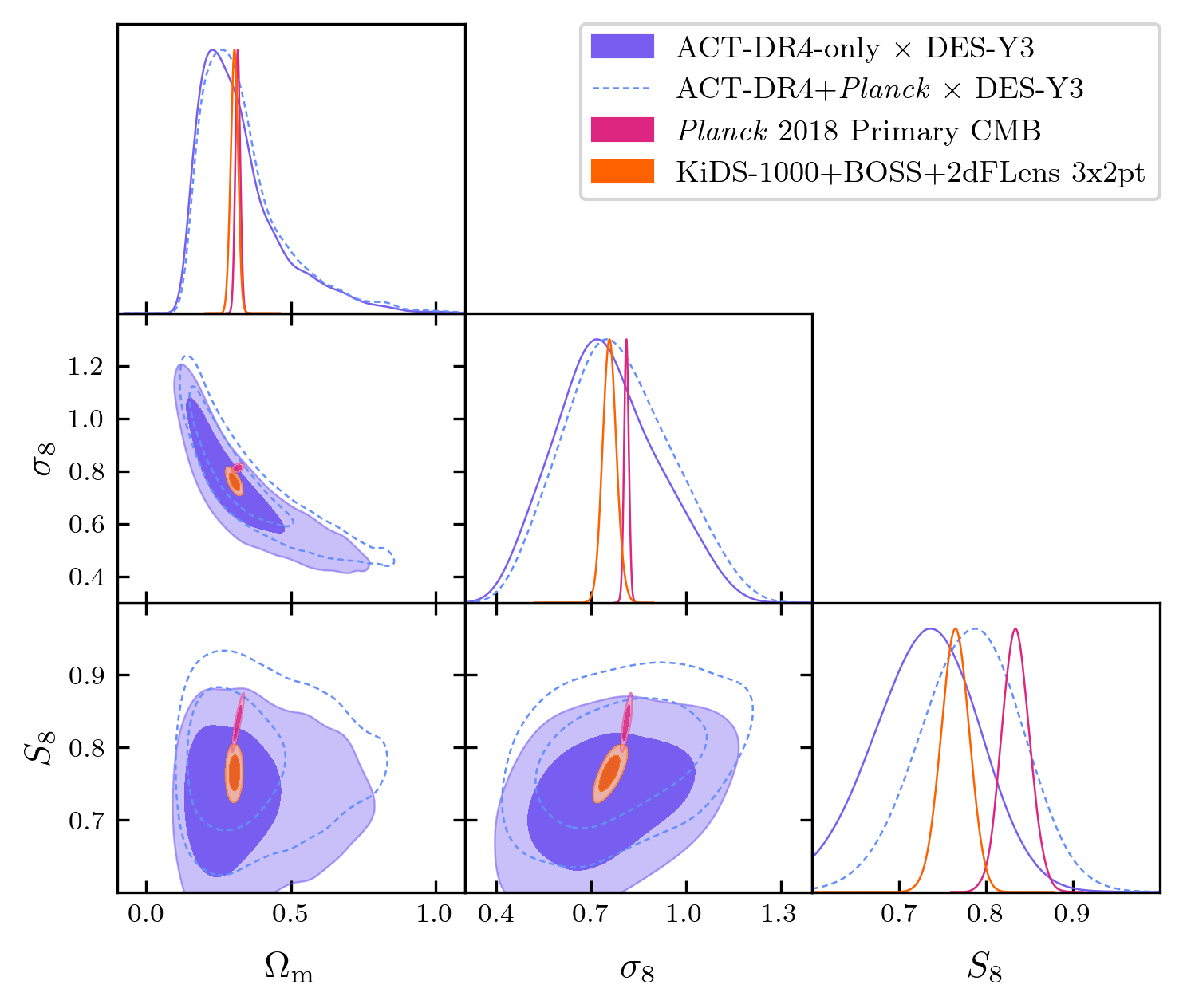}\\
    \includegraphics[width=0.45\textwidth]{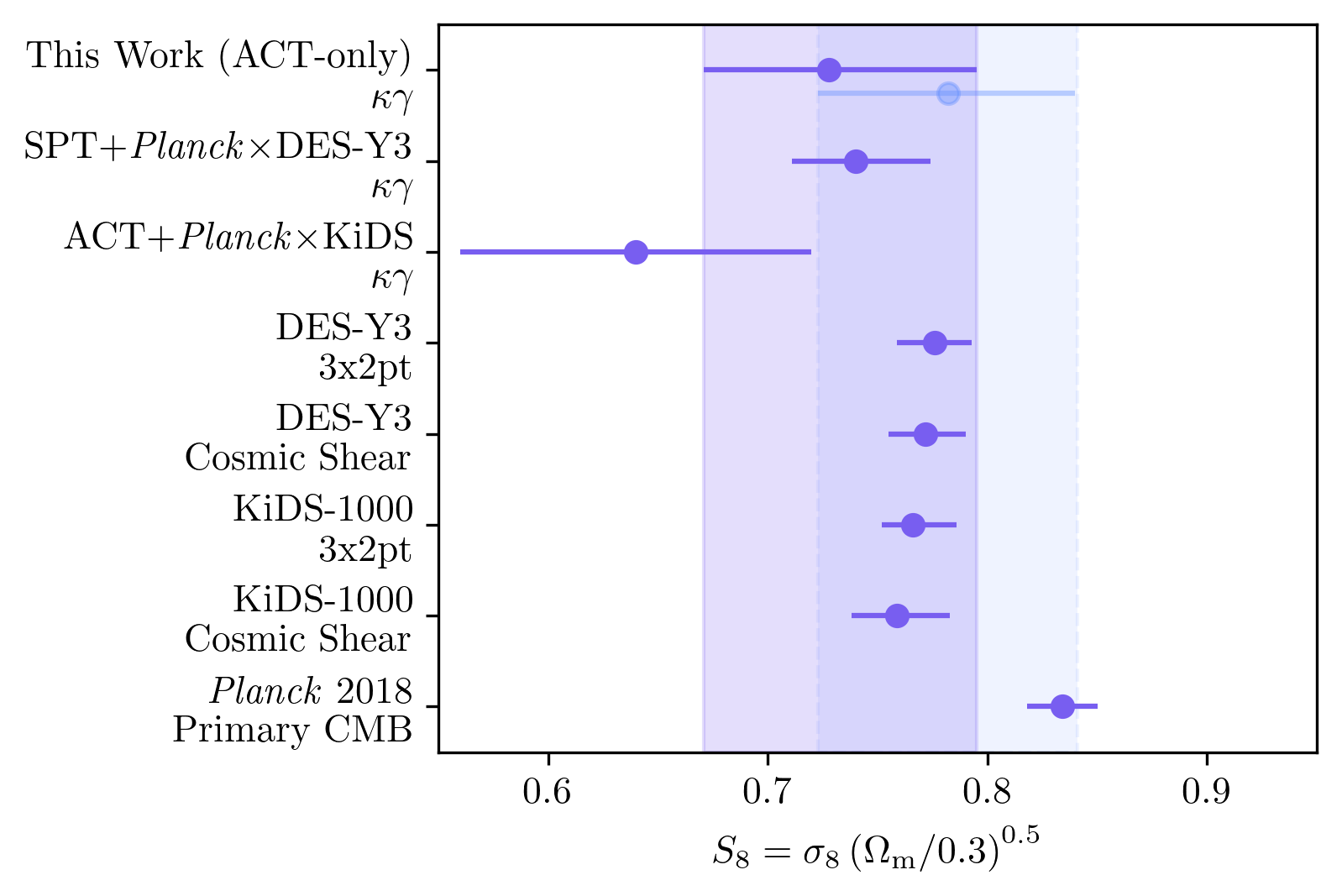}
    \caption{\emph{Top:} The inferred distribution of cosmological parameters $\Omegam, \sigma_8, S_8$ from our ACT-only \clkg data vector alongside two other measurements (the result from the fiducial ACT+\planck data vector is shown as unfilled dashed contours). The ACT-only posterior has $S_{8} = 0.728^{+0.067}_{-0.057}$. \emph{Bottom:} ACT-only $S_8$ alongside a number of other measurements (with the result from the fiducial ACT+\planck data vector shown in a lighter shade). Note that the ACT+\planck$\times$KiDS measurement uses the \kcmb of the BN region of ACT-DR4 data \citep{2021A&A...649A.146R}.}
    \label{fig:s8_compilation_actonly}
\end{figure}

\begin{figure}
    \centering
    \includegraphics[width=0.45\textwidth]{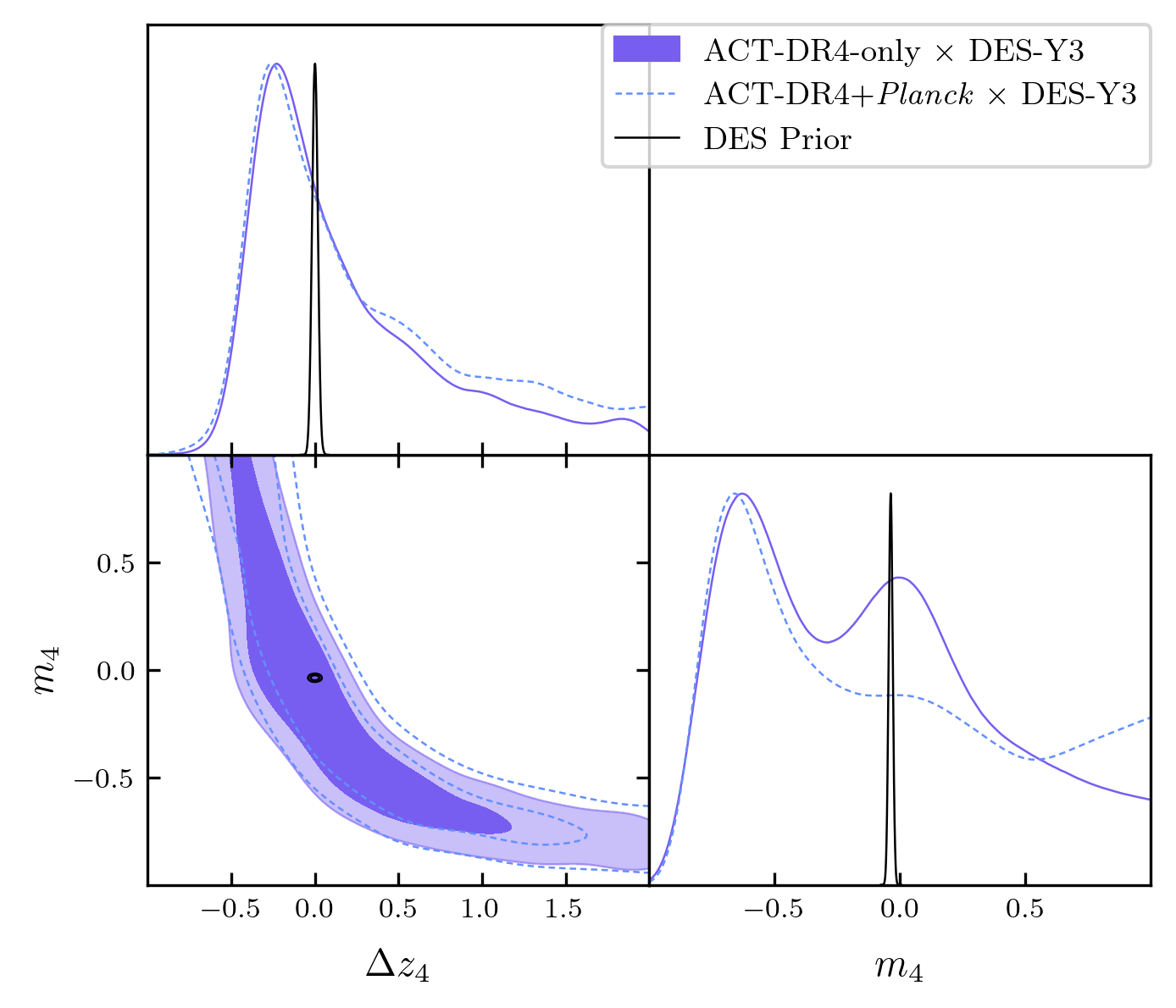}
    \caption{Constraints on the galaxy weak lensing nuisance parameters in DES-Y3 tomographic bin 4 ($0.87 < z < 2.0$) from ACT-only data vector (the result from the fiducial ACT+\planck data vector is shown as the dashed contour). The DES-Y3 prior used in the main analysis is shown as unfilled contours; note the consistency with the filled contours obtained with fixed cosmological parameters but broad priors on nuisance parameters.}
    \label{fig:results_nuisance_actonly}
\end{figure}

The parameter inference with ACT-only \clkg is performed with identical modelling choices as that of ACT+\planck \clkg. \cref{fig:model_stability_actonly} shows the results with simulations of ACT-only \clkg indicating the stability of recovery of the $S_8$ parameter for different models used for inference. \cref{fig:split_stability_actonly} shows the consistency of the $S_8$ inference for different data combinations similar to the case of ACT+\planck as depicted by points with the lighter shade. We show the posterior distribution of the cosmological parameters in \cref{fig:s8_compilation_actonly}. We find that the $\sigma_8$ and $S_8$ inferred with ACT-only \clkg are somewhat smaller than those inferred from ACT+\planck \clkg. This is consistent with what we observe in \cref{fig:compare_ACTonly_AP_data}. ACT-only \clkg has a smaller amplitude than ACT+\planck \clkg at lower multiples, which are multipole bins with higher SNR, resulting in lower $S_8$ inference. The constraints on the shear calibration and photometric uncertainty parameters are similar to those obtained with ACT+\planck \kcmb shown in \cref{fig:results_nuisance_actonly}.

\section{Further Illustrative Posterior Plots}
\label{app:posterior_zooms}
\begin{figure}
    \centering
    \includegraphics[width=0.475\textwidth]{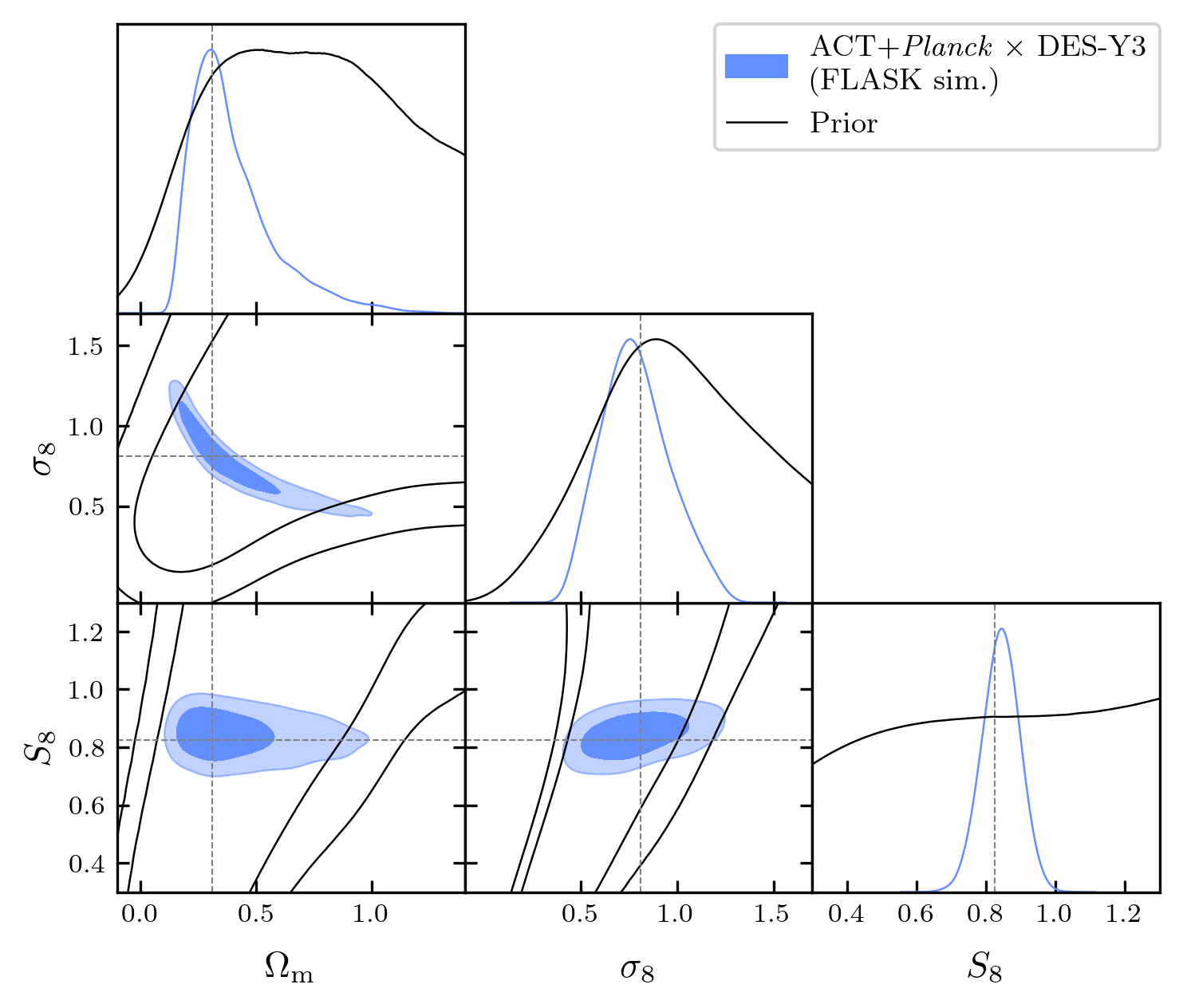}\\
    \caption{Recovery of input model parameters using our pipeline as described in \cref{sec:validation} alongside the 68\% and 95\% contours of the prior (unfilled black contours). These are the same marginalised 1D and 2D posterior distributions shown in \cref{fig:wideia_prior}, but display only the cosmological parameters to increase the clarity of details.}
    \label{fig:model_recovery_cosm}
\end{figure}

\begin{figure}
    \centering
    \includegraphics[width=0.475\textwidth]{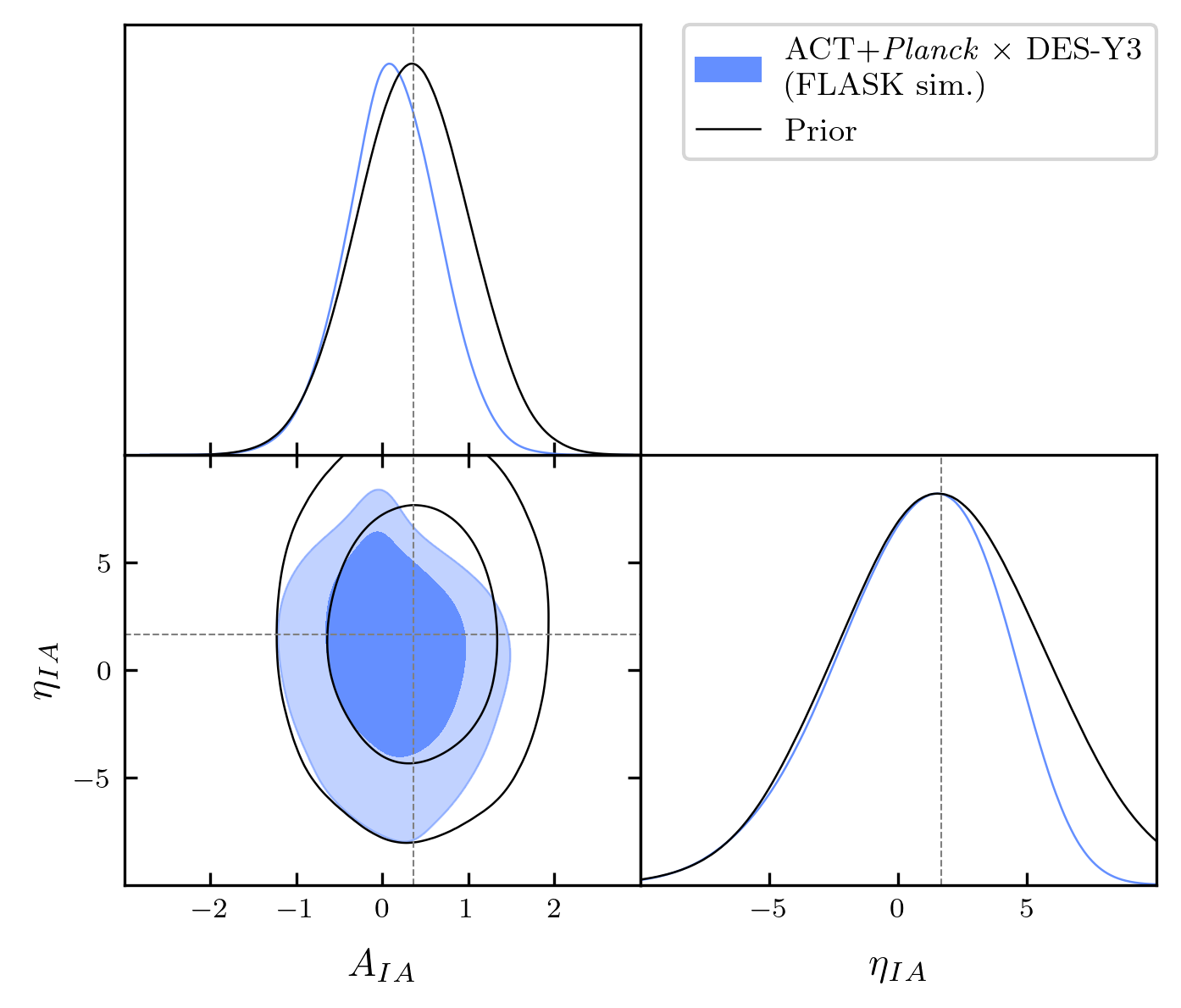}\\
    \includegraphics[width=0.475\textwidth]{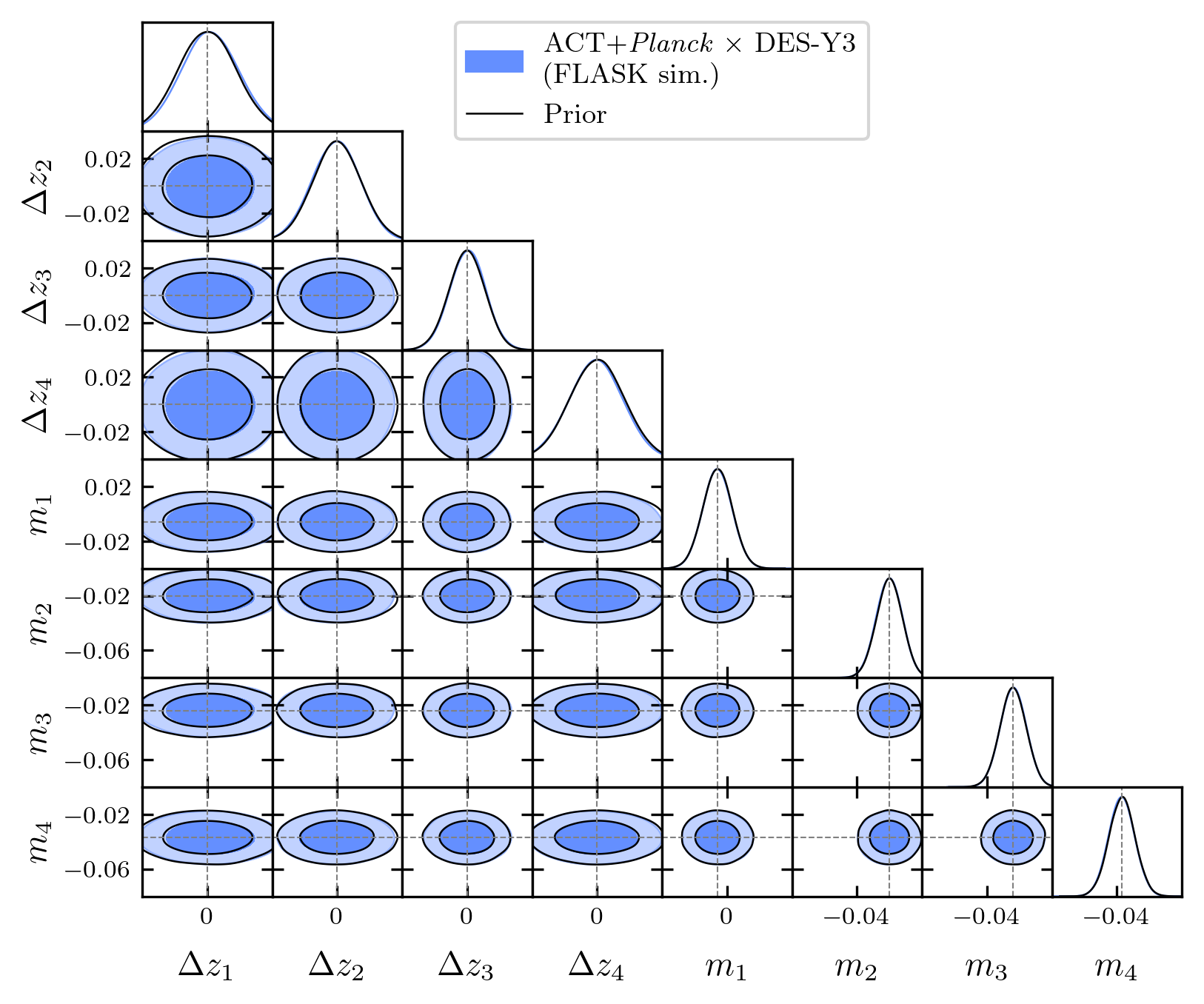}
    \caption{Recovery of input model parameters using our pipeline as described in \cref{sec:validation} alongside the 68\% and 95\% contours of the prior (unfilled black contours). These are the same marginalised 1D and 2D posterior distributions shown in \cref{fig:wideia_prior} but display only the intrinsic alignment parameters (\emph{top}) and weak lensing nuisance parameters (\emph{bottom}) to increase the clarity of details.}
    \label{fig:model_recovery_nuisance}
\end{figure}

\begin{figure*}
\centering
\includegraphics[width = 0.975\textwidth]{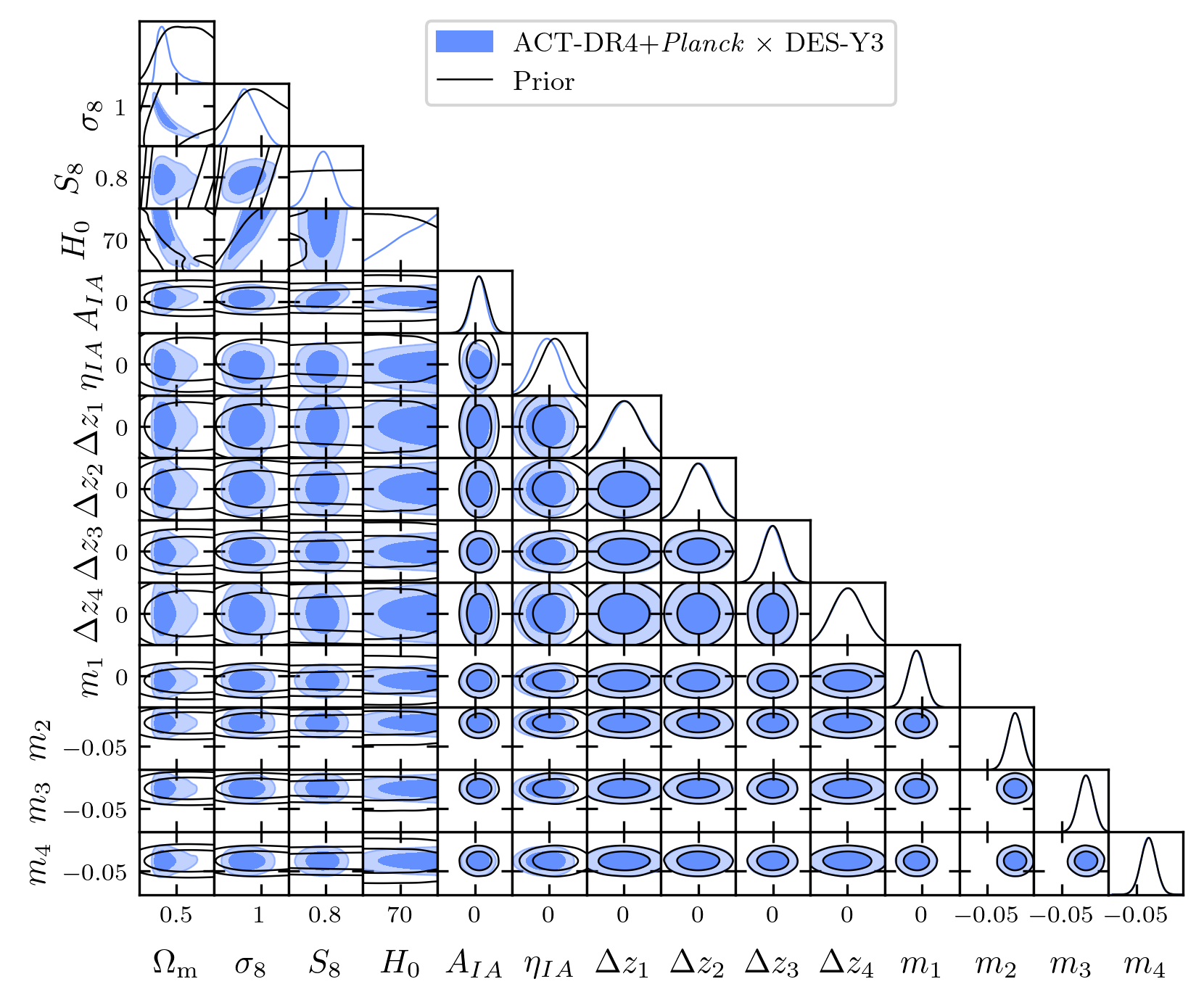}
\caption{The full posterior from our baseline analysis is already shown in part in \cref{fig:result_cosmology}. Black unfilled contours show the priors specified in \cref{tab:params} as they appear in this output parameter space, and blue-filled contours show the recovered posterior constraints.}
\label{fig:results_cosmology_full}
\end{figure*}

In \cref{fig:model_recovery_cosm,fig:model_recovery_nuisance}, we present a complete version of the posterior shown in \cref{fig:wideia_prior} (for the simulated data vector), but in blocks of panels containing only cosmological parameters, galaxy intrinsic alignment parameters and weak lensing nuisance parameters. When presented here, axis limits are changed in order to get a full view of the relevant prior distributions. In \cref{fig:results_cosmology_full}, we show the full posterior from our baseline analysis of the ACT+\planck $\times$ DES-Y3 data, a part of which is already shown in \cref{fig:result_cosmology}.

\bsp	
\label{lastpage}
\end{document}